\documentclass[12pt]{article}
\pdfoutput=1
\usepackage{epsfig,amssymb,amsmath,psfrag,multirow,epstopdf,color,graphicx,caption,subcaption}
\usepackage{breqn}
\usepackage{placeins}

\numberwithin{equation}{section}

\newcommand{\be}{\begin{equation}}
\newcommand{\ee}{\end{equation}}
\newcommand{\bea}{\begin{eqnarray}}
\newcommand{\eea}{\end{eqnarray}}
\newcommand{\eqn}[1]{eq.~\eqref{#1}}
\def\fig#1{fig.~{\ref{#1}}}

\def\sect#1{section~{\ref{#1}}}
\def\eqn#1{eq.~(\ref{#1})}
\def\Eqn#1{Equation~(\ref{#1})}
\def\eqns#1#2{eqs.~(\ref{#1}) and~(\ref{#2})}
\def\Eqns#1#2{Eqs.~(\ref{#1}) and~(\ref{#2})}
\def\tab#1{table~{\ref{#1}}}

\def\Eqn#1{Equation~(\ref{#1})}
\def\eqn#1{eq.~(\ref{#1})}
\def\eqns#1#2{eqs.~(\ref{#1}) and~(\ref{#2})}
\def\Eqns#1#2{Eqs.~(\ref{#1}) and~(\ref{#2})}

\def\cO{{\mathcal O}}

\def\ws{{w^\ast}}
\def\Su{{\mathcal{S}_u}}

\def\to{\rightarrow}
\def\lr{\leftrightarrow}

\def\Qep{Q_{\rm ep}}

\def\beq{\begin{equation}}
\def\eeq{\end{equation}}

\def\bsp#1\esp{\begin{split}#1\end{split}}

\newcommand{\LPA}[2]{[L^+_{#1}]^#2}

\newcommand{\LSPA}[1]{L^+_{#1}}
\newcommand{\LSPB}[2]{L^+_{#1,#2}}
\newcommand{\LSPC}[3]{L^+_{#1,#2,#3}}

\newcommand{\LSPE}[5]{L^+_{#1,#2,#3,#4,#5}}

\newcommand{\LMA}[2]{[L^-_{#1}]^#2}
\newcommand{\LMB}[3]{[L^-_{#1,#2}]^#3}

\newcommand{\LSMA}[1]{L^-_{#1}}
\newcommand{\LSMB}[2]{L^-_{#1,#2}}
\newcommand{\LSMC}[3]{L^-_{#1,#2,#3}}
\newcommand{\LSMD}[4]{L^-_{#1,#2,#3,#4}}

\newcommand{\Enun}[1]{{E_{\nu,n}^#1}}
\newcommand{\EnunOne}[0]{{E_{\nu,n}}}
\newcommand{\dE}[1]{{D_\nu^#1 E_{\nu,n}}}
\newcommand{\dEOne}[0]{{D_\nu E_{\nu,n}}}
\newcommand{\dEP}[2]{{[D_\nu^#1 E_{\nu,n}]^#2}}
\newcommand{\dEPOne}[1]{{[D_\nu E_{\nu,n}]^#1}}

\newcommand{\Ord}{{\cal O}}

\DeclareMathOperator{\sgn}{sgn}

\newcommand{\dnu}[0]{{D_{\nu}}}

\newfont{\scyr}{wncyr10 scaled 550}

\newcommand{\cS}{{\cal S}}

\def\beq{\begin{equation}}
\def\eeq{\end{equation}}
 
\begin{document}

\catcode`\@=11
\font\manfnt=manfnt
\def\Watchout{\@ifnextchar [{\W@tchout}{\W@tchout[1]}}
\def\W@tchout[#1]{{\manfnt\@tempcnta#1\relax%
  \@whilenum\@tempcnta>\z@\do{%
    \char"7F\hskip 0.3em\advance\@tempcnta\m@ne}}}
\let\foo\W@tchout
\def\dubious{\@ifnextchar[{\@dubious}{\@dubious[1]}}
\let\enddubious\endlist
\def\@dubious[#1]{%
  \setbox\@tempboxa\hbox{\@W@tchout#1}
  \@tempdima\wd\@tempboxa
  \list{}{\leftmargin\@tempdima}\item[\hbox to 0pt{\hss\@W@tchout#1}]}
\def\@W@tchout#1{\W@tchout[#1]}
\catcode`\@=12


\thispagestyle{empty}

\null\vskip-60pt \hfill
\begin{minipage}[t]{92mm}
IPPP/14/09\quad DCPT/14/18\quad SLAC--PUB--15902\\ 
LAPTH-010/14\quad CERN-PH-TH/2014-027\\
\end{minipage}
\vspace{5mm}

\begingroup\centering
{\Large\bfseries\mathversion{bold}
The four-loop remainder function\\
and multi-Regge behavior at NNLLA\\
\vskip0.2cm
in planar ${\cal N}=4$ super-Yang-Mills theory}%

\vspace{7mm}

\begingroup\scshape\large
Lance~J.~Dixon$^{(1)}$, James~M.~Drummond$^{(2,3,4)}$,\\
Claude Duhr$^{(5)}$ and Jeffrey Pennington$^{(1)}$\\
\endgroup
\vspace{5mm}
\begingroup\small
$^{(1)}$\emph{SLAC National Accelerator Laboratory,
Stanford University, Stanford, CA 94309, USA} \\
$^{(2)}$\emph{CERN, Geneva 23, Switzerland} \\
$^{(3)}$\emph{School of Physics and Astronomy, University of Southampton\\
Highfield, Southampton, SO17 1BJ, U.K.} \\
$^{(4)}$\emph{LAPTH, CNRS et Universit\'e de Savoie,
F-74941 Annecy-le-Vieux Cedex, France}\\
$^{(5)}$  \emph{Institute for Particle Physics Phenomenology, 
University of Durham,\\ Durham, DH1 3LE, U.K.}
\endgroup

\vspace{0.4cm}
\begingroup\small
E-mails:\\
{\tt lance@slac.stanford.edu}, {\tt drummond@cern.ch}, \\
{\tt claude.duhr@durham.ac.uk}, {\tt jpennin@stanford.edu}\endgroup
\vspace{0.7cm}

\textbf{Abstract}\vspace{5mm}\par
\begin{minipage}{14.7cm}
We present the four-loop remainder function for six-gluon scattering
with maximal helicity violation in planar ${\cal N}=4$ super-Yang-Mills
theory, as an analytic function of three dual-conformal cross ratios.
The function is constructed entirely from its analytic properties,
without ever inspecting any multi-loop integrand.  We employ the
same approach used at three loops, writing an ansatz in terms
of {\it hexagon functions}, and fixing coefficients in the ansatz
using the multi-Regge limit and the operator product expansion in the
near-collinear limit.  We express the result in terms of multiple
polylogarithms, and in terms of the coproduct for
the associated Hopf algebra.  From the remainder function,
we extract the BFKL eigenvalue
at next-to-next-to-leading logarithmic accuracy (NNLLA),
and the impact factor at N$^3$LLA.  We plot the remainder
function along various lines and on one surface, studying
ratios of successive loop orders.  As seen previously through
three loops, these ratios are surprisingly constant over large
regions in the space of cross ratios, and they are not far from the
value expected at asymptotically large orders of perturbation theory.
\end{minipage}\par
\endgroup

\newpage
\tableofcontents

\newpage

\section{Introduction}

Recently there has been great interest in studying perturbative
scattering amplitudes in $\mathcal{N}=4$ super-Yang-Mills theory, both
for their own sake and as prototypes for the kinds of mathematical
functions that will be encountered at the multi-loop level in QCD and
other theories.  In the planar limit of a large number of colors,
$\mathcal{N}=4$ super-Yang-Mills amplitudes are dual to Wilson loops
for closed polygons with light-like edges, and possess a dual
conformal symmetry~\cite{DualPseudoConformal,AMStrong,DualConformal}.
This symmetry is anomalous~\cite{Drummond2007cf},
but the anomaly, as well as various infrared divergences, can be
removed by factoring out the BDS ansatz~\cite{Bern2005iz}.  For the
case of the maximally-helicity-violating (MHV) configuration of
external gluon helicities, the finite
{\it remainder function}~\cite{Bern2008ap} that is left behind is a
function only of the dual conformally invariant cross ratios.
The first scattering amplitude to have nontrivial cross ratios and a
nonvanishing remainder function is the six-point case, corresponding to
a hexagonal Wilson loop, for which there are three such cross ratios.

The remainder function is expected to be a pure transcendental
function with a transcendental {\it weight} $2L$ at loop order $L$.
Examples of transcendental functions include the logarithm (weight 1),
the classical polylogarithms ${\rm Li}_k$ (weight $k$), and products
thereof.  A more general class of transcendental functions is provided
by iterated integrals~\cite{Chen}, or multiple
polylogarithms~\cite{FBThesis,Gonch}.  Other types of functions, such
as elliptic integrals, can appear in scattering amplitudes.  The
two-loop equal-mass sunrise integral is elliptic~\cite{Laporta2004rb}, as
is an integral entering a particular 10-point scattering amplitude in
planar ${\cal N}=4$ super-Yang-Mills theory~\cite{CaronHuot2012ab}.
However, based on a novel form of the planar loop
integrand~\cite{ArkaniHamed2012nw}, and also a recent twistor-space
formulation~\cite{LipsteinMason}, it is expected that all six-point
amplitudes are non-elliptic and can be described in terms of multiple
polylogarithms.

The purpose of this paper is to use the six-point amplitude to demonstrate
the power of a bootstrap~\cite{Dixon2011pw,Dixon2011nj,Dixon2013eka}
for scattering amplitudes in planar
$\mathcal{N}=4$ super-Yang-Mills theory.  This bootstrap operates
at the level of integrated scattering amplitudes, not loop integrands.
It imposes physical constraints at this level, in terms of the
external kinematics alone, in order to uniquely determine the final answer.
The critical assumption is that the amplitude belongs to a certain
space of functions that can be identified at low loop order. 
In the present case it will be a particular class of iterated integrals.
Suppose one can enumerate all such functions and characterize their properties
in the kinematic limits that are needed to impose the physical
constraints.  Then one can write an ansatz for the amplitude as a
linear combination of the functions with unknown coefficients (which
should all be rational numbers).  Physical constraints provide
simple linear equations relating the coefficients.

If the basic ansatz is correct, then the only other question of principle
is whether there is enough ``boundary data''; that is, whether one has
enough physical constraints to fix all the coefficients.\footnote{%
There is also a nontrivial computational question, namely how to most
efficiently generate and impose a large number of constraints on expressions
that can be rather bulky.}
Fortunately, there is a great deal of data indeed.  Much of it comes
from the operator product expansion (OPE) for Wilson loops,
which corresponds to the near-collinear limit of scattering amplitudes.
The OPE was first analyzed by Alday, Gaiotto, Maldacena, Sever and
Vieira~\cite{Alday2010ku,Gaiotto2010fk,Gaiotto2011dt,Sever2011da}.
More recently, even more powerful OPE information has become available
via integrability~\cite{MinahanZarembo,BKS,Beisert2006ez,IntegrabilityReview}.
The application of integrability to the relevant system of flux tube
excitations has been pioneered by Basso, Sever and Vieira
(BSV)~\cite{Basso2010in,Basso2013vsa,Basso2013aha,Basso2014T2}.
We will show that when this data is combined with that from the
multi-Regge limit~\cite{Bartels2008ce,Bartels2008sc,Schabinger2009bb,%
Lipatov2010qg,Lipatov2010ad,Bartels2010tx,Dixon2011pw,Fadin2011we,%
Dixon2012yy},
it is exceedingly powerful, uniquely determining the six-point
remainder function through at least four loops.

The need for a remainder function beginning at six points and two
loops was first identified in the study of the multi-Regge
limit~\cite{Bartels2008ce}, and also from direct numerical evaluation
of the amplitude and hexagonal Wilson loop at finite values of
the cross ratios~\cite{Bern2008ap}.  (There were also previous indications
at strong coupling that a remainder function would be required, at least
in the limit of a large number of external legs~\cite{AMComments}.)
The two-loop hexagon Wilson loop integrals were performed
analytically in terms of multiple polylogarithms~\cite{DelDuca2009au},
and then simplified dramatically to classical polylogarithms using the
notion of the {\it symbol} of a transcendental
function~\cite{Goncharov2010jf}.

Based on the form of the two-loop symbol, it was
conjectured~\cite{Dixon2011pw,Dixon2011nj} that for six-point amplitudes to
all loop orders the transcendental functions entering the
remainder function (and also the next-to-MHV ratio
function~\cite{Dixon2011nj}) should be polylogarithmic functions
whose symbols are made from an alphabet of nine letters, corresponding to
nine projectively-inequivalent differences $z_{ij}$ of projective variables
$z_i$~\cite{Goncharov2010jf}. These letters can also be represented in terms
of momentum twistors~\cite{Hodges2009hk}.
For any weight, there are a finite number of such functions.
Using the symbol, one can enumerate them all, and then impose physical
constraints on a generic linear combination of them.
In this way, the symbol for the three-loop six-point remainder function
was obtained, up to two undetermined parameters~\cite{Dixon2011pw}
which were fixed~\cite{CaronHuot2011kk} using a dual supersymmetry
``anomaly'' equation~\cite{CaronHuot2011kk,Bullimore2011kg}. 

However, the symbol does not determine the full function.
Lower-weight functions multiplied by constant Riemann $\zeta$ values
give rise to pure functions but vanish at the level of the symbol.
In ref.~\cite{Dixon2013eka} it was shown how to identify and fix these
parameters at the level of the full three-loop remainder function.
In this paper, we will follow the same general strategy at four loops.

In fact, two separate strategies were pursued in ref.~\cite{Dixon2013eka}.
One strategy was to pick a particular region in the space of cross ratios,
and promote the symbol to an explicit linear combination of multiple
polylogarithms.  The additional beyond-the-symbol parameters multiply
products of Riemann $\zeta$ values with multiple polylogarithms of lower
weight.  Knowledge of the limiting behavior of the multiple polylogarithms
on certain boundaries of this region can then be used to impose the physical
constraints.  A second strategy is to characterize the remainder
function by its {\it coproduct}.  The coproduct is part of the
Hopf algebra conjecturally satisfied by multiple  
polylogarithms~\cite{Gonch3,Gonch2,Brown2011ik}.
It has been applied to a number of different physical problems
recently~\cite{Duhr2011zq,Duhr2012fh,Chavez2012kn,Drummond2013nda,%
vonManteuffel2013uoa,Schlotterer2012ny}.
In particular, the ``$\{k-1,1\}$''
element of the coproduct of a weight $k$ function specifies all
of its first derivatives in terms of weight $k-1$ transcendental
functions.  One can iterate in the weight, and define a candidate
remainder function in terms of a set of coupled first-order differential
equations.  In the limits relevant for the physical constraints,
the coupled equations can be solved in terms of a simpler set
of transcendental functions, involving harmonic polylogarithms
in a single variable~\cite{Remiddi1999ew}.  In the present work,
we use the multiple-polylogarithm approach to constrain all of the
parameters, and both strategies to examine the limiting behavior of the
uniquely-determined function.

Besides certain standard symmetry and parity constraints,
and the physical constraints to be described shortly,
we also impose a constraint on the final entry of the symbol.
The final entry should be expressible in terms of only six letters
rather than all nine. This constraint comes from a supersymmetric
formulation of the polygonal Wilson loop~\cite{CaronHuot2011ky}
and also from examining the differential equations obeyed by
one-loop~\cite{Drummond2010cz,Dixon2011ng,DelDuca2011wh}
and multi-loop integrals~\cite{Dixon2011pw,Dixon2011nj}
related to $\mathcal{N}=4$ super-Yang-Mills scattering amplitudes.
The final-entry constraint on the symbol corresponds to a differential
constraint we shall impose at function level, which also has a
simple description in terms of the coproduct of the function.

The two limiting regions in which we impose physical constraints on 
the remainder function are the near-collinear limit and the multi-Regge
limit.  In the near-collinear limit, one of the cross ratios vanishes
and the sum of the other two ratios approaches one.
Because the remainder function has a total $S_3$ permutation symmetry
under exchange of the three cross ratios, it does not matter which
cross ratio we take to vanish.  Let's call this variable $v$ for
definiteness, and let $v=T^2+\Ord(T^4)$ as $v$ and $T \to 0$.
Because there is no remainder function
at the five-point level, the six-point remainder function must vanish
as $v\to0$.  The precise way in which it vanishes is controlled
by the OPE.  The first OPE information to be
determined~\cite{Alday2010ku,Gaiotto2010fk,Gaiotto2011dt,Sever2011da}
concerned the leading-discontinuity terms, which correspond to just the
maximum allowed power of $\ln T$ ($\ln^{L-1}T$ at $L$ loops).
Terms with arbitrary power suppression in $T$ can be determined,
as long as they have $L-1$ powers of $\ln T$.
These terms are dictated by the one-loop anomalous dimensions of the operators
corresponding to excitations of the Wilson line, or flux tube.
Higher-loop corrections to anomalous dimensions and OPE coefficients only
generate terms with fewer logarithms of $T$.  At two loops, the
leading discontinuity is the only discontinuity, and it suffices to
completely determine the remainder function~\cite{Gaiotto2010fk}.
At three loops~\cite{Dixon2011pw}, and particularly at four loops,
more information is required.

Recently, Basso, Sever and Vieira~\cite{Basso2013vsa,Basso2013aha,Basso2014T2}
were able to exploit integrability in order to provide much more
OPE information.  They partition a generic polygonal Wilson loop
into a number of ``pentagon transitions'' between flux tube excitations.
They find that certain bootstrap consistency conditions for
the pentagon transitions can be solved in terms of factorizable $S$ matrices
for two-dimensional scattering of the flux tube excitations.
These $S$ matrices are known for finite coupling, and they can be
expanded out in perturbation theory to any desired order.
The powers of $T$ in the OPE expansion correspond to the number
of flux tube excitations.
In their initial papers~\cite{Basso2013vsa,Basso2013aha}, the leading
nonvanishing OPE terms, $\Ord(T^1)$, were described,
corresponding to single excitations.
The $\Ord(T^1)$ information, combined with the multi-Regge
limits and an assumption about the final entry of the symbol, was enough to
completely fix the three-loop remainder function~\cite{Dixon2013eka}.
However, it is not enough at four loops.  Fortunately,
Basso, Sever and Vieira~\cite{Basso2014T2} have also been able
to determine the contributions to the OPE of two flux-excitation states,
and thereby obtain the $\Ord(T^2)$ terms.\footnote{We thank them
for making these results available to us prior to
publication~\cite{BSVPrivate}.}

The $\Ord(T^2)$ terms from the OPE were found to agree perfectly with those
extracted from the three-loop remainder function~\cite{Dixon2013eka}.
Because there were no free parameters in this comparison --- all parameters
had been fixed at $\Ord(T^1)$ --- the agreement is a powerful check
on the assumptions underlying both approaches.
At four loops, we will need to use some of the $\Ord(T^2)$ information,
supplied to us by BSV, to fix a small number of remaining parameters in
the four-loop remainder function --- four parameters in the symbol,
and then one more at the level of the full function. However, there is
considerably more information in the $\Ord(T^2)$ OPE expansion, and
so the fact that it agrees between our approach and BSV's at four loops
is certainly a strong indication that both approaches are correct.

The other physical limit which can be used to constrain the remainder
function is the multi-Regge limit.  In this limit, two incoming gluons
scatter into four gluons that are well separated in rapidity.
Whereas the near-collinear OPE limit can be approached in the Euclidean
region, this kinematical configuration is in Minkowski space.  Coming from
the Euclidean region, one first needs to analytically continue to
Minkowski space by rotating the phase of one of the cross ratios,
let's call it $u$, by $2\pi$. 
Then $u$ should be taken to unity at the same rate
that the other two cross ratios, $v$ and $w$, vanish.
The analytic continuation in $u$ generates an imaginary part, as well as
a real part from a double discontinuity.  Both the imaginary
and real parts diverge as powers of $\ln(1-u)$ as $u\to1$.
The leading logarithmic approximation (LLA) has a behavior
proportional to $\ln^{L-1}(1-u)$ at $L$ loops, and it is pure
imaginary~\cite{Bartels2008ce,Bartels2008sc,Schabinger2009bb,%
Lipatov2010qg}.

It has been proposed that factorization in the multi-Regge
limit can be extended to subleading logarithmic
accuracy~\cite{Lipatov2010ad,Fadin2011we,CaronHuot2013fea}.
In ref.~\cite{Fadin2011we} the functions that should control
the factorization were computed directly through the
next-to-leading-logarithmic approximation (NLLA).  In
ref.~\cite{CaronHuot2013fea} a closely-related form of multi-Regge
factorization was proposed, based on the hypotheses
of rapidity factorization and the completeness of a description
in terms of undecorated, null, infinite Wilson lines.
In principle, if these hypotheses are true, then the factorization could
hold to arbitrary subleading logarithmic order, up to
terms that are power-suppressed like $\Ord(1-u)$.
In this paper, we will assume that the factorization holds through
arbitrary subleading logarithmic accuracy.  In practice, our four-loop results
are sensitive to at most N$^3$LLA.  The fact that we find a consistent solution
provides evidence in favor of factorization beyond NLLA.

The assumption of factorization
makes it possible to bootstrap multi-Regge information
from one loop order to the next.  That is, the leading-logarithmic behavior
of the remainder function is present already at two
loops~\cite{Bartels2008ce,Bartels2008sc} and can be used to
predict the LLA $\ln^{L-1}(1-u)$ behavior at three~\cite{Lipatov2010ad}
and higher loops~\cite{Dixon2012yy}.
Similarly, the NLLA
behavior~\cite{Lipatov2010ad,Fadin2011we}
first appears fully at three loops, and can be used to
predict the $\ln^{L-2}(1-u)$ behavior at four and higher
loops~\cite{Dixon2012yy}.

The factorization takes place in variables which are related
to the original variables by a Fourier-Mellin
transform~\cite{Fadin2011we}.
Two functions control the expansion:  the BFKL eigenvalue
and the impact factor.  Each function has an expansion in the coupling;
successive orders in the expansion are needed for higher accuracy
in the logarithmic expansion.   The N$^k$LLA term in the impact factor makes
its first appearance in the remainder function in the $\ln^0(1-u)$ term
at $k+1$ loops; whereas the N$^k$LLA term in the BFKL eigenvalue
appears one loop order later, at $k+2$ loops, accompanied by
one power of $\ln(1-u)$.

In ref.~\cite{Dixon2012yy} it was observed that in the multi-Regge limit
the coefficients in the expansion of remainder function in powers
of $\ln(1-u)$ are single-valued harmonic polylogarithms (SVHPLs),
first introduced by Brown~\cite{BrownSVHPLs}.
Based on this observation, techniques for performing the inverse
Fourier-Mellin transform were developed, in order to efficiently
find the consequences of the N$^k$LLA approximation for the
remainder function at a given loop order.
Furthermore, part of the program of this paper to determine
the four-loop remainder function was carried out there~\cite{Dixon2012yy}:
Several constraints were applied to the relevant space of symbols:
$S_3$ symmetry, parity, the OPE leading discontinuity and the final-entry
condition were applied.  These constraints left 113 symbol-level parameters
undetermined.  However, in the multi-Regge limit only one symbol-level
parameter, called $a_0$, survived.  This allowed the NNLLA BFKL
eigenvalue and N$^3$LLA impact factor to be almost completely
constrained at symbol level.  At function level, however, there were
an additional 26 undetermined rational numbers in the multi-Regge limit.

In the continuation of this program in the present paper, we apply
additional multi-Regge constraints from NLLA~\cite{Fadin2011we} that we did
not impose earlier, in order to fix 33 of the 113 remaining
symbol-level parameters. Then we match the $\Ord(T^1)$ and $\Ord(T^2)$ 
behavior to the OPE~\cite{Basso2013vsa,Basso2013aha,Basso2014T2}, to fix
the final 80 symbol-level parameters.  We then account for 68 additional
beyond-the-symbol parameters, and fix them all using the same OPE information,
which we now implement at the level of full functions using the
multiple-polylogarithmic representation.  With the remainder function
uniquely determined, we return to the Minkowski multi-Regge limit and
determine the values of the 27 parameters we had previously introduced.
This completes the determination of the NNLLA BFKL eigenvalue and
N$^3$LLA impact factor begun in ref.~\cite{Dixon2012yy}.
We find that the NNLLA BFKL eigenvalue has a very suggestive form that
is closely related to the spectrum of anomalous dimensions
for flux tube excitations~\cite{Basso2010in}.

We then study the quantitative behavior of the four-loop remainder
function in various regions, including special lines in the space
of cross ratios where it collapses to linear combinations of harmonic
polylogarithms of a single variable.  We will explore various numerical
observations made at three loops in ref.~\cite{Dixon2013eka} about
the sign and constancy of ratios of successive loop orders.
We will find that these observations remain true, and are even
reinforced at four loops.  We will also discuss how close the remainder
function at four loops might be, in a certain region, to its expected
behavior at large perturbative orders.

The remainder of this paper is organized as follows.  In
\sect{sec:construction} we describe the construction of
the four-loop remainder function.  In \sect{sec:MRK} we
describe its behavior in the multi-Regge limit and extract the
NNLLA BFKL eigenvalue and N$^3$LLA impact factor.
In \sect{sec:quant} we explore the sign of the four-loop remainder
function in a certain ``positive'' region.
We plot the ratio of successive loop orders on a two-dimensional surface,
and on various lines where its functional form simplifies considerably,
as well as discussing expectations for large perturbative orders.
Finally, in \sect{sec:conclusions} we conclude and discuss
avenues for future research.
We include one appendix on the coproduct representation,
and a second one characterizing logarithmic divergences of the remainder
function on two particular boundaries of the Euclidean region.

Many of the analytic results in this paper are too lengthy to present in
the manuscript.  Instead we provide a set of ancillary files in
computer readable format. 

\section{The construction}
\label{sec:construction}


\subsection{Hexagon functions}
\label{sec:hexagon}

The six-point remainder function $R_6$ is defined by factoring off
the BDS ansatz from the MHV planar amplitude,
\beq
A_6^{\rm MHV} = A_6^{\rm BDS} \times \exp(R_6) \,.
\label{removeBDS}
\eeq
The BDS ansatz accounts for all of the amplitude's infrared divergences,
or ultraviolet divergences in the case of the Wilson loop interpretation.
It also absorbs the (related) anomaly in dual conformal transformations.
Because $R_6$ is invariant under such transformations,
it can only depend on the dual conformal cross ratios,
\beq\label{eq:uvw_def}
u = \frac{x_{13}^2\,x_{46}^2}{x_{14}^2\,x_{36}^2}\,, 
\qquad v = \frac{x_{24}^2\,x_{51}^2}{x_{25}^2\,x_{41}^2}\,, \qquad
w = \frac{x_{35}^2\,x_{62}^2}{x_{36}^2\,x_{52}^2}\,,
\eeq
where $x_i$, $1\le i\le6$, denote dual coordinates, related to the external
momenta by $p_i = x_{i}-x_{i+1}$. The six-point remainder function admits the
perturbative expansion, starting at two loops,\footnote{%
Beginning at four loops, it is important to specify whether or not $R_6$ is
exponentiated in the definition~(\ref{removeBDS}), because the two
alternative definitions would differ by $\tfrac{1}{2}[R_6^{(2)}]^2$ at this
order.}
\beq
R_6(u,v,w) = \sum_{L=2}^\infty a^L\,R_6^{(L)}(u,v,w)\,,
\eeq
where $a=g_{\textrm{YM}}^2N_c/(8\pi^2)$ is the 't Hooft coupling
constant, $g_{\textrm{YM}}$ is the Yang-Mills coupling constant and
$N_c$ is the number of colors.

The coefficients $R_6^{(L)}(u,v,w)$ are expected to be pure functions
of transcendental weight $2L$, \emph{i.e.}, they should be
$\mathbb{Q}$-linear combinations of polylogarithmic functions of
weight $2L$. For this reason, it is convenient to consider the symbol
of $R_6^{(L)}(u,v,w)$. The symbol of a transcendental function $f^{(k)}$
of weight $k$ can most conveniently be defined as follows: if the
total differential of $f^{(k)}$ can be written as a finite
sum of the form
\beq
df^{(k)} = \sum_{r} f_r^{(k-1)} \, d\ln\phi_r\,,
\eeq
where the $\phi_r$ are rational functions and the
$f_r^{(k-1)}$ are transcendental functions of weight $k-1$,
then the symbol of $f^{(k)}$ can be defined recursively by,
\beq
\cS(f^{(k)}) = \sum_r \cS(f_r^{(k-1)}) \otimes \phi_r \,.
\eeq
The six-point remainder function for arbitrary values of the cross
ratios is currently known at two~\cite{DelDuca2009au,Goncharov2010jf}
and three loops~\cite{Dixon2011pw,Dixon2013eka}. One of the main
results of this paper is to present the fully analytic answer for the
four-loop remainder function $R_6^{(4)}(u,v,w)$. The construction of
the result will be performed following closely the ideas of
ref.~\cite{Dixon2013eka}, which allow us to bootstrap the four-loop
answer without ever inspecting the multi-loop integrand.
This bootstrap will be described in the remainder of this section.

In ref.~\cite{Dixon2013eka}, a set of polylogarithmic functions 
called {\it hexagon functions} were introduced. Their symbols are 
built out of the nine letters,
\be
\Su = \{u,v,w,1-u,1-v,1-w,y_u,y_v,y_w\}\,,
\label{nineletters}
\ee
where
\be
y_u = \frac{u-z_+}{u-z_-}\,, \qquad y_v = \frac{v-z_+}{v-z_-}\,, 
\qquad y_w = \frac{w - z_+}{w - z_-}\,,
\label{yfromu}
\ee
and
\be
z_\pm = \frac{1}{2}\Bigl[-1+u+v+w \pm \sqrt{\Delta}\Bigr]\,, 
\qquad \Delta = (1-u-v-w)^2 - 4 uvw\,.
\label{zDeltadef}
\ee
(We sometimes also use the labeling $u_1=u$, $u_2=v$, $u_3=w$,
$y_1=y_u$, $y_2=y_v$, $y_3=y_w$.)
The branch cut locations for hexagon functions are restricted to the
points where the cross ratios $u_i$ either vanish or approach infinity.
In terms of the symbol, this implies~\cite{Gaiotto2011dt}
that the first entry must be one of the cross ratios $u,v,w$.

In ref.~\cite{Dixon2013eka}, a method based on the coproduct on
multiple polylogarithms
(or, equivalently, a corresponding set of first-order partial
differential equations) was developed that allows for the construction
of hexagon functions at arbitrary weight. Using this method, the
three-loop remainder function was determined as a particular
weight-six hexagon function.
In this article, we extend the analysis and construct the four-loop
remainder function, which is a hexagon function of weight eight. 


\subsection{Constraints at symbol level}
\label{sec:symbolconstraints}

As in the three-loop case, we begin by constructing the symbol. Referring
to the discussion in ref.~\cite{Dixon2012yy}, the symbol may be written as
\beq\label{eq:4-loop-symbol2}
\cS(R_6^{(4)}) = \sum_{i=1}^{113}\alpha_i\,S_i\,,
\eeq
where $\alpha_i$ are undetermined rational numbers. The $S_i$ are
drawn from the complete set of eight-fold tensor products
(\emph{i.e.}~symbols of weight eight) that satisfy the first-entry condition.
They also are required to obey the following properties:
\begin{enumerate}
\setcounter{enumi}{-1}
\item All entries in the symbol are drawn from the set
$\{u_i,1-u_i,y_i\}_{i=1,2,3}$.
\item The symbol is integrable
(\emph{i.e.}~it is the symbol of some function).
\item The symbol is totally symmetric under $S_3$ permutations
of the three cross ratios $u_i$.
\item The symbol is invariant under the parity transformation $y_i\to 1/y_i$. 
\item The symbol vanishes in the collinear limit $u_2\to0$, $u_1+u_3\to1$.
(The other two collinear limits follow from the $S_3$ symmetry.)
\item In the near-collinear limit, the symbol agrees with the predictions
of the leading discontinuity terms in the OPE~\cite{Alday2010ku}.
We implement this condition exactly as was done at three
loops~\cite{Dixon2011pw}.
\item The final entry of the symbol is drawn from the set 
$\{u_i/(1-u_i), y_i\}_{i=1,2,3}$.
\end{enumerate}
Imposing the above constraints on the most general ansatz of all
$9^8$ possible words will yield \eqn{eq:4-loop-symbol2}; however,
performing the linear algebra on such a large system is
challenging. Therefore, it is useful to employ the shortcuts described
in refs.~\cite{Dixon2012yy,Dixon2013eka}: the first- and second-entry
conditions\footnote{%
The second entry must be drawn from the set $\{u_i,\,1-u_i\}$.
This restriction follows from the first-entry condition and the requirement
that the symbol be integrable, when the integrability condition on pairs
of adjacent entries is applied to the first two
entries~\cite{Gaiotto2011dt,Dixon2011pw,Dixon2013eka}.}
reduce somewhat the size of the initial ansatz, and
applying the integrability condition iteratively softens the
exponential growth of the ansatz with the weight. Even still, the
computation requires a dedicated method, since out-of-the-box linear
algebra packages cannot handle such large systems. We implemented a
batched Gaussian elimination algorithm, performing the back
substitution with {\tt FORM}~\cite{Vermaseren2000nd}, similar to the
method described in ref.~\cite{Blumlein2009cf}.

As discussed in ref.~\cite{Dixon2012yy}, the factorization formula of
Fadin and Lipatov~\cite{Fadin2011we} in the multi-Regge limit
(see \sect{sec:NNLLBFKL}) provides additional constraints on the
113 parameters entering~\eqn{eq:4-loop-symbol2},
\begin{enumerate}
\setcounter{enumi}{6}
\item The symbol agrees with the prediction from BFKL factorization
at NLL~\cite{Fadin2011we}.
\end{enumerate}

We may also apply constraints in the near-collinear limit by matching
onto the recent predictions by BSV based on
the OPE for flux tube excitations~\cite{Basso2013vsa},
\begin{enumerate}
\setcounter{enumi}{7}
\item The symbol is in agreement to order $T^1$ with the OPE prediction
of the near-collinear expansion~\cite{Basso2013vsa,Basso2013aha}.
\item The symbol is in agreement to order $T^2$ with the OPE prediction
of the near-collinear expansion~\cite{Basso2014T2,BSVPrivate}.
\end{enumerate}
The dimension of the ansatz for the symbol after applying each of these
constraints successively is summarized in
table~\ref{tab:multi_loop_symbol}.  In this table, we also provide
the corresponding numbers at two and three loops, so that one can
appreciate the increased computational complexity of the four-loop
problem. It is worth noting that some constraints become even stronger
when promoted to function-level properties, not only fixing beyond-the
symbol terms, but also implying additional relations on the symbol-level
parameters. An example of this was already seen at three
loops~\cite{Dixon2011pw} where, ultimately, only a single free parameter
remained to be determined by the $O(T^1)$ near-collinear
limit~\cite{Dixon2013eka}.  \\

\begin{table}[!ht]
\begin{center}
\begin{tabular}{|l|c|c|c|}
\hline\hline
\multicolumn{1}{|c|}{Constraint} &\multicolumn{1}{c|}{$L=2$ Dim.}
&\multicolumn{1}{c|}{$L=3$ Dim.} &\multicolumn{1}{c|}{$L=4$ Dim.} \\
\hline\hline
1. Integrability & 75 & 643 & 5897  \\
\hline
2. Total $S_3$ symmetry & 20 & 151 & 1224 \\
\hline
3. Parity invariance & 18 & 120 & 874 \\
\hline
4. Collinear vanishing ($T^0$) & 4 & 59 & 622 \\
\hline
5. OPE leading discontinuity & 0 & 26 & 482 \\
\hline
6. Final entry & 0 & 2 & 113  \\
\hline
7. Multi-Regge limit & 0 & 2 & 80 \\
\hline
8. Near-collinear OPE ($T^1$) & 0 & 0 & 4\\
\hline
9. Near-collinear OPE ($T^2$) & 0 & 0 & 0\\
\hline\hline
\end{tabular}
\caption{\label{tab:multi_loop_symbol} For loop order $L=2,3,4$, we
tabulate the dimensions of the space of symbols with weight $2L$ and
first entry belonging to $\{u,v,w\}$, after applying the various
constraints successively. The final four-loop symbol is uniquely determined,
including normalization, after applying the final constraint, so the vector
space of possible solutions has dimension zero.}
\end{center}
\end{table}

In ref.~\cite{Dixon2013eka}, the last two constraints were applied at
function level to fully determine the three-loop remainder
function. In fact, we will soon apply them at function level in the
four-loop case as well, but first we will apply them at symbol level
in order to determine the constants not fixed by the first seven
constraints. For this purpose, it is necessary to expand the
symbol $\cS(R_6^{(4)})$ in the near-collinear limit $v\to0$, $u+w\to1$.
Because we are comparing to OPE information from ref.~\cite{Basso2013vsa},
it is convenient to adopt the parametrization used there
in terms of variables $F$, $S$, and $T$, which are related to the $u_i$
and $y_i$ variables by,
\be
\bsp
u &= \frac{FS^2}{(1+T^2)(F+FS^2+ST+F^2ST+FT^2)} \,,\\
v &= \frac{T^2}{1+T^2} \,,\\
w &= \frac{F}{F+FS^2+ST+F^2ST+FT^2} \,,\\
y_u &= \frac{FS+T}{F(S+FT)} \,,\\
y_v &= \frac{(S+FT)(1+FST+T^2)}{(FS+T)(F+ST+FT^2)} \,,\\
y_w &= \frac{F+ST+FT^2}{F(1+FST+T^2)} \,.
\esp
\label{BSVparam}
\ee
The near-collinear limit is the limit $T\to0$ for fixed $F$ and $S$.


\subsection{Expanding the symbol in a limit}
\label{sec:symbolexpansion}

We wish to expand symbols and functions in a particular
kinematic limit, which in the present case is $T\to0$.
To this end, we formulate the expansion of an arbitrary pure
function $F(T)$ in a manner that can easily be extended to the symbol. 
The function may contain arbitrary dependence on $S$ and $F$, which
is not shown explicitly.  The expansion is not entirely trivial because
it will in general contain powers of $\ln T$, as well as powers of $T$,
and some care must be taken to keep track of them. 
Let us explicitly separate the power-law behavior from the logarithmic
behavior by writing,
\beq
F(T) = \Big[F(T)\Big]_0 + \Big[F(T)\Big]_1 + \Big[F(T)\Big]_2 + \ldots,
\label{FTi}
\eeq
where $[\cdot]_i$ indicates the $T^i$ power-law term of the expansion
of $F(T)$ around $T=0$.  For example, if
\beq
F(T) = \ln^2T + \ln T \ln S + T F \Bigl( \ln T + \ln S \Bigr)
 + T^2 \ln S + \ldots,
\label{FTex}
\eeq
then we have
\beq\bsp
\Big[F(T)\Big]_0 &\,= \ln^2T + \ln T \ln S\,, \label{FT0ex}\\
\Big[F(T)\Big]_1 &\,= T F \Bigl( \ln T + \ln S \Bigr)\,,\\
\esp\eeq
and so forth.

Now consider a pure function $F(T)$ for which $F(0) = [F(T)]_0 = 0$. 
The function can contain powers of $\ln T$ in the expansion around $T=0$,
as long as they are accompanied by positive powers of $T$ so that the limit
as $T\to 0$ vanishes.  Because symbols provide information about the
derivatives of functions in a convenient way, we write $F$
as the integral of its derivative, to leading order in the expansion
around $T=0$,
\begin{equation}
\label{recurs}
\Big[F(T)\Big]_1 = \int_{0}^{T} dT_1\, \Big[F'(T_1)\Big]_0 \,,
\end{equation}
Owing to the presence of logarithms, it is possible that in
evaluating $[F'(T)]_0$ we might generate a pole in $T$. We let
\begin{equation}
F'(T)=\frac{f_{-1}(T)}{T} + f_0(T) + \mathcal{O}(T^1)\,,
\end{equation}
where the first term comes from differentiating explicit $\ln T$
factors in $F(T)$.  Then we can write the expansion
of the integrand in \eqn{recurs} as
\begin{equation}
\Big[F'(T)\Big]_0 = \frac{1}{T}\,\Big[f_{-1}(T)\Big]_1 + \Big[f_0(T)\Big]_0\,.
\end{equation}
Notice that $f_{-1}(0)=0$ (since otherwise $F(0)\neq 0$), so we can
calculate $[f_{-1}(T)]_1$ by again applying eq.~(\ref{recurs}), this
time with $F\to f_{-1}$. Therefore~\eqn{recurs} defines a recursive
procedure for extracting the first term in the expansion around
$T=0$. The recursion will terminate after a finite number of steps for
a pure function.

The only data necessary to execute this procedure are the ability to
evaluate the function when $T=0$, and the ability to take
derivatives. Since both of these operations carry over to the symbol,
we can apply this method directly to $\cS(R_6^{(4)})$.  To give a flavor
of how the recursion works, we expand the symbol in the following way,%
\begin{equation}
\label{R64symbexpand}
\cS(R_6^{(4)})= [\hat{A}_0\otimes R_0] + [\hat{A}_1\otimes R_1] \otimes T
+ [\hat{A}_2\otimes R_2] \otimes T\otimes T
+ [\hat{A}_3\otimes R_3] \otimes T\otimes T \otimes T + \ldots \,,
\end{equation}
where we write schematically $[\hat{A}_i\otimes R_i]$ for a sum of
terms of the form $\hat{A}_i\otimes R_i$ in which $R_i\neq T$ is defined
to have length one and the $\hat{A}_i$ have length $7-i$.
There are terms with up to six consecutive $T$ entries
in the final slots. Although we have made explicit the $T$ entries at the
back end of the symbol, there may be other $T$ entries hidden inside the
$\hat{A}_i$. Applying eq.~(\ref{recurs}), we obtain,
\begin{equation}
\label{R64_1}
\begin{split}
\Big[\mathcal{S}(R_6^{(4)})\Big]_1 &=
\int_0^T dT_0\, \Big[\frac{R_0'(T_0)}{R_0(T_0)}A_0(T_0)\Big]_0 
\; + \; \int_0^T \frac{dT_0}{T_0}
\int_0^{T_0} dT_1 \,\Big[\frac{R_1'(T_1)}{R_1(T_1)}A_1(T_1)\Big]_0\\
&\quad + \int_0^T \frac{dT_0}{T_0}
\int_0^{T_0} \frac{dT_1}{T_1} 
\int_0^{T_1} dT_2\,\Big[\frac{R_2'(T_2)}{R_2(T_2)}A_2(T_2)\Big]_0\\
&\quad+\int_0^T \frac{dT_0}{T_0}
\int_0^{T_0}\frac{dT_1}{T_1} \int_0^{T_1}\frac{dT_2}{T_2} 
\int_0^{T_2} dT_3 \,\Big[\frac{R_3'(T_3)}{R_3(T_3)}A_3(T_3)\Big]_0 + \ldots\,,
\end{split}
\end{equation}
where the $A_i$ schematically denote functions whose symbols
are the $\hat{A}_i$.
As indicated by the brackets $[.]_0$, the integrands should be
expanded around $T=0$ to order $T^0$.

The coefficients $[A_i(T_i)]_0$ in \eqn{R64_1} are functions of $S$ and $T_i$,
which are obtained from the symbols $\hat{A}_i$ in \eqn{R64symbexpand}
as follows:  One first separates out all the explicit $T$ entries
in $\hat{A}_i$, which all originate from $v$ entries after making
the substitution~(\ref{BSVparam}).  Then one
sets $T$ to zero everywhere in $\hat{A}_i$ except for the explicit $T$
entries.  The explicit factors of $T$ in the symbol
give rise to logarithms of $T_i$ in the function $[A_i(T_i)]_0$.
They appear in the symbol shuffled (summed over appropriate permutations)
together with functions of $S$.  (The variable $F$ disappears from
the symbol when $T$ is set to zero.)  For example,
\bea
\cS(\tfrac{1}{2} \, \ln T \, \ln^2 S) 
= S \otimes S \otimes T + S \otimes T \otimes S 
+ T \otimes S \otimes S\,.
\label{shuffle_example}
\eea
It is straightforward to extract the powers of $\ln T$ by
reversing such relations, {\it i.e.}~unshuffling the factors of $T$
from $[\hat{A}_i]_0$.
Performing this extraction, and setting $T\to T_i$, we obtain 
the functions $[A_i(T_i)]_0$.  At this point the integrations over $T_i$ can
be performed. 
It should be clear how to extend \eqn{R64_1} to the terms in $\cS(R_6^{(4)})$
that have more factors of $T$ on the back end.
Notice that the innermost integrals
have no $1/T_i$ in the measure, and as such they will generate terms
of mixed transcendentality.  The mixed transcendentality is not
surprising; indeed it is typical whenever one expands a function of
uniform transcendentality to subleading order in a given limit.
For example, $\ln x + \ln(1-x) = \ln x - x + \Ord(x^2)$ as $x\to0$.

The extension of \eqn{R64_1} gives the expansion of
$\mathcal{S}(R_6^{(4)})$ to order $T^1$.   One can easily generalize
this method to extract more terms in the $T$ expansion. To obtain the
$T^n$ term, we first subtract off the
expansion through order $T^{n-1}$ and divide by $T^{n-1}$, yielding a
function that vanishes when $T=0$. Then we can proceed as above and
calculate the $T^1$ term, which will correspond to the $T^n$ term of
the original function.

Proceeding in this manner, we obtain the expansion of the symbol of
$R_6^{(4)}$ through order $T^2$. To compare this expansion with the
data from the OPE, we must first disregard all terms containing
factors of $\pi$ or $\zeta_n$, since these constants are not captured
by the symbol. We must also convert from the remainder
function to the logarithm of the specific Wilson loop ratio considered
by BSV.  Both expressions are
finite and dual conformal invariant, but they differ by 
a simple additive function:\footnote{%
A version of this equation in ref.~\cite{Dixon2013eka} contained
a spurious ``1'', which is corrected here.}
\be
\label{eq:WL_convert}
\ln \mathcal{W}_{\textrm{hex}}(a/2)
\ =\ R_6(a)\ +\ \frac{\gamma_K(a)}{8} \, X(u,v,w)\,,
\ee
where the cusp anomalous dimension is
\beq
\gamma_K(a) = \sum_{L=1}^\infty a^L\,\gamma_K^{(L)} = 4a - 4 \zeta_2\,a^2
+ 22\zeta_4\,a^3
- 4 \biggl( \frac{219}{8} \zeta_6 + (\zeta_3)^2 \biggr) a^4 
+ \mathcal{O}(a^5)\,,
\label{cuspdef}
\eeq
and the function $X(u,v,w)$ is given by
\be
X(u,v,w)\ =\ 
- H^u_2 - H^v_2 - H^w_2 - \ln\biggl(\frac{uv}{w(1-v)}\biggr)\ln (1-v)
- \ln u \ln w + 2 \zeta_2 \,,
\label{Xuvw}
\ee
where $H^u_2 = H_{0,1}(1-u) = {\rm Li}_2(1-u)$ denotes a harmonic
polylogarithm (HPL)~\cite{Remiddi1999ew}.
The conventional loop expansion parameter for the Wilson loop, $g^2$,
is related to our expansion parameter by $g^2=a/2$.

Performing the comparison in \eqn{eq:WL_convert} at four loops,
we find that the information
at order $T^1$ is sufficient to fix all but four of the remaining
parameters. At order $T^2$, all four of these constants are determined
and many additional cross-checks are satisfied. The final expression
for the symbol of $R_6^{(4)}$ has 1,544,205 terms and can be downloaded in a
computer-readable file from~\cite{R64website}.


\subsection{Constraints at function level}
\label{sec:functionconstraints}

We now turn to the problem of promoting the symbol to a function. In
principle, the procedure is identical to that described in
ref.~\cite{Dixon2013eka}; indeed, with enough computational power we
could construct the full basis of hexagon functions at weight seven
(or even eight), and replicate the analysis of
ref.~\cite{Dixon2013eka}. In practice, it is difficult to build the
full basis of hexagon functions beyond weight five or six, and so we
briefly describe a more efficient procedure that requires only a
subset of the full basis.

To begin, we wish to construct a function-level ansatz for the $\{5,1,1,1\}$
components of the coproduct of $R_6^{(4)}$, denoted by
$\Delta_{5,1,1,1}(R_6^{(4)})$.  In general, the $\{n-k,1,1,\ldots,1\}$
components of the coproduct of a pure transcendental function $f$ of
weight $n$ (where there are $k$ 1's in the list) are defined iteratively by
differentiation.  Given that the differential of $f$ can be written as
\be
df = \sum_{s_k\in\Su} f^{s_k} \, d\ln s_k \,,
\label{df_def}
\ee
where $f^{s_i}$ are pure functions of weight $n-1$,
the $\{n-1,1\}$ element of the coproduct of $f$ is defined by
\be
\label{fnm11}
\Delta_{n-1,1}(f) = \sum_{s_k\in \Su} f^{s_k} \otimes \ln s_k \,.
\ee
(In contrast to the symbol, it is conventional in the coproduct
to keep the explicit ``$\ln$'' present in \eqn{fnm11}, because other
components of the coproduct, such as $\{n-m,m\}$ for $m>1$, require
different transcendental functions in all entries.)
To obtain the $\{n-2,1,1\}$ coproduct components $f^{s_j,s_k}$, we
differentiate each of the functions $f^{s_k}$, and expand their differentials
in terms of $d\ln s_j$,
\be
df^{s_k} = \sum_{s_j\in\Su} f^{s_j,s_k} \, d\ln s_j \,,
\label{dfk_def}
\ee
thereby defining
\be
\label{fnm211}
\Delta_{n-2,1,1}(f) = \sum_{s_j,s_k\in \Su} f^{s_j,s_k} 
\otimes \ln s_j \otimes \ln s_k \,.
\ee
If we were to iterate this procedure $n$ times, we would arrive at
the symbol.  However, here we wish to stop after the third iteration,
because the $\{n-3,1,1,1\}$ coproduct components for $n=8$ are weight-five
functions, and a full basis of hexagon functions already
exists~\cite{Dixon2013eka} at this weight.  We can match these functions to
functions derived from the symbol for $R_6^{(4)}$.

The $\{5,1,1,1\}$ coproduct of the ansatz for $R_6^{(4)}$ is a four-fold tensor
product whose first slot is a weight-five function and whose last
three slots are logarithms. The symbols of the weight-five functions
can be read off of the symbol of $R_6^{(4)}$, by clipping off the
last three entries.  They can then be identified with
functions in the weight-five hexagon basis. Therefore we can
immediately write down,
\be
\label{eq:R645111anz}
\Delta_{5,1,1,1}(R_6^{(4)}) = \sum_{s_i,s_j,s_k\in \Su} [R_6^{(4)}]^{s_i,s_j,s_k}
\otimes \ln s_i \otimes \ln s_j \otimes \ln s_k \,,
\ee
where $[R_6^{(4)}]^{s_i,s_j,s_k}$ are the most general linear
combinations of weight-five hexagon functions with the correct symbol
and correct parity. There will be many arbitrary parameters, all
of which are associated with $\zeta$ values multiplying lower-weight
functions.

Many of these parameters can be fixed by demanding that
$\sum_{s_i \in \Su} [R_6^{(4)}]^{s_i,s_j,s_k} \otimes \ln s_i $
be the $\{5,1\}$ component of the
coproduct for some weight-six function for every choice of $j$ and
$k$. This is simply the integrability constraint, discussed
extensively in ref.~\cite{Dixon2013eka}, applied to the first
two slots of the four-fold tensor product in~\eqn{eq:R645111anz}. We
also require that each weight-six function has the proper branch cut
structure; again, this constraint may be applied using the techniques
discussed in ref.~\cite{Dixon2013eka}. Finally, we must
guarantee that the weight-six functions have all of the symmetries
exhibited by their symbols. For example, if a particular coproduct
entry vanishes at symbol level, we require that it vanish at
function level as well. We also demand that the function have definite
parity since the symbol-level expressions have this property.

After imposing these mathematical consistency conditions, we will have
constructed the $\{5,1\}$ component of the coproduct for each of the
weight-six functions entering $\Delta_{6,1,1}(R_6^{(4)})$, as well as
all the integration constants necessary to define corresponding
integral representations (see section 4 of
ref.~\cite{Dixon2013eka}). There are many undetermined parameters, but
they all correspond to $\zeta$ values multiplying lower-weight hexagon
functions, so they cannot be fixed at this stage.

It is also also straightforward to represent
$\Delta_{6,1,1}(R_6^{(4)})$ directly in terms of multiple
polylogarithms in a particular subspace of the Euclidean region,
called Region I in ref.~\cite{Dixon2013eka}:
\be
\textrm{Region I}:\quad
\left\{
\begin{array}{l}
\Delta > 0\,,\quad 0<u_i<1\,, \quad~\textrm{and}~\quad u_1+u_2+u_3<1,\\
0< y_i < 1 \, .
\end{array}
\right.
\label{RegionIDef}
\ee
The fact that the $y_i$ are all real and between $0$ and $1$ facilitates
a representation in terms of multiple polylogarithms,
as discussed in ref.~\cite{Dixon2013eka}.  This region
is also of interest because it corresponds to positive external
kinematical data in $(2,2)$ signature.

To this end, we now describe how to integrate
directly the $\{n-1,1\}$ component of the coproduct of a weight-$n$
function in terms of multiple polylogarithms.
The method~\cite{Chen,FBThesis,Bogner2012dn}
is very similar to the integral given in eq.~(3.8) of
ref.~\cite{Dixon2013eka}, which maps symbols directly into multiple
polylogarithms. Instead of starting from the symbol, we start from the
$\{n-1,1\}$ coproduct component, and therefore we only have to perform
one integration, corresponding to the final iteration of the $n$-fold
iterated integration in eq.~(3.8) of ref.~\cite{Dixon2013eka}. 
As discussed in ref.~\cite{Dixon2013eka}, we are free to integrate along
a contour that goes from the origin $t_i=0$ to the point $t_i=y_i$
sequentially along the directions $t_u$, $t_v$ and $t_w$. The
integration is over $\omega = d\ln\phi$ with $\phi\in
\mathcal{S}_y$, where $\mathcal{S}_y$ is the set of 10 letters in the $y_i$
variables~\cite{Dixon2013eka}.  The integrand is a combination of
weight-$(n\!-\!1)$ multiple polylogarithms in Region I.  Together,
these two facts imply that the integral may always be evaluated
trivially by invoking the recursive definition of multiple
polylogarithms, $G(z)=1$, and
\be
\label{eq:main_G_def}
G(a_1,\ldots,a_n; z) = 
\int_0^z\; \frac{dt}{t-a_1}\,G(a_2,\ldots,a_n;t)\,, 
\quad\quad G(\underbrace{0,\ldots,0}_{p}; z) = \frac{\ln^p z}{p!} \,.
\ee
(Many of the properties of multiple polylogarithms are reviewed in
appendix~A of ref.~\cite{Dixon2013eka}.)

Applying this method to the case at hand, we obtain an expression for
$\Delta_{6,1,1}(R_6^{(4)})$ in terms of multiple polylogarithms in
Region I. Again, we enforce mathematical consistency by requiring
integrability in the first two slots, proper branch cut locations, and
well-defined parity. We then integrate the expression using the same
method, yielding an expression for $\Delta_{7,1}(R_6^{(4)})$. Finally,
we iterate the procedure once more and obtain a representation for
$R_6^{(4)}$ itself. At each stage we keep track of all the
undetermined parameters. Any parameter that survives all the way to
the weight-eight ansatz for $R_6^{(4)}$ must be associated with a
$\zeta$ value multiplying a lower-weight hexagon function with the
proper symmetries, branch-cut locations, and the function-level
analog of the final-entry condition. There are 68 such
functions. The counting of parameters is presented in
table~\ref{tab:R64_bts}.

\begin{table}[!ht]
\begin{center}
\begin{tabular}{|c|c|c|c|}
\hline\hline
$k$ & MZVs of weight $k$ &Functions of weight $8-k$ & Total parameters\\
\hline\hline
2 & $\zeta_2$ & $38$ & $38$\\
\hline
3 & $\zeta_3$ & $14$ & $14$\\
\hline
4 & $\zeta_4$ & $6$ & $6$\\
\hline
5 & $\zeta_2 \zeta_3$, $\zeta_5$ & $2$ & $4$\\
\hline
6 & $(\zeta_3)^2$, $\zeta_6$ & $1$ & $2$\\
\hline
7 & $\zeta_2\zeta_5$, $\zeta_3\zeta_4$, $\zeta_7$ & $0$ & $0$\\
\hline
8 & $\zeta_2(\zeta_3)^2$, $\zeta_3\zeta_5$, $\zeta_8$, $\zeta_{5,3}$ & $1$ & $4$\\
\hline\hline
\multicolumn{1}{|c|}{} &\multicolumn{2}{c|}{Total} 
&\multicolumn{1}{c|}{68} \\
\hline\hline
\end{tabular}
\caption{\label{tab:R64_bts} Characterization of the beyond-the-symbol
ambiguities in $R_6^{(4)}$ after imposing all mathematical consistency
conditions.}
\end{center}
\end{table}

It is straightforward to expand our 68-parameter ansatz for
$R_6^{(4)}$ in the near-collinear limit. Indeed, the methods discussed
in ref.~\cite{Dixon2013eka} can be applied directly to this case. We
carried out this expansion through order $T^3$, though even at order
$T^1$ the result is too lengthy to present here. The expansion 
(after fixing all parameters)
is available in a computer-readable format from~\cite{R64website}.

Demanding that our ansatz vanish in the strict collinear limit fixes
all but ten of the beyond-the-symbol constants.  Consistency with the
OPE at order $T^1$, corresponding to contributions of 
single (gluonic) flux-tube excitation, fixes nine of the ten remaining
constants. The final constant is fixed at order $T^2$, corresponding to
double flux-tube excitations, as well as twist-two bound-state
contributions~\cite{Basso2014T2}.  The rest of the data at order $T^2$
provides many nontrivial consistency checks of the result.

In slightly more detail, we can characterize the contributions
at a given order in the $T$ expansion by their dependence on $F$,
or equivalently on an azimuthal angle $\phi$, introduced by letting
$F = e^{i\phi}$.  As discussed in ref.~\cite{Basso2014T2}, the $F$
dependence is correlated with the helicity of the excitations.
The order $T^1$ term in the near-collinear expansion of the
$L$-loop remainder function always has the form,
\be
\Bigl[ R_6^{(L)} \Bigr]_1 = T (F+F^{-1}) \sum_{k=0}^{L-1} \ln^k T\, c_k(S)\,,
\label{twist1form}
\ee
where $c_k(S)$ is a linear combination of HPLs~\cite{Remiddi1999ew}
of the form $H_{\vec{m}}(-S^2)$, $m_i\in\{0,1\}$,
multiplied by simple rational functions of $S$.  The weight of
the HPLs is at most $2L-k$, but can be lower, in accordance with
the mixed transcendentality of the $T$ expansion mentioned earlier.
The single powers of $F$ and $F^{-1}$ correspond to the helicity $\pm1$
gluonic excitations, which have equal contributions due to parity.
The expansion of the three-loop remainder function at this order was
given explicitly in ref.~\cite{Dixon2013eka}.

At order $T^2$, the expansion has the form,
\be
\Bigl[ R_6^{(L)} \Bigr]_2 = T^2 \Bigl[
(F^2+F^{-2}) \sum_{k=0}^{L-1} \ln^k T\, d_k(S)
+ \sum_{k=0}^{L-1} \ln^k T\, f_k(S) \Bigr] \,,
\label{twist2form}
\ee
where $d_k(S)$ and $f_k(S)$, like $c_k(S)$, are linear combinations
of HPLs multiplied by rational functions of $S$ (more complicated
ones than appear in $c_k(S)$).
The terms in \eqn{twist2form} that have the $F^{\pm2}$ prefactors
come entirely from gluonic excitations --- either pairs of
single excitations, or the contribution of a twist-two
gluonic bound-state, either of which can have helicity $\pm2$;
whereas the $T^2F^0$ terms can come from excitations of pairs
of gluons, fermions or scalars~\cite{Basso2014T2}.
All of the constraints at order $T^2$ that were needed to fix the
five parameters at that stage (four parameters
at symbol level and one beyond-the-symbol) came from matching
the $T^2F^2$ contributions.  Hence the comparison of the 
$T^2 F^0$ terms, which tests the scalar and fermion contributions
as well as gluonic ones, was completely rigid, with no free parameters.

In practice the comparison to the OPE predictions was done after expanding
the functions of $S$ in an expansion around $S=0$.  For the $T^2F^2$
comparison we matched the terms through $S^{20}$; for the $T^2F^0$
comparison, through $S^{10}$.  Certainly higher orders could be matched
if desired; on the OPE side this just amounts to evaluating more residues
in the complex rapidity plane~\cite{Basso2013vsa,Basso2013aha,Basso2014T2}.
In some cases one can also perform the residue sums to all orders,
see {\it e.g.} ref.~\cite{Papathanasiou2013uoa}.

The final expression for $R_6^{(4)}$ in terms of multiple polylogarithms
in Region I is available from~\cite{R64website} in a computer-readable format.
We also provide a coproduct-based description of it; see
appendix~\ref{sec:coproduct}.


\section{Multi-Regge limit}
\label{sec:MRK}

\subsection{Fixing constants at four loops}
\label{sec:MRK4}

In the limit of multi-Regge kinematics (MRK), the cross
ratios~$u_1$, $u_2$ and $u_3$ approach the values
\beq
u_1 \to 1\,,\qquad u_2,u_3 \to 0\,,
\eeq
with the ratios
\beq
\frac{u_2}{1-u_1}\equiv \frac{1}{(1+w)\,(1+\ws)} {\rm~~and~~} 
\frac{u_3}{1-u_1}\equiv \frac{w\ws}{(1+w)\,(1+\ws)}
\label{wdef}
\eeq
held fixed\footnote{The (complex) variable $w$ defined in
eq.~\eqref{wdef} should not be confused with the cross ratio $w=u_3$.}.
While the remainder function vanishes in Euclidean MRK, this is no longer
the case once it is analytically continued to a different Riemann sheet,
according to $u_1 \to e^{-2\pi i}\,|u_1|$~\cite{Bartels2008ce}.
On this Riemann sheet we can write, 
\beq
R_{6}\big|_{\textrm{MRK}} = 
2\pi i \, \sum_{L=2}^\infty \sum_{n=0}^{L-1} \, a^L\,\ln^n(1-u_1) \, 
\Bigl[ g_n^{(L)}(w,\ws) + 2\pi i\,h_n^{(L)}(w,\ws) \Bigr]\,.
\label{MRKexpansion}
\eeq
The LLA series of coefficients has $n=L-1$.  The coefficients
$h_{L-1}^{(L)}(w,\ws)$ vanish trivially, while the coefficients
$g_{L-1}^{(L)}(w,\ws)$ are known to all orders in perturbation
theory~\cite{Dixon2012yy,Pennington2012zj}.
At NLLA ($n=L-2$), results for the coefficients $g_{L-2}^{(L)}(w,\ws)$ and
$h_{L-2}^{(L)}(w,\ws)$ have been given up to nine
loops~\cite{Lipatov2010ad,Fadin2011we,Dixon2011pw,Dixon2012yy}.

At NNLLA ($n=L-3$), only the three-loop coefficients are
known~\cite{Lipatov2010ad,Dixon2011pw,Dixon2013eka}.
In ref.~\cite{Dixon2012yy}, the four-loop coefficients at NNLLA and
N$^3$LLA, $g_1^{(4)}(w,\ws)$ and $g_0^{(4)}(w,\ws)$, respectively,
were heavily constrained and their functional form was completely
determined, up to 27 rational numbers
$a_i$, $b_j$, $i\in\{0,\ldots,8\}$, $j\in\{1,\ldots,18\}$.
As mentioned in the introduction, $a_0$ is a parameter that
enters at the level of the symbol.  The remaining 26 parameters
are beyond-the-symbol; they appear with $\zeta$ values multiplying them.
Since we have now a complete and unique analytic expression for the
four-loop remainder function in general kinematics, the coefficients
$g_1^{(4)}$ and $g_0^{(4)}$ can be extracted by using the techniques described in
ref.~\cite{Dixon2013eka}.  Appendix~\ref{sec:coproduct} gives
a brief description of how the coproduct representation of $R_6^{(4)}$
may be used for this purpose.

In this way, we find expressions for the two previously-undetermined MRK
coefficients at four loops,
\be
\bsp
g^{(4)}_1(w,\ws)&\, = \frac{19}{8}\,\LSPA{1}\,\LSPA{5}
+ \frac{1}{4}\,\LSMA{0}\,\LSMB{4}{1} + \frac{5}{4}\,\LSPA{1}\,\LSPC{3}{1}{1}
+ \frac{1}{2}\,\LSPA{1}\,\LSPC{2}{2}{1}
- \frac{3}{4}\,\LSMA{0}\,\LSMD{2}{1}{1}{1} \\
&\, - \left( \frac{29}{64}\,\LMA{0}{2}
       + \frac{17}{48}\,\LPA{1}{2} \right) \LSPA{1}\,\LSPA{3}
+ \frac{1}{96}\,\LMA{0}{3}\,\LSMB{2}{1}
+ \frac{5}{32}\,\LPA{3}{2} - \frac{1}{8}\,\LMB{2}{1}{2} \\
&\, - \frac{1}{4} \left( \LSMA{4} - \LSMC{2}{1}{1} \right) \LSMA{2}
+ \frac{3}{128} \left( \LMA{0}{2} -  4\, \LPA{1}{2} \right) \LMA{2}{2}  \\
&\, - \frac{11}{30720}\,\LMA{0}{6} + \frac{73}{1536}\,\LMA{0}{4}\,\LPA{1}{2}
+ \frac{19}{384}\,\LMA{0}{2}\,\LPA{1}{4} + \frac{11}{480}\,\LPA{1}{6} \\
&\,+ \zeta_2 \left( \frac{3}{2}\,\LSPA{1}\LSPA{3}
 + \frac{1}{2}\,\LSMA{0}\LSMB{2}{1} - \frac{3}{8}\,\LMA{2}{2}
 - \frac{11}{768}\,\LMA{0}{4} - \frac{1}{4}\,\LMA{0}{2}\LPA{1}{2}
 - \frac{7}{16}\,\LPA{1}{4} \right) \\
&\,- \frac{1}{8} \zeta_3 \left( \LSPA{3} - \frac{15}{4}\,\LMA{0}{2}\LSPA{1}
 - \LPA{1}{3} \right)
- \frac{27}{32} \, \zeta_4 \left( \LMA{0}{2} -  4\,\LPA{1}{2} \right) \\
&\,- \left( \frac{3}{2}\,\zeta_5 - \zeta_2\zeta_3 \right) \LSPA{1}
+ \frac{1}{8}\,(\zeta_3)^2 \, ,
\esp
\ee
and
\be
\bsp
g_0^{(4)}(w,\ws) &\,= -\frac{125}{8}\,\LSPA{7} + 5\,\LSPC{5}{1}{1}
+ \frac{11}{4}\,\LSPC{4}{2}{1} + \frac{1}{2}\,\LSPC{4}{1}{2}
+ \frac{3}{4}\,\LSPC{3}{3}{1} - 4\,\LSPE{3}{1}{1}{1}{1}
- \frac{3}{2}\,\LSPE{2}{2}{1}{1}{1} \\
&\, - \frac{1}{2}\,\LSPE{2}{1}{2}{1}{1}
+ \left( \frac{129}{64}\,\LMA{0}{2}
      + \frac{25}{16}\,\LPA{1}{2} \right) \LSPA{5}
+ \frac{1}{4}\,\LSMA{0}\,\LSPA{1} 
  \left( \LSMB{4}{1} - \LSMD{2}{1}{1}{1} \right) \\
&\, + \left( \frac{3}{32}\,\LMA{0}{2} + \frac{7}{8}\,\LPA{1}{2} \right)
   \LSPC{3}{1}{1}
- \frac{1}{16} \left( \LMA{0}{2} - 4\,\LPA{1}{2} \right) \LSPC{2}{2}{1}
- \frac{1}{8}\,\LSMA{0}\,\LSPA{3}\,\LSMB{2}{1} \\
&\, - \left( \frac{5}{24}\,\LMA{0}{4} + \frac{21}{64}\,\LMA{0}{2}\,\LPA{1}{2}
       + \frac{7}{48}\,\LPA{1}{4} \right) \LSPA{3}
- \left( \frac{7}{192}\,\LMA{0}{2} + \frac{1}{16}\,\LPA{1}{2} \right)
  \LSMA{0}\,\LSPA{1}\,\LSMB{2}{1} \\
&\,+ \frac{1007}{46080}\,\LMA{0}{6}\,\LSPA{1}
+ \frac{7}{144}\,\LMA{0}{4}\,\LPA{1}{3} + \frac{9}{320}\,\LMA{0}{2}\,\LPA{1}{5}
+ \frac{1}{210}\,\LPA{1}{7} \\
&\, - \frac{1}{4} \left( \LSPA{1} \left( \LSMA{4} - \LSMC{2}{1}{1} \right)
   + \frac{5}{4}\,\LSMA{0}\,\LSPB{3}{1} \right) \LSMA{2}
+ \left( \frac{5}{64}\,\LMA{0}{2} - \frac{1}{16}\,\LPA{1}{2} \right)
   \LSPA{1}\,\LMA{2}{2} \\
&\,- \zeta_2 \biggl( \frac{21}{4}\,\LSPA{5} + 3\,\LSPC{3}{1}{1}
   + \frac{3}{2}\,\LSPC{2}{2}{1}
   - \left( \frac{25}{32}\,\LMA{0}{2} + \frac{15}{8}\,\LPA{1}{2} \right)
      \LSPA{3}
   - \LSMA{0}\,\LSPA{1}\,\LSMB{2}{1} \\
&\qquad\quad \, + \frac{19}{192}\,\LMA{0}{4}\,\LSPA{1}
   + \frac{19}{48}\,\LMA{0}{2}\,\LPA{1}{3} + \frac{1}{5}\,\LPA{1}{5}
            \biggr)
   \nonumber
   \esp\eeq
\beq\bsp
\phantom{g_0^{(4)}(w,\ws)}
&\,+ \zeta_3 \left( - \frac{3}{4}\,\LSPA{1}\,\LSPA{3}
   + \frac{1}{4}\,\LMA{2}{2} + \frac{7}{256}\,\LMA{0}{4}
   + \frac{1}{2}\,\LMA{0}{2}\,\LPA{1}{2} + \frac{7}{48}\,\LPA{1}{4} \right)\\
&\,+ \zeta_4 \left( - \frac{15}{2}\,\LSPA{3}
    + \frac{11}{16}\,\LMA{0}{2}\,\LSPA{1}
    + \frac{9}{4}\,\LPA{1}{3} \right)
+ \zeta_5 \left( \frac{17}{16}\,\LMA{0}{2} - \frac{5}{2}\,\LPA{1}{2} \right) \\
&\, + \zeta_2 \zeta_3 \left( - \frac{9}{16}\,\LMA{0}{2} 
                         + \frac{5}{4}\,\LPA{1}{2} \right)
+ \frac{3}{2} \, (\zeta_3)^2 \, \LSPA{1} 
+ \frac{25}{4}\,\zeta_7 + \frac{3}{4} \, \zeta_2 \, \zeta_5 \,.
\esp
\ee
The functions $L^\pm_{\vec m}$ appearing in these expressions are
single-valued harmonic polylogarithms (SVHPLs)~\cite{BrownSVHPLs}.
They appear in a basis defined in ref.~\cite{Dixon2012yy},
which diagonalizes the $\mathbb{Z}_2 \times \mathbb{Z}_2$
action of inversion and conjugation of the variables $(w,\ws)$.

The expressions above match with those of eqs.~(7.14) and~(7.15) of
ref.~\cite{Dixon2012yy}, provided that the constants defined
in that reference take the values,
\be
\bsp
 a_0 &\,= 0, \quad a_1 = -\frac{1}{6}, \quad a_2 = -5, \quad a_3 = 1, \quad
 a_4 = \frac{4}{3}, \\
 a_5 &\,= -\frac{4}{3}, \quad a_6 = \frac{17}{180}, \quad a_7 = \frac{15}{4}, 
 \quad a_8 = -29\,,
\label{aivalues}
\esp
\ee
and
\be
\bsp
b_1 &\,= \frac{97}{1220}, \quad b_2 = \frac{127}{3660},
\quad b_3 = \frac{1720}{183}, \quad b_4 = \frac{622}{183},
\quad  b_5 = \frac{644}{305}, \quad b_6 = \frac{2328}{305}, \\
b_7 &\,= -1, \quad b_8 =-\frac{554}{305}, \quad b_9 = -\frac{10416}{305},
\quad b_{10} = \frac{248}{3}, \quad b_{11} = -\frac{11}{6}, 
\quad b_{12} =49, \\
b_{13} &\,= -112, \quad b_{14} = \frac{83}{12},
\quad b_{15} = -\frac{1126}{61}, \quad b_{16} = \frac{849}{122}, 
\quad b_{17} = \frac{83}{6}, \quad b_{18} = -10.
\label{bivalues}
\esp
\ee

The coefficient functions $h_n^{(L)}$ entering the real part in
\eqn{MRKexpansion} are completely determined by the functions
$g_n^{(L)}$ entering the imaginary part.  The LLA and NLLA functions
were given in eq.~(2.19) of ref.~\cite{Dixon2012yy}, but we provide
them here for completeness,
\beq\bsp
h_3^{(4)}(w,\ws)&\, = 0\,,\\
h_2^{(4)}(w,\ws)&\, = \frac{3}{2} \, g^{(4)}_3(w,\ws) 
- \frac{1}{2} \left[ g^{(2)}_1(w,\ws) \right]^2
- \frac{1}{8} \, \gamma_K^{(1)} \, L_1^+ \, g^{(3)}_2(w,\ws) \,,
\label{hLLANLLA}
\esp\eeq
where $\gamma_K^{(L)}$ are the $L$-loop coefficients
of the cusp anomalous dimension defined in \eqn{cuspdef},
and the lower loop $g$ coefficients are given in ref.~\cite{Dixon2012yy}.

The four-loop NNLLA and N$^3$LLA real-part coefficients are given by,
\beq\bsp
h_1^{(4)}(w,\ws)&\, = g^{(4)}_2(w,\ws) - g^{(2)}_0(w,\ws) \, g^{(2)}_1(w,\ws) \\
&\, - \frac{1}{8} \, L_1^+ \left( \gamma_K^{(1)} \, g^{(3)}_1(w,\ws)
             + \gamma_K^{(2)} \, g^{(2)}_1(w,\ws) \right) \,, 
\label{h4_1}
\esp\eeq
and
\beq\bsp
h_0^{(4)}(w,\ws) &\, = \frac{1}{2} \, g^{(4)}_1(w,\ws)
- \frac{1}{2} \left[ g^{(2)}_0(w,\ws) \right]^2
+ \pi^2 \, g^{(4)}_3(w,\ws) \\
&\, - \frac{1}{8} \, L_1^+ \left( \gamma_K^{(1)} \, g^{(3)}_0(w,\ws)
   + \gamma_K^{(2)} \, g^{(2)}_0(w,\ws)
   + 2 \, \pi^2 \, \gamma_K^{(1)} \, g^{(3)}_2(w,\ws) \right) \\
&\, - \frac{\pi^2}{393216} \, \left[\gamma_K^{(1)}\right]^4
\left( [L_0^-]^4 - 24 \, [L_0^-]^2 \, [L_1^+]^2 + 80 \, [L_1^+]^4 \right) \\
&\, + \frac{1}{512} \left( \left[\gamma_K^{(2)}\right]^2
          + 2 \, \gamma_K^{(1)}\, \gamma_K^{(3)}\right) 
   \left( [L_0^-]^2 - 4 \, [L_1^+]^2 \right) \\
&\, + \frac{\pi^2}{32} \left[\gamma_K^{(1)}\right]^2 
   \, [L_1^+]^2 \, g^{(2)}_1(w,\ws) \,.
\label{h4_0}
\esp\eeq
We checked explicitly that our result for $R_6^{(4)}$ correctly reproduces
all the real-part coefficient functions in the multi-Regge limit,
from $h_3^{(4)}(w,\ws)$ through $h_0^{(4)}(w,\ws)$.


\subsection{The NNLL BFKL eigenvalue and N$^3$LL impact factor}
\label{sec:NNLLBFKL}

The functions $g_1^{(4)}(w,\ws)$ and $g_0^{(4)}(w,\ws)$, in turn, determine the
NNLLA BFKL eigenvalue and N$^3$LLA impact factor,
through a master equation~\cite{Fadin2011we},
\bea
e^{R+i\pi\delta}|_{\textrm{MRK}}
 &=& \cos\pi\omega_{ab}
 + i \, {a\over 2}\sum_{n=-\infty}^\infty(-1)^n
 \,\left({w\over \ws}\right)^{{n\over 2}}\int_{-\infty}^{+\infty}
 {d\nu\over \nu^2+{n^2\over 4}}\,|w|^{2i\nu}
  \,\Phi_{\textrm{Reg}}(\nu,n) \nonumber\\
&&\hskip2cm\null
 \times \exp\left[ -\omega(\nu,n) \left( \ln(1-u_1) + i\pi
        + L_1^+ \right) \right] \,,
\label{eq:MHV_MRK_2}
\eea
where
\bea
\omega_{ab} &=& \frac{1}{8}\,\gamma_K(a) \, L_0^- \,, 
\label{omegaabdef}\\
\delta &=& \frac{1}{4}\,\gamma_K(a)\, L_1^+ \,,
\label{deltadef}
\eea
recalling that $L_0^- = \ln|w|^2$ and
$L_1^+ = \tfrac{1}{2} \ln(|w|^2/|1+w|^4)$.
The BFKL eigenvalue $\omega(\nu,n)$ and the impact factor 
$\Phi_{\textrm{Reg}}(\nu,n)$ can be expanded perturbatively,
\beq\bsp
\omega(\nu,n) &\,= 
- a \left(E_{\nu,n} + a\,E_{\nu,n}^{(1)}+ a^2\,E_{\nu,n}^{(2)}+\cO(a^3)\right)\,,\\
\Phi_{\textrm{Reg}}(\nu,n)&\, = 1 + a \, \Phi_{\textrm{Reg}}^{(1)}(\nu,n)
 + a^2 \, \Phi_{\textrm{Reg}}^{(2)}(\nu,n)
 + a^3 \, \Phi_{\textrm{Reg}}^{(3)}(\nu,n)+\cO(a^4)\,.
\label{expandomegaPhi}
\esp\eeq
We remark that an alternate version of the master equation
has recently been found in ref.~\cite{CaronHuot2013fea}.
In contrast to \eqn{eq:MHV_MRK_2}, the denominator $\nu^2 + n^2/4$
contains an additional term proportional to the square of
the cusp anomalous dimension.  It also lacks the explicit Regge pole
contribution (the $\cos\pi\omega_{ab}$ term), although this contribution
can be recovered by evaluating the $n=0$ term and $\nu=0$ residue in
the integral at finite coupling.  Then the two factorization forms
become equivalent, up to a different definition of the impact factor.
In this paper, we will continue to use the form~(\ref{eq:MHV_MRK_2}).

The first two nontrivial orders in the expansion of the BFKL eigenvalue
and the impact factor were known previously~\cite{Fadin2011we,%
Bartels2008sc,Lipatov2010qg,Dixon2013eka,Dixon2012yy},
\bea
E_{\nu,n} &=& -{1\over2}\,{|n|\over \nu^2+{n^2\over 4}}
+\psi\left(1+i\nu+{|n|\over2}\right) +\psi\left(1-i\nu+{|n|\over2}\right) 
- 2\psi(1)\,, \label{E_0}\\
E_{\nu,n}^{(1)} &=& - {1\over4} \, \dE{2}
 + {1\over2} \, V \, \dnu E_{\nu,n} - \zeta_2 \, E_{\nu,n} 
- 3 \, \zeta_3 \,, \label{E_1}\\
 \Phi^{(1)}_{\textrm{Reg}}(\nu,n) &=& - {1\over2}E_{\nu,n}^2 
- {3\over8} N^2 - \zeta_2\,, \label{Phi_Reg1}\\
 \Phi^{(2)}_{\textrm{Reg}}(\nu,n) &=&
{1\over2}\left[\Phi^{(1)}_{\textrm{Reg}}(\nu,n)\right]^2 - E^{(1)}_{\nu,n} \, E_{\nu,n}
+ \frac{1}{8}\,[D_\nu E_{\nu,n}]^2 
+ \frac{5}{64}\,N^2 \, ( N^2 + 4\,V^2) \nonumber\\
&&\hskip-0.7cm\null  
- \frac{\zeta_2}{4} \Bigl( 2\,E_{\nu,n}^2 + N^2 + 6\,V^2 \Bigr)
+ \frac{17}{4}\,\zeta_4 \,, \label{Phi_Reg2}
\eea
where $\psi(z) = {d\over dz}\ln\Gamma(z)$ is the digamma function, 
$\psi(1)=-\gamma_E$ is the Euler-Mascheroni constant,
and $V$ and $N$ are given by,
\beq\bsp
V &\equiv - \frac{1}{2}
  \left[ \frac{1}{i\nu+\frac{|n|}{2}} - \frac{1}{-i\nu+\frac{|n|}{2}} \right]
  = \frac{i \nu}{\nu^2+\frac{|n|^2}{4}},\\
N &\equiv \sgn(n)
  \left[ \frac{1}{i\nu+\frac{|n|}{2}} + \frac{1}{-i\nu+\frac{|n|}{2}} \right]
  = \frac{n}{\nu^2+\frac{|n|^2}{4}}\,,
\label{VNdef}
\esp\eeq
with $\dnu\equiv-i\partial_\nu \equiv -i\,\partial/\partial\nu$.

After expanding the master equation~(\ref{eq:MHV_MRK_2}) to the relevant
order in $a$ and $\ln(1-u)$, one has to match the resulting combinations
of SVHPLs in $(w,\ws)$ against the inverse Fourier-Mellin transforms
of suitable functions of $\nu$ and $n$.   This was carried out
in ref.~\cite{Dixon2012yy}, in terms of the then-undetermined
$a_i$ and $b_i$ constants.  Inserting the values~(\ref{aivalues})
and (\ref{bivalues}) into the respective expressions, we obtain,
\begin{eqnarray}
E^{(2)}_{\nu,n} &=& \frac{1}{8} \biggl\{ \frac{1}{6} \, \dE{4}
 - V \, \dE{3} + ( V^2 + 2 \zeta_2 ) \, \dE{2}
 - V \, ( N^2 + 8 \zeta_2 ) \dEOne \nonumber\\
&&\hskip0.3cm\null
 + \zeta_3 ( 4 \, V^2 + N^2 )
 + 44 \zeta_4 \, E_{\nu,n}
 + 16 \zeta_2 \zeta_3 + 80 \zeta_5 \biggr\} \,,
\label{E_2}
\end{eqnarray}
and
\be\bsp
\Phi_{\textrm{Reg}}^{(3)} &= 
-\frac{1}{48}\biggl\{\Enun{6}+\frac{9}{4}\Enun{4}N^2
+\frac{57}{16}\Enun{2}N^4 + \frac{189}{64}N^6
+\frac{15}{2}\Enun{2} N^2 V^2+\frac{123}{8} N^4 V^2\\
&\quad\quad\quad + 9 N^2 V^4 - 3\Bigl(4 \Enun{3}V 
+ 5 \EnunOne N^2 V\Bigr)\dEOne \\
&\quad\quad\quad+3 \Bigl(\Enun{2}+\frac{3}{4}N^2
+ 2 V^2\Bigr) \dEPOne{2}
+ 6\EnunOne\Bigl(\Enun{2}+\frac{3}{4}  N^2 +  V^2\Bigr)\dE{2}\\
&\quad\quad\quad - 12 V [\dEOne][\dE{2}] - 6 \EnunOne V \dE{3}
+ 2\,[\dEOne][\dE{3}]\\
&\quad\quad\quad+2\, \dEP{2}{2}+ \EnunOne \dE{4}\biggr\}\\
&\quad -\frac{1}{8} \zeta_2\Bigl[3 \Enun{4}+2 \Enun{2} N^2
-\frac{1}{16}N^4 - 6 \Enun{2} V^2 - 16 N^2 V^2 - 12 \EnunOne V \dEOne\\
&\quad\quad\quad\quad+\dEPOne{2}+4 \EnunOne \dE{2}\Bigr]\\
&\quad -\frac{1}{2}\zeta_3 \Bigl[3\Enun{3}+\frac{5}{2}\EnunOne N^2
+ \EnunOne V^2 - 3 V \dEOne+\frac{13}{6}\dE{2}\Bigr]\\
&\quad - \frac{1}{4} \zeta_4 \Bigl[27 \Enun{2}+N^2 -45 V^2\Bigr]
- 5 \bigl(2 \zeta_5+\zeta_2\zeta_3) \EnunOne
- \frac{219}{8}\zeta_6 - \frac{14}{3} (\zeta_3)^2\,.
\label{Phi_Reg3}
\esp\ee
\Eqns{Phi_Reg2}{E_2} allow the master equation~(\ref{eq:MHV_MRK_2})
to be evaluated at NNLL accuracy.  \Eqn{Phi_Reg3}, together with the
N$^3$LL BFKL eigenvalue $E^{(3)}_{\nu,n}$ (when the latter becomes available),
will permit an evaluation at N$^3$LLA --- assuming that the factorization
continues to hold at this order.

In ref.~\cite{Dixon2012yy} it was observed that $E^{(2)}_{\nu,n}$ in \eqn{E_2}
has a nonvanishing limit $\nu\to0$ (after setting $n=0$),
\beq\label{eq:E2nuto0}
\lim_{\nu\to0}E^{(2)}_{\nu,0} = -\frac{1}{2}\pi^2\,\zeta_3\,,
\eeq
even though $E_{\nu,n}$ and $E^{(1)}_{\nu,n}$ vanish in this
limit~\cite{Fadin2011we}.
This limit of $E^{(2)}_{\nu,n}$ held independently of all the constants
in \eqns{aivalues}{bivalues}, which were unknown at that time.
The reason it was independent of the constants was that the four-loop
remainder function was required to vanish in the collinear corner of the
MRK limit, $|w|^2 \to 0$.  This limit in the $(w,\ws)$ plane in turn
controls the $n=0$, $\nu\to0$ limit of the BFKL eigenvalue $\omega(\nu,n)$.
In ref.~\cite{CaronHuot2013fea}, the general constraints imposed by collinear
triviality of the remainder function were derived at finite coupling,
and \eqn{eq:E2nuto0} was obtained as a byproduct.


\subsection{NNLL coefficient functions at five loops}
\label{sec:MRK5}

The MRK factorization implicit in the master equation lets us
bootstrap higher-loop coefficients in the MRK limit.
We simply insert the results for the BFKL eigenvalue and the impact
factor through NNLLA into the master equation~(\ref{eq:MHV_MRK_2}).
We then use the techniques of ref.~\cite{Dixon2012yy} to perform
the inverse Fourier-Mellin transform from $(\nu,n)$ space back
to $(w,\ws)$ space.  This transform is facilitated by having a
complete basis of SVHPLs at the appropriate transcendental weight.
The inverse Fourier-Mellin transform leads to double sums, which
can either be summed explicitly, or truncated and then matched to
a Taylor expansion of the SVHPL basis.  In this way we can obtain explicit
expressions for $R_6$ in MRK at NNLLA, just as was done at LLA and NLLA
in ref.~\cite{Dixon2012yy}.  These data will be important in order to help
constrain the functional form of the remainder function at higher loop orders.

As an example, we present here the result for the five-loop six-point
remainder function at NNLLA. For the imaginary part, we find,
\beq\bsp
g_2^{(5)}(w,\ws) &\, = - 4\,\LSPA{7} - \frac{105}{32}\,\LSPC{5}{1}{1}\
- \frac{17}{8}\,\LSPC{4}{2}{1} - \frac{13}{16}\,\LSPC{4}{1}{2}
- \frac{15}{16}\,\LSPC{3}{3}{1} - \frac{1}{2}\,\LSPC{3}{2}{2} \\
&\, + \frac{19}{8}\,\LSPE{3}{1}{1}{1}{1} + \frac{3}{4}\,\LSPE{2}{2}{1}{1}{1}
+ \frac{1}{4}\,\LSPE{2}{1}{2}{1}{1} 
+ \left( \frac{147}{256}\,\LMA{0}{2}
       + \frac{5}{4}\,\LPA{1}{2} \right) \LSPA{5} \\
&\, + \LSMA{0}\,\LSPA{1} \left( \frac{29}{64}\,\LSMB{4}{1}
   + \frac{3}{16}\,\LSMB{3}{2} + \frac{5}{16}\,\LSMD{2}{1}{1}{1} \right)
+ \left( \frac{5}{16}\,\LMA{0}{2} 
       - \frac{3}{16}\,\LPA{1}{2} \right) \LSPC{3}{1}{1} \\
&\, + \frac{5}{32}\,\LMA{0}{2}\,\LSPC{2}{2}{1}
+ \frac{5}{32}\,\LSPA{1}\,\LPA{3}{2}
+ \frac{5}{32}\,\LSMA{0}\,\LSPA{3}\,\LSMB{2}{1}
+ \frac{1}{8}\,\LSPA{1}\,\LMB{2}{1}{2} \\
&\, - \left( \frac{23}{384}\,\LMA{0}{4}
           + \frac{35}{128}\,\LMA{0}{2}\,\LPA{1}{2}
           + \frac{25}{192}\,\LPA{1}{4} \right) \LSPA{3} \\
&\, - \left( \frac{11}{96}\,\LMA{0}{2}
       + \frac{7}{64}\,\LPA{1}{2} \right) \LSMA{0}\,\LSPA{1}\,\LSMB{2}{1}
+ \frac{23}{3840}\,\LMA{0}{6}\,\LSPA{1}
+ \frac{167}{4608}\,\LMA{0}{4}\,\LPA{1}{3} \\
&\, + \frac{31}{960}\,\LMA{0}{2}\,\LPA{1}{5} + \frac{29}{3360}\,\LPA{1}{7}
- \left( \frac{7}{32}\,\LSPA{1}\,\LSMA{4} + \frac{1}{32}\,\LSMA{0}\,\LSPB{3}{1}
       + \frac{3}{8}\,\LSPA{1}\,\LSMC{2}{1}{1} \right) \LSMA{2} \\
&\, + \frac{1}{16}\,\LSPA{3}\,\LMA{2}{2}
+ \left( \frac{1}{64}\,\LMA{0}{2}
       + \frac{1}{12}\,\LPA{1}{2} \right) \LSPA{1}\,\LMA{2}{2} \\
&\, + \zeta_2 \biggl(
- \frac{173}{32}\,\LSPA{5} - \frac{9}{2}\,\LSPC{3}{1}{1} - 3\,\LSPC{2}{2}{1}
+ \left( \frac{13}{16}\,\LMA{0}{2} + \frac{1}{4}\,\LPA{1}{2} \right) \LSPA{3}\\
&\qquad + \frac{3}{8}\,\LSMA{0}\,\LSPA{1}\,\LSMB{2}{1}
+ \frac{1}{4}\,\LSPA{1}\,\LMA{2}{2}
- \frac{55}{768}\,\LMA{0}{4}\,\LSPA{1} + \frac{11}{96}\,\LMA{0}{2}\,\LPA{1}{3}
- \frac{17}{40}\,\LPA{1}{5} \biggr) \\
&\, + \zeta_3 \left(
- \frac{5}{32}\,\LSPA{1}\,\LSPA{3} - \frac{3}{8}\,\LSMA{0}\,\LSMB{2}{1}
+ \frac{1}{16}\,\LMA{2}{2} + \frac{15}{256}\,\LMA{0}{4}
+ \frac{1}{16}\,\LMA{0}{2}\,\LPA{1}{2} - \frac{7}{96}\,\LPA{1}{4} \right) \\
&\, + \zeta_4 \left( - 3\,\LSPA{3} + 2\,\LPA{1}{3} \right)
+ \zeta_5 \left( - \frac{3}{16}\,\LMA{0}{2}
                 + \frac{35}{32}\,\LPA{1}{2} \right) \\
&\, - \frac{3}{4}\,\zeta_2\,\zeta_3\, \left( 2\,\LMA{0}{2} - \LPA{1}{2} \right)
- \frac{3}{16}\,(\zeta_3)^2\,\LSPA{1} + \frac{1}{4}\,\zeta_7 \,.
\label{g5_2}
\esp\eeq
The corresponding results at LLA and NLLA were given in
ref.~\cite{Dixon2012yy}.

The NNLL real-part coefficient is related to the imaginary parts
at NLLA and at lower loop orders; it is given by
\beq\bsp
h_2^{(5)}(w,\ws) &\,= \frac{3}{2}\,g_3^{(5)}(w,\ws)
- g_1^{(2)}(w,\ws)\,g_1^{(3)}(w,\ws) - g_0^{(2)}(w,\ws)\,g_2^{(3)}(w,\ws) \\ 
&\, - \frac{1}{8} \, L_1^+
  \left[ \gamma_K^{(1)} \, g_2^{(4)}(w,\ws)
       + \gamma_K^{(2)} \, g_2^{(3)}(w,\ws) \right] \,.
\label{h5_2}
\esp\eeq
Finally, we give the N$^3$LL real-part coefficient, which
is related to \eqn{g5_2} and to imaginary parts at lower logarithmic
or lower loop orders,
\beq\bsp
h_1^{(5)}(w,\ws) &\,= g_2^{(5)}(w,\ws)
- g_1^{(2)}(w,\ws)\,g_0^{(3)}(w,\ws) - g_0^{(2)}(w,\ws)\,g_1^{(3)}(w,\ws) \\
&\,
+ 4\,\pi^2 \left( g_4^{(5)}(w,\ws) - g_1^{(2)}(w,\ws)\,g_2^{(3)}(w,\ws) \right) \\
&\, - \frac{1}{8} \, L_1^+ 
\left[ \gamma_K^{(1)} \, g_1^{(4)}(w,\ws) + \gamma_K^{(2)} \, g_1^{(3)}(w,\ws)
     + \gamma_K^{(3)} \, g_1^{(2)}(w,\ws) \right] \\
&\, - \frac{\pi^2}{4} \biggl\{ 
 L_1^+ \, \gamma_K^{(1)} \left( 3 \, g_3^{(4)}(w,\ws)
                        - 2 \, \left[ g_1^{(2)}(w,\ws) \right]^2 \right) \\
&\qquad\quad - \frac{1}{4} \, \LPA{1}{2}
               \, \left[\gamma_K^{(1)}\right]^2 \, g_2^{(3)}(w,\ws) 
 + \frac{1}{96} \, \LPA{1}{3} 
    \, \left[\gamma_K^{(1)}\right]^3 \, g_1^{(2)}(w,\ws) \biggr\} \,.
\label{h5_1}
\esp\eeq
Although the real parts are related by analyticity to the imaginary
parts, they still provide useful additional constraints on ans\"{a}tze
for the remainder function.

\subsection{Connection between BFKL and the flux tube spectrum?}
\label{sec:BFKLfluxtube}

We conclude this section by noting that the result for the BFKL eigenvalue
at NNLLA suggests an intriguing connection between the BFKL
eigenvalues $E_{\nu,n}$, $E^{(1)}_{\nu,n}$, and $E^{(2)}_{\nu,n}$ and
the weak-coupling expansion of the energy $E(u)$ of a gluonic
excitation of the GKP string as a function of its rapidity $u$, given in
ref.~\cite{Basso2010in}. First we rewrite the expressions for
$E_{\nu,n}$, $E^{(1)}_{\nu,n}$, and $E^{(2)}_{\nu,n}$ explicitly in
terms of $\psi$ functions and their derivatives,
\be
\bsp
\EnunOne &= \psi(\xi^+)+\psi(\xi^-)-2\psi(1)-\frac{1}{2}\sgn (n)N \,,\\
E^{(1)}_{\nu,n} &=-\frac{1}{4}\Bigl[\psi^{(2)}(\xi^+) + \psi^{(2)}(\xi^-) 
- \sgn(n) N \Bigl( \frac{1}{4}N^2+V^2\Bigr)\Bigr]\\
&\quad+ \frac{1}{2}V\Bigl[\psi^{(1)}(\xi^+)-\psi^{(1)}(\xi^-)\Bigr] 
- \zeta_2 \EnunOne  - 3 \zeta_3 \,,\\
E^{(2)}_{\nu,n} &= \frac{1}{8} \biggl\{ \frac{1}{6} \Bigl[
\psi^{(4)}(\xi^+) + \psi^{(4)}(\xi^-)
- 60 \sgn(n) N \Bigl( V^4 + \frac{1}{2} V^2 N^2 + \frac{1}{80} N^4 \Bigr)
  \Bigr]
\\
&\hskip0.3cm\null
 - V \Bigl[ \psi^{(3)}(\xi^+) - \psi^{(3)}(\xi^-)
 - 3 \sgn(n) V N ( 4 V^2 + N^2 ) \Bigr]
\\
&\hskip0.3cm\null
 + ( V^2 + 2 \zeta_2 )
     \Bigl[ \psi^{(2)}(\xi^+) + \psi^{(2)}(\xi^-)
           - \sgn(n) N \Bigl( 3 V^2 + \frac{1}{4}  N^2 \Bigr) \Bigr]
\\
&\hskip0.3cm\null
 - V ( N^2 + 8 \zeta_2 ) \bigl[ \psi^{(1)}(\xi^+) - \psi^{(1)}(\xi^-)
                         - \sgn(n) V N \bigr]
 + \zeta_3 \, ( 4 V^2 + N^2 )
\\
&\hskip0.3cm\null
 + 44 \,\zeta_4 \EnunOne
 + 16 \,\zeta_2 \zeta_3 + 80 \, \zeta_5 \biggr\} \,,
\label{ENNLLA_psi_form}
\esp
\ee
where $\xi^\pm \equiv 1 \pm i\nu + \tfrac{|n|}{2}$.

Next, we keep only the pure $\psi$ (and $\zeta$) terms, dropping anything
with a $V$ or an $N$,
\be
\bsp
E_{\nu,n}\Big|_{\psi\ {\rm only}} 
&=  \psi(\xi^+)+\psi(\xi^-)-2\psi(1) \,,\\
E^{(1)}_{\nu,n}\Big|_{\psi\ {\rm only}} 
&= -\frac{1}{4}\Bigl[\psi^{(2)}(\xi^+) + \psi^{(2)}(\xi^-)\Bigr]
- \zeta_2 \Bigl[\psi(\xi^+)+\psi(\xi^-)-2\psi(1)\Bigr]  - 3 \zeta_3 \,,\\
E^{(2)}_{\nu,n}\Big|_{\psi\ {\rm only}} 
&= \frac{1}{8} \biggl\{ \frac{1}{6} \Bigl[
\psi^{(4)}(\xi^+) + \psi^{(4)}(\xi^-) \Bigr]
 + 2 \, \zeta_2
     \Bigl[ \psi^{(2)}(\xi^+) + \psi^{(2)}(\xi^-) \Bigr]
\\
&\hskip0.3cm\null
 \quad\quad + 44 \,\zeta_4 \bigl[ \psi(\xi^+) + \psi(\xi^-) - 2 \psi(1) \bigr]
 + 16 \,\zeta_2 \zeta_3 + 80 \, \zeta_5 \biggr\} \,.
\label{ENNLLA_psi_only}
\esp
\ee
Finally we write,
\be\label{eq:omega_psi_only}
-\omega(\nu,n)\Big|_{\psi\ {\rm only}} = a \Bigl( E_{\nu,n}\Big|_{\psi\ {\rm only}}  
+ a E^{(1)}_{\nu,n}\Big|_{\psi\ {\rm only}} +a^2 E^{(2)}_{\nu,n}\Big|_{\psi\ {\rm only}}
+ \cdots\Bigr)\,.
\ee
Now we compare this formula to equation (4.21) of ref.~\cite{Basso2010in}
for the energy $E(u)$ of a gauge field ($\ell=1$) and its bound state
($\ell > 1$),
\be
\bsp
\label{E_Basso}
E(u) &= \ell +
\frac{1}{2}\,\gamma_K(2g^2) \, \Bigl[ \psi_0^{(+)}(s,u) - \psi(1) \Bigr]
 - 2 g^4 \Bigl[ \psi_2^{(+)}(s,u) + 6\zeta_3 \Bigr]
\\
&\hskip0cm\null
 + \frac{g^6}{3} \biggl[ \psi_4^{(+)}(s,u)  + 2\pi^2 \psi_2^{(+)}(s,u) 
 + 24\zeta_3 \psi_1^{(+)}(s-1,u) + 8 \Bigl( \pi^2 \zeta_3 + 30\zeta_5 \Bigr)
 \biggr]\\
&\hskip0cm\null
 + {\cal O}(g^8)\,,
\esp
\ee
where $g^2=a/2$ is the loop expansion parameter, $s=1+\ell/2$,
and
\begin{equation}
 \psi_n^{(\pm)}(s,u) \equiv \frac{1}{2} 
\Bigl[ \psi^{(n)}(s+iu) \pm \psi^{(n)}(s-iu) \Bigr] \,.
\end{equation}
Neglecting the constant offset at order $a^0$ (the classical operator
scaling dimension), \eqn{E_Basso} matches 
perfectly with~\eqn{eq:omega_psi_only} at order $a^1$ and $a^2$,
provided that we identify,
\begin{equation}
 \ell = |n|, \qquad u = \nu.
\end{equation}
The correspondence continues to order $a^3$ if we also drop the 
term $24\,\zeta_3\, \psi_1^{(+)}(s-1,u)$. It would be very interesting
to understand the origin of this correspondence, and whether there is
a physical meaning to the operation of dropping all terms with 
a $N$ or a $V$. We leave this question to future work and return
our attention to the quantitative behavior of the four-loop
remainder function.

\section{Quantitative behavior}
\label{sec:quant}

In this section we investigate the quantitative behavior of the
four-loop remainder function in the Euclidean region where
all three cross ratios are positive.  It will prove particularly
instructive to plot the ratios of successive loop orders,
$R_6^{(3)}/R_6^{(2)}$ and $R_6^{(4)}/R_6^{(3)}$.
It was observed in ref.~\cite{Dixon2013eka} that the former
ratio was quite stable along large portions of a line and a
two-dimensional surface where it was examined.
We will find that the stability of such ratios extends to four
loops, {\it i.e.} to the latter ratio,
and to a number of different lines and one two-dimensional
surface, as long as the cross ratios are not too large or too small.
We will also examine certain limiting behavior analytically,
where it can sometimes shed light on the remarkable stability
of the ratios.  Finally, we will discuss how perturbation theory
is doing with respect to the approach to large orders.


\subsection{Region I}
\label{sec:RegI}

\begin{figure}[!t]
\centering
\includegraphics[width=0.7\linewidth]{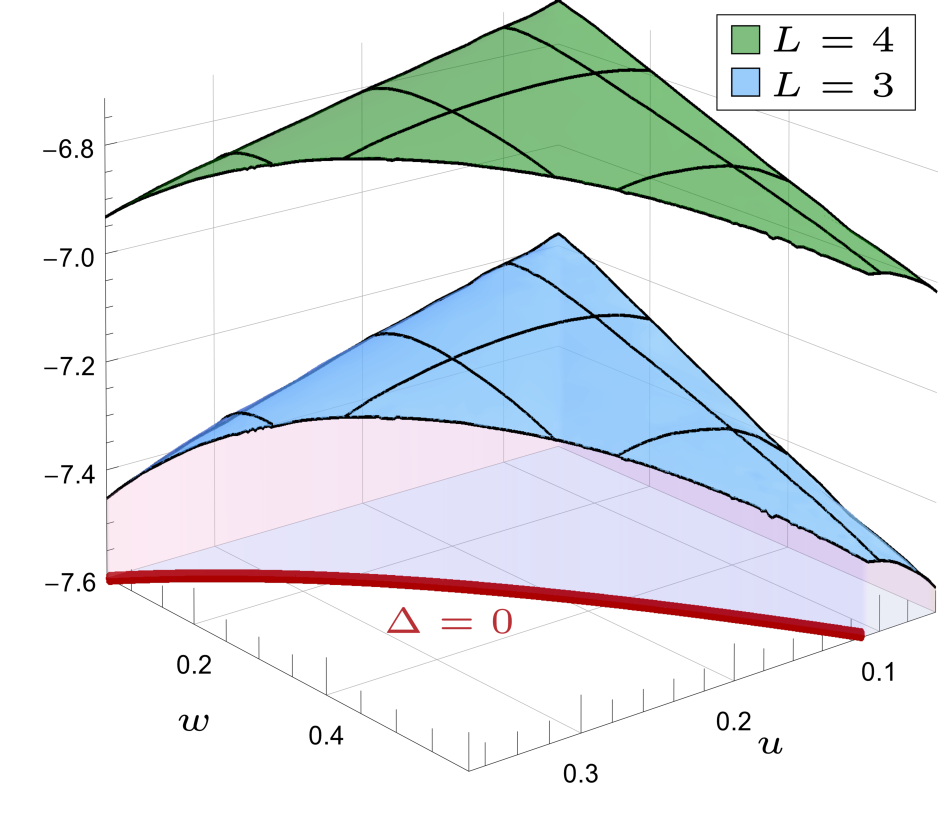}
\caption{\label{fig:plot_uw} The ratio $R^{(L)}_6(u,u,w)/R^{(L-1)}_6(u,u,w)$ 
for $L=3$ (blue) and $L=4$ (green) in Region I. The solid red line represents
the curve $\Delta(u,u,w)=0$. At small values of $(u,w)$, the plot is cut
off at $u=0.06$ or $w=0.06$.}
\centering
\end{figure}

While the full function $R_6^{(4)}$ is too lengthy to be shown here,
its representation in terms of multiple polylogarithms
can easily be evaluated numerically in Region I, defined in
\eqn{RegionIDef}, using {\sc GiNaC}~\cite{Bauer2000cp, Vollinga2004sn}.
In table~\ref{tab:numerics}, we show the value of the four-loop remainder
function for five reference points. In addition, in
fig.~\ref{fig:plot_uw} we plot the ratio $R^{(L)}_6/R^{(L-1)}_6$ for
$L=3$ and $L=4$ in Region I, restricted to the two-dimensional surface
$u=v$.  It is remarkable that the ratio of $R^{(4)}_6/R^{(3)}_6$ is
essentially flat throughout Region I.  The value of the ratio is close
to $-7$.  The ratio $R^{(3)}_6/R^{(2)}_6$ has a very similar behavior,
offset by about $0.5$ from the former ratio throughout most of the plot.

\begin{center}
\begin{table}[!th]
\begin{center}
\begin{tabular}{|c|c|}
\hline\hline
$(u,v,w)$ & $R_6^{(4)}$\\
\hline\hline
(0.214,\,0.214,\,0.184) &  ~97.251 \\
(0.333, 0.039, 0.286)   &  103.975 \\
(0.206, 0.008, 0.652)   &  ~53.664 \\
(0.617, 0.090, 0.043)   &  ~85.383 \\
(0.743, 0.002, 0.216)   &  ~19.752\\
\hline\hline
\end{tabular}
\caption{\label{tab:numerics} Numerical evaluation of the four-loop
remainder function at a selection of points in Region I.}
\end{center}
\end{table}
\end{center}

For $u=v$, the boundary of Region I in the interior
of the Euclidean region is defined by $\Delta(u,u,w)=0$,
where $\Delta$ is given in \eqn{zDeltadef}; this parabola
$w=(1-2u)^2$ is shown as the red line in the plot.
We restrict the plot to stay slightly away from the boundaries
of the Euclidean region, taking $u,w>0.06$.  At these boundaries,
$R_6(u,u,w)$ diverges logarithmically, order by order in perturbation
theory, whenever one of the cross ratios becomes very small and the
other one is kept finite.  At a given loop order, the degree of the
logarithmic divergence is one power lower when $w\to0$ with $u$ fixed,
than it is for the opposite case when $u\to0$ with $w$ fixed:
\beq\bsp
\label{eq:UWLLdef}
R^{(L)}_6(u,u,w) &\,\sim
\sum_{k=0}^{L-1} \mathcal{U}^{(L)}_k(y_u) \, \ln^{k}w\,,\quad\,\, w\to 0\,,
\,\, u\textrm{ finite}\,,\\
R^{(L)}_6(u,u,w) &\,\sim 
\sum_{k=0}^L \mathcal{W}^{(L)}_k(w) \, \ln^{k}u\,,\quad u\to 0\,,
\,\, w\textrm{ finite}\,.
\esp\eeq
The coefficients $\mathcal{W}^{(L)}_k(w)$ can be expressed in terms of
HPLs whose weight vectors are built entirely out of $0$ and $1$,
with argument $w$.  The coefficients $\mathcal{U}^{(L)}_k(y_u)$, in contrast,
require HPLs with argument $y_u$ rather than $u$, and
the weight vectors require $-1$ as well as $0$ and $1$.


The analytic expressions for the coefficients $\mathcal{U}^{(L)}_k$
and $\mathcal{W}^{(L)}_k$ are quite lengthy, so we do not show them here.
We list the results for the coefficients of just the leading logarithmic
divergence up to four loops in appendix~\ref{app:LLUV}.  Because the leading
logarithm increases by one with each additional loop, the ratios
plotted in fig.~\ref{fig:plot_uw} diverge like a single logarithm
as either boundary is approached.  However, the leading logarithms
in the numerator and denominator of the ratio are far from dominant
at the boundaries of the plot where $u$ or $w=0.06$.  If one keeps all
subleading logarithms, and neglects the power-suppressed terms,
one gets quite close to the exact numerical value of the ratio at either
boundary of the plot.

It was recently conjectured~\cite{AHCHTPrivate} that the remainder
function should have a uniform sign in Region I, which corresponds to
the kinematic regime of positive external momentum twistor kinematics.
Recent formulations of the planar scattering amplitude loop
integrand~\cite{ArkaniHamed2010kv,ArkaniHamed2012nw,Amplituhedra}
lead to manifestly positive integrands in this region.
On the other hand, an infinite subtraction is required to pass
to the remainder function.  Nevertheless, it was observed that this
conjecture indeed holds at two loops~\cite{AHCHTPrivate} and also
at three loops~\cite{Dixon2013eka}. Given that $R_6^{(3)}$
is negative in Region I~\cite{Dixon2013eka}, it is obvious from
fig.~\ref{fig:plot_uw} that $R_6^{(4)}$ has a uniform (positive) sign
in Region I, at least on the surface $u=v$. In total we checked more
than 1000 points in Region I, both on and off the $u=v$ surface;
for all points checked, the value of $R_6^{(4)}$ is positive,
in agreement with the conjecture.

In the rest of this section we focus on the remainder function restricted to 
certain one-dimensional subspaces where the functional form simplifies
drastically.  These lines may prove useful in trying to find
a form for the remainder function that is valid to all loop orders,
\emph{i.e.} at finite coupling, beyond what is presently known
in the OPE limit~\cite{Basso2013vsa,Basso2013aha,Basso2014T2}.
The first line we discuss has one endpoint which intersects the OPE limit.
Perhaps this proximity could allow the knowledge of the OPE limit to
anchor such a finite-coupling construction.
The other two lines never approach the OPE limit, although they have
other interesting properties.


\subsection{The line $(u,u,1)$}
\label{sec:uu1sec_R64}

As noted in ref.~\cite{Dixon2013eka}, the two- and three-loop
remainder functions can be expressed solely in terms of harmonic
polylogarithms (HPLs) of a single argument, $1-u$, on the line $(u,u,1)$,
and we use the notation $H^u_{\vec m}\equiv H_{\vec m}(1-u)$.  
The same is true at four loops, although the resulting expression is
rather lengthy.  It can be obtained by taking the limit of the general
coproduct representation described in appendix~\ref{sec:coproduct} onto
the line $(u,u,1)$.  The quantity $\Delta$ defined in \eqn{zDeltadef}
vanishes on this line.  As a consequence, all parity-odd functions
vanish on the line too.  The derivatives of weight $n$ parity-even functions
can be expressed using \eqn{eq:der_F} in terms of parity-even and
parity-odd coproduct components of weight $n-1$.  The vanishing
of the parity-odd functions as one approaches the line is fast enough
that they can be neglected in computing the derivative along the line.
Then one obtains from \eqn{eq:der_F},
\be
 \frac{dF(u,u,1)}{du} = \frac{F^u(u,u,1)+F^v(u,u,1)}{u}
   - \frac{F^{1-u}(u,u,1)+F^{1-v}(u,u,1)}{1-u} \,,
\label{uu1_derivative}
\ee
which is easily integrated in terms of the functions $H^u_{\vec m}$,
given that the coproduct components $F^u$, $F^v$, $F^{1-u}$ and $F^{1-v}$
are also expressible in this form.

Because the four-loop expression is still rather lengthy, in order
to save space we first expand all products of HPLs using the shuffle
algebra.  The resulting ``linearized'' representation will have HPL
weight vectors $\vec m$ consisting entirely of $0$'s and $1$'s,
which we can interpret as binary numbers. Finally, we
can write these binary numbers in decimal, making sure to keep track
of the length of the original weight vector, which we write as a
superscript.  For example,
\be
H_1^u H_{2,1}^u = H_{1}^u H _{0,1,1}^u
= 3 H_{0,1,1,1}^u + H_{1,0,1,1}^u \to 3 h^{[4]}_7 + h^{[4]}_{11}\, .
\ee
In this notation, $R_6^{(2)}(u,u,1)$ and $R_6^{(3)}(u,u,1)$ read,
\bea
R_6^{(2)}(u,u,1) & = &  h^{[4]}_{1}  -   h^{[4]}_{3} 
+   h^{[4]}_{9}  -   h^{[4]}_{11} -\frac{5}{2}\zeta_4 \, ,
\label{R62uu1}\\
R_6^{(3)}(u,u,1) & = & - 3  h^{[6]}_{1}  + 5  h^{[6]}_{3}  
+ \frac{3}{2}  h^{[6]}_{5}  - \frac{9}{2}  h^{[6]}_{7}
- \frac{1}{2}  h^{[6]}_{9}  - \frac{3}{2}  h^{[6]}_{11}
-   h^{[6]}_{13}  - \frac{3}{2}  h^{[6]}_{17}\nonumber \\
&&+ \frac{3}{2}  h^{[6]}_{19}  - \frac{1}{2}  h^{[6]}_{21}
- \frac{3}{2}  h^{[6]}_{23}  - 3  h^{[6]}_{33}  + 5  h^{[6]}_{35}
+ \frac{3}{2}  h^{[6]}_{37}  - \frac{9}{2}  h^{[6]}_{39}  \nonumber\\
&&- \frac{1}{2}  h^{[6]}_{41}  - \frac{3}{2}  h^{[6]}_{43}  -   h^{[6]}_{45}
- \frac{3}{2}  h^{[6]}_{49}  + \frac{3}{2}  h^{[6]}_{51}
- \frac{1}{2}  h^{[6]}_{53}  - \frac{3}{2}  h^{[6]}_{55} \label{R63uu1} \\
&& + \zeta_2\Bigl[ -   h^{[4]}_{1}  + 3  h^{[4]}_{3}  + 2  h^{[4]}_{5}
-   h^{[4]}_{9}  + 3  h^{[4]}_{11}  + 2  h^{[4]}_{13}\Bigr] \nonumber\\
&& + \zeta_4\Bigl[-2 h^{[2]}_{1} -2 h^{[2]}_{3}\Bigr] + (\zeta_3)^2
+ \frac{413}{24}\zeta_6\,, \nonumber
\eea
and the four-loop remainder function on the line $(u,u,1)$ is,
\be
\bsp\label{R64uu1}
R_6^{(4)}(u,u,1) &= 15  h^{[8]}_{1}  - 41  h^{[8]}_{3}  - \frac{31}{2}  h^{[8]}_{5}
+ \frac{105}{2}  h^{[8]}_{7}  - \frac{7}{2}  h^{[8]}_{9}
+ \frac{53}{2}  h^{[8]}_{11}  + 12  h^{[8]}_{13}  - 42  h^{[8]}_{15}  \\
&\quad+ \frac{5}{2}  h^{[8]}_{17}  + \frac{11}{2}  h^{[8]}_{19}
+ \frac{9}{2}  h^{[8]}_{21}  - \frac{41}{2}  h^{[8]}_{23}  +   h^{[8]}_{25}
- 13  h^{[8]}_{27}  - 7  h^{[8]}_{29}  - 5  h^{[8]}_{31}  \\
&\quad+ 6  h^{[8]}_{33}  - 11  h^{[8]}_{35}  - 3  h^{[8]}_{37}  + 3  h^{[8]}_{39}
- 4  h^{[8]}_{43}  - 4  h^{[8]}_{45}  - 11  h^{[8]}_{47}  + \frac{3}{2}  h^{[8]}_{49}
- \frac{3}{2}  h^{[8]}_{51}  \\
&\quad- 3  h^{[8]}_{53}  - 5  h^{[8]}_{55}  + \frac{3}{2}  h^{[8]}_{57}
- \frac{3}{2}  h^{[8]}_{59}  + 9  h^{[8]}_{65}  - 25  h^{[8]}_{67}  - 9  h^{[8]}_{69}
+ 27  h^{[8]}_{71}  - 2  h^{[8]}_{73}  \\
&\quad+ 9  h^{[8]}_{75}  + 2  h^{[8]}_{77}  - 23  h^{[8]}_{79}  + 2  h^{[8]}_{81}
-   h^{[8]}_{85}  - 8  h^{[8]}_{87}  + 2  h^{[8]}_{89}  - 3  h^{[8]}_{91}
+ \frac{5}{2}  h^{[8]}_{97}  \\
&\quad- \frac{7}{2}  h^{[8]}_{99}  - \frac{1}{2}  h^{[8]}_{101}
+ \frac{5}{2}  h^{[8]}_{103}  + \frac{1}{2}  h^{[8]}_{105}
+ \frac{1}{2}  h^{[8]}_{107}  + \frac{1}{2}  h^{[8]}_{109}
- \frac{5}{2}  h^{[8]}_{111}  + 15  h^{[8]}_{129} \\
&\quad  - 41  h^{[8]}_{131}  - \frac{31}{2}  h^{[8]}_{133}
+ \frac{105}{2}  h^{[8]}_{135}  - \frac{7}{2}  h^{[8]}_{137}
+ \frac{53}{2}  h^{[8]}_{139}+ 12  h^{[8]}_{141}  - 42  h^{[8]}_{143}  \\
&\quad  + \frac{5}{2}  h^{[8]}_{145}  + \frac{11}{2}  h^{[8]}_{147}
+ \frac{9}{2}  h^{[8]}_{149}  - \frac{41}{2}  h^{[8]}_{151}  + h^{[8]}_{153}
- 13  h^{[8]}_{155}  - 7  h^{[8]}_{157}  \\
&\quad  - 5  h^{[8]}_{159}  + 6  h^{[8]}_{161}  - 11  h^{[8]}_{163}
- 3  h^{[8]}_{165}  + 3  h^{[8]}_{167}  - 4  h^{[8]}_{171}  - 4  h^{[8]}_{173}  \\
&\quad- 11  h^{[8]}_{175}  + \frac{3}{2}  h^{[8]}_{177}
- \frac{3}{2}  h^{[8]}_{179}  - 3  h^{[8]}_{181}  - 5  h^{[8]}_{183}
+ \frac{3}{2}  h^{[8]}_{185}  - \frac{3}{2}  h^{[8]}_{187}  \\
&\quad+ 9  h^{[8]}_{193}  - 25  h^{[8]}_{195}  - 9  h^{[8]}_{197}
+ 27  h^{[8]}_{199}  - 2  h^{[8]}_{201}  + 9  h^{[8]}_{203}  + 2  h^{[8]}_{205}
- 23  h^{[8]}_{207} \\
&\quad  + 2  h^{[8]}_{209}  -   h^{[8]}_{213}  - 8  h^{[8]}_{215}  + 2  h^{[8]}_{217}
- 3  h^{[8]}_{219}  + \frac{5}{2}  h^{[8]}_{225}  - \frac{7}{2}  h^{[8]}_{227}
- \frac{1}{2}  h^{[8]}_{229} \\
&\quad  + \frac{5}{2}  h^{[8]}_{231}  + \frac{1}{2}  h^{[8]}_{233}
+ \frac{1}{2}  h^{[8]}_{235}  + \frac{1}{2}  h^{[8]}_{237}
- \frac{5}{2}  h^{[8]}_{239} \\
&\quad + \zeta_2\Bigl[2  h^{[6]}_{1}  - 14  h^{[6]}_{3}
- \frac{15}{2}  h^{[6]}_{5}  + \frac{37}{2}  h^{[6]}_{7}
- \frac{5}{2}  h^{[6]}_{9}  + \frac{25}{2}  h^{[6]}_{11}  + 7  h^{[6]}_{13}
- \frac{1}{2}  h^{[6]}_{17}  
\esp\eeq
\beq\bsp\nonumber
\phantom{R_6^{(4)}(u,u,1) }
&\quad\qquad+ \frac{5}{2}  h^{[6]}_{19}  + \frac{7}{2}  h^{[6]}_{21}
+ \frac{9}{2}  h^{[6]}_{23}  - 3  h^{[6]}_{25}  + 3  h^{[6]}_{27}  + 2  h^{[6]}_{33}
- 14  h^{[6]}_{35}  - \frac{15}{2}  h^{[6]}_{37}  \\
&\quad\qquad+ \frac{37}{2}  h^{[6]}_{39}  - \frac{5}{2}  h^{[6]}_{41}
+ \frac{25}{2}  h^{[6]}_{43}  + 7  h^{[6]}_{45}  - \frac{1}{2}  h^{[6]}_{49}
+ \frac{5}{2}  h^{[6]}_{51}  + \frac{7}{2} h^{[6]}_{53}  \\
&\quad\qquad+ \frac{9}{2}  h^{[6]}_{55}  - 3  h^{[6]}_{57}
+ 3  h^{[6]}_{59}\Bigr]\\
&\quad+\zeta_4\Bigl[\frac{15}{2}  h^{[4]}_{1}  - \frac{55}{2}  h^{[4]}_{3}
- \frac{41}{2}  h^{[4]}_{5}  + \frac{15}{2}  h^{[4]}_{9}
- \frac{55}{2}  h^{[4]}_{11}  - \frac{41}{2}  h^{[4]}_{13} \Bigr]\\
&\quad  + \Bigl(\zeta_2\zeta_3-\frac{5}{2}\zeta_5\Bigr)
\Bigl[h^{[3]}_{3}  +   h^{[3]}_{7}\Bigr]
- \Bigl( (\zeta_3)^2 - \frac{73}{4}\zeta_6\Bigr)
\Bigl[h^{[2]}_{1}  +   h^{[2]}_{3}\Bigr] \\
&\quad -\frac{3}{2}\zeta_2(\zeta_3)^2 - \frac{5}{2}\zeta_3 \zeta_5
- \frac{471}{4} \zeta_8 + \frac{3}{2}\zeta_{5,3}\,.
\esp
\ee

The remainder function $R_6^{(4)}(u,v,w)$, as a function of three variables,
satisfies a differential constraint, corresponding to the
final-entry condition imposed on the symbol. As discussed in
appendix~\ref{sec:coproduct},
this means that the $\{7,1\}$ components of the coproduct
obey $R_6^{(4)\, 1-u_i} = - R_6^{(4)\, u_i}$.  This property of
the partial derivatives does not necessarily extend
to the ordinary derivatives along a generic line.  However,
from \eqn{uu1_derivative} it is easy to see that it must
hold along the line $(u,u,1)$, where it implies
that
\be
\frac{dR_6^{(L)}(u,u,1)}{du} = 
\left( \frac{1}{u} + \frac{1}{1-u} \right) \times \hbox{pure function.}
\label{uu1_pure}
\ee
It is easy to check that the property (\ref{uu1_pure}) holds for the expressions for
$R_6^{(L)}(u,u,1)$ in eqs.~(\ref{R62uu1}), (\ref{R63uu1}) and (\ref{R64uu1}),
by verifying their symmetry under the operation,
\be\label{eq:check_ep}
h^{[n]}_{m} \to h^{[n]}_{m+2^{n-1}}\,,
\ee
where the lower index is taken mod $2^n$. This operation exchanges
$0\leftrightarrow1$ in the initial term of the weight vectors,
which, according to the definition of the HPLs, pairs the $1/u$ and
$1/(1-u)$ terms in \eqn{uu1_pure}.

\begin{figure}
\begin{center}
\includegraphics[width=5.0in]{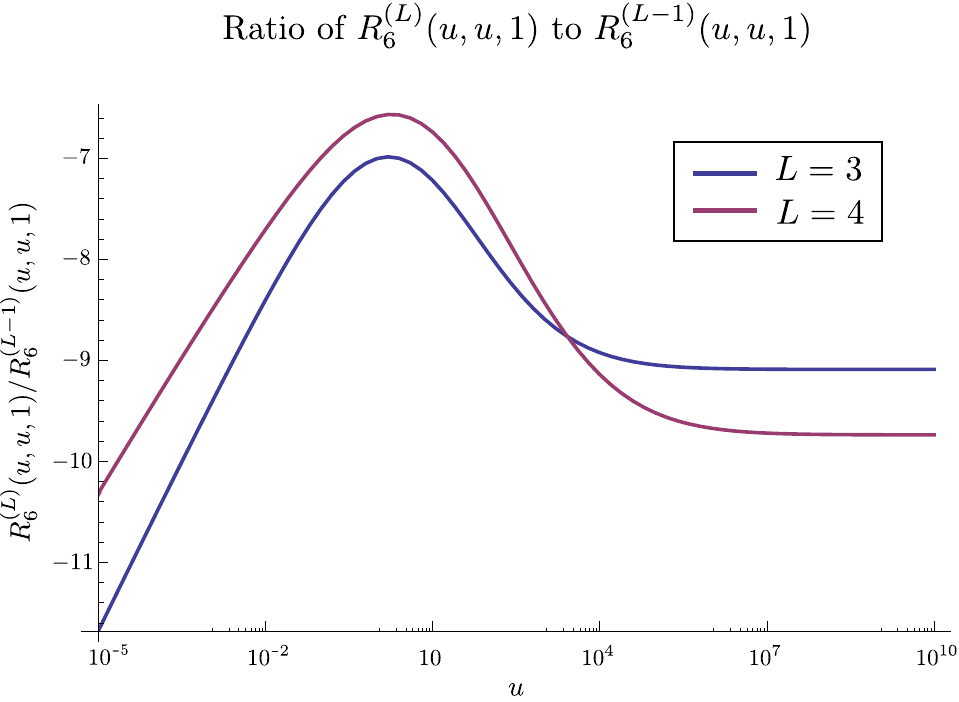}
\end{center}
\caption{The successive ratios $R_6^{(L)}/R_6^{(L-1)}$ on the line $(u,u,1)$.}
\label{fig:uu1_R64}
\end{figure}

Setting $u=1$ in the above formulas leads to
\be
\bsp
R_6^{(2)}(1,1,1) &= - (\zeta_2)^2 = - \frac{5}{2} \zeta_4
= -2.7058080842778\ldots \,, \\
R_6^{(3)}(1,1,1) &= \frac{413}{24} \, \zeta_6 + (\zeta_3)^2
= 18.951719323416\ldots \, \\
R_6^{(4)}(1,1,1) &= -\frac{3}{2}\zeta_2(\zeta_3)^2 - \frac{5}{2}\zeta_3 \zeta_5
- \frac{471}{4} \zeta_8 + \frac{3}{2}\zeta_{5,3} = -124.85491111408\ldots\, .
\esp
\ee
Note that $R_6^{(4)}(1,1,1)$ contains the multiple $\zeta$ value (MZV)
$\zeta_{5,3}$. It follows from standard conjectures on
MZVs~\cite{doubleshuffle} that $\zeta_{5,3}$ cannot be expressed in terms of
ordinary $\zeta$ values. While it is known that MZVs can appear in the
results for individual master integrals, this is one of the first examples
where an MZV enters the final result for a field theoretic quantity.

We remark that the point $(1,1,1)$ is the unique six-point kinematics
which can be considered as a two-dimensional scattering
configuration~\cite{AM2D,HK}.  At strong coupling~\cite{AMStrong},
using the AdS/CFT correspondence, the string world-sheet configuration
lies in three-dimensional anti-de Sitter space, AdS$_3$.
From \eqn{R6strong_R64} below, the strong-coupling value of the remainder
function at this point is
\be
R_6^{(\infty)}(1,1,1) = \frac{\pi}{6} - \frac{\pi^2}{12}
= -0.2988682578258\ldots.
\label{R6strong_R64_111}
\ee
We will explore the relation between weak-coupling and strong-coupling
behavior more thoroughly in section~\ref{sec:uuusec_R64}.

The numerical values of the $L$-loop to the $(L-1)$-loop ratios at the point
$(1,1,1)$ are remarkably close,
\be\label{ratiouuu_u1_R64}
\bsp
\frac{R_6^{(3)}(1,1,1)}{R_6^{(2)}(1,1,1)}\ =\ -7.004088513718\ldots \, ,\\
\frac{R_6^{(4)}(1,1,1)}{R_6^{(3)}(1,1,1)}\ =\ -6.588051932566\ldots \, .
\esp
\ee
In fact, the ratios are also similar away from this point,
as can be seen in~\fig{fig:uu1_R64}. The logarithmic scale for $u$
highlights how little the ratios vary over a broad range in $u$,
as well as how the $u$-dependence differs minimally between the
successive ratios.

We also give the leading term in the expansion of $R_6^{(4)}(u,u,1)$
around $u=0$,
\be
\bsp
R_6^{(4)}(u,u,1) &= u \biggl[ -\frac{5}{48} \ln^4u
 + \Bigl(\frac{3}{4} \zeta_2 + \frac{5}{3} \Bigr) \ln^3u
 - \Bigl(\frac{27}{4}\zeta_4 -\frac{1}{2}\zeta_3+5 \zeta_2
 + \frac{25}{2}\Bigr) \ln^2u\\
&\quad\quad
 + \Bigl(15\zeta_4 -3\zeta_3+13 \zeta_2 +50\Bigr) \ln u\\
&\quad\quad
+\frac{219}{8}\zeta_6 + (\zeta_3)^2 + 5\zeta_5+\zeta_2 \zeta_3
-\frac{71}{8}\zeta_4+6\zeta_3-10 \zeta_2-\frac{175}{2}\biggr]\\
&\hskip0.5cm + {\cal O}(u^2)\,.
\label{R64uu1_smallu}
\esp
\ee
We note the intriguing observation that the maximum-transcendentality
piece of the $u^1 \ln^0 u$ term is proportional to the four-loop
cusp anomalous dimension, 
$\frac{219}{8}\zeta_6 + (\zeta_3)^2=-\frac{1}{4}\gamma_K^{(4)}$.
In fact, the corresponding pieces of the two- and three-loop results,
given in ref.~\cite{Dixon2013eka}, can be checked to similarly
correspond to $-\frac{1}{4}\gamma_K^{(2)}$ and $-\frac{1}{4}\gamma_K^{(3)}$.

In the limit $u\to0$, the line $(u,u,1)$ touches the end
of the collinear line $v=0$, $u+w=1$.  So one could ask
where the cusp anomalous dimension seen in \eqn{R64uu1_smallu}
originates in the near-collinear
limit from the OPE perspective.  Actually, it is not there
at all in the limiting behavior $S\to0$, $T\to0$ of the Wilson loop
ratio ${\cal W}_{\rm hex}$ employed in
refs.~\cite{Basso2013vsa,Basso2013aha,Basso2014T2}.
To see this, first recall from \eqn{BSVparam} that to leading order in $T$,
$u=S^2/(1+S^2)$, $v=T^2$, and $w=1/(1+S^2)$.  Hence the line $(u,u,1)$ for
$u\to0$ matches the $S\to0$, $T\to0$ limit, after making the identification
$u=S^2=T^2$, to leading order.
Now let's inspect the additive term $\tfrac{1}{8} \, \gamma_K(a) \, X(u,v,w)$
in \eqn{eq:WL_convert} relating $R_6$ to $\ln{\cal W}_{\rm hex}$.
The function $X(u,v,w)$ defined in \eqn{Xuvw}
is suppressed by a power of $u$ in this limit,
\be
X(u,u,1)\ =\ 2 \, u + \Ord(u^2),
\label{Xuu1_0}
\ee
as $u\to0$.  This limiting behavior has the precise form and value
to cancel the $-\frac{1}{4}\gamma_K(a)\cdot u$ in $R_6(u,u,1)$
in passing to $\ln {\cal W}_{\rm hex}(a/2)$ via \eqn{eq:WL_convert}.

Suppose, however, that we look at the other end of the collinear line
$v=0$, $u+w=1$; namely the line $(1,u,u)$ as $u\to0$.
This line matches the $S\to\infty$, $T\to0$ limit, with the identification
$u=1/S^2=T^2$ to leading order.  The $S_3$ permutation symmetry of the
remainder function implies that $R_6(1,u,u)=R_6(u,u,1)$.  However, the
function $X$ has a different behavior in this limit,
\be
X(1,u,u)\ =\ 2 \, u ( 1 - \ln u ) + \Ord(u^2).
\label{X1uu_0}
\ee
The logarithmic term implies that in the $S\to\infty$, $T\to0$ limit
the cusp anomalous dimension is visible in the OPE.  The difference
between the two limits (or more generally, the lack of symmetry of the
Wilson loop ratio) is related to changing the ``framing'' of the
hexagonal Wilson loop, by making the other possible choice of pentagons
and box to remove the ultraviolet divergences.  This change of frame
always involves the cusp anomalous dimension~\cite{Basso2014T2}.
It may be useful to study the limiting behavior of the $(u,u,1)$
and $(1,u,u)$ lines in more detail, as an avenue along which the OPE
might potentially be resummable at finite coupling.

Comparing~\eqn{R64uu1_smallu} with the corresponding results for
$R_6^{(2)}$ and $R_6^{(3)}$~\cite{Dixon2013eka}, we see that the ratios
$R_6^{(L)}/R_6^{(L-1)}$ both diverge logarithmically as $u\to0$
along this line:
\be
\bsp
\frac{R_6^{(3)}(u,u,1)}{R_6^{(2)}(u,u,1)}&\ \sim\ \frac{1}{2} \ln u, \qquad
\hbox{as $u\to0$}\,,\\
\frac{R_6^{(4)}(u,u,1)}{R_6^{(3)}(u,u,1)}&\ \sim\ \frac{5}{12} \ln u, \qquad
\hbox{as $u\to0$.}
\esp
\ee
The slight difference in these coefficients is reflected in the
slight difference in slopes in the region of small $u$
in~\fig{fig:uu1_R64}.

As $u\to\infty$, the leading behavior at four loops is,
\be
\bsp
R_6^{(4)}(u,u,1) &= -\frac{88345}{144}\zeta_8
- \frac{19}{4} \zeta_2 (\zeta_3)^2-\frac{63}{4}\zeta_3 \zeta_5
+\frac{5}{4}\zeta_{5,3}\\
&\quad
+ \frac{1}{u} \biggl[\frac{1}{42}\ln^7u+\frac{1}{6}\ln^6u
+\Bigl(1+\frac{4}{5}\zeta_2\Bigl) \ln^5u
-\Bigl(\frac{11}{12}\zeta_3-4\zeta_2-5\Bigr)\ln^4u \\
&\quad\hskip0.9cm\null
+ \Bigl(\frac{605}{24}\zeta_4 -\frac{11}{3}\zeta_3 + 16 \zeta_2 +20 \Bigr) 
\ln^3 u\\
&\quad\hskip0.9cm\null
 - \Bigl(7\zeta_5 +9\zeta_2\zeta_3-\frac{605}{8}\zeta_4+11\zeta_3
 -48\zeta_2-60\Bigr) \ln^2 u\\
&\quad\hskip0.9cm\null
+ \Bigl( \frac{6257}{32} \zeta_6 +\frac{13}{4}(\zeta_3)^2 - 14\zeta_5 
-18 \zeta_2\zeta_3 +\frac{605}{4}\zeta_4 -22 \zeta_3 \\
&\quad\hskip1.7cm\null
+ 96 \zeta_2 + 120 \Bigr) \ln u \\
&\quad\hskip0.9cm\null
-\frac{13}{2}\zeta_7 -25 \zeta_2\zeta_5-\frac{173}{4}\zeta_3\zeta_4 
+\frac{6257}{32}\zeta_6+\frac{13}{4}(\zeta_3)^2-14 \zeta_5\\
&\quad\hskip0.9cm\null
-18 \zeta_2\zeta_3+\frac{605}{4}\zeta_4-22\zeta_3+96\zeta_2+120
\biggr]\\
&\quad + {\cal O}\left(\frac{1}{u^2}\right)\,.
\esp
\label{R64uu1_largeu}
\ee
Just like at two and three loops, $R_6^{(4)}(u,u,1)$ approaches a constant
as $u\to\infty$. Comparing with eq.~(7.17) of ref.~\cite{Dixon2013eka}, 
we find
\be
\bsp
\frac{R_6^{(3)}(u,u,1)}{R_6^{(2)}(u,u,1)}&\ \sim\ - 9.09128803107\ldots,
\qquad \hbox{as $u\to\infty$.}\\
\frac{R_6^{(4)}(u,u,1)}{R_6^{(3)}(u,u,1)}&\ \sim\  -9.73956178163\ldots,
\qquad \hbox{as $u\to\infty$.}
\label{ratiouu1_largeu_R64}
\esp
\ee
These values are not very different from the ratios at $(1,1,1)$
presented in \eqn{ratiouuu_u1_R64}.


\subsection{The line $(u,1,1)$}
\label{sec:u11_section_R64}

Next we consider the line $(u,1,1)$, which, due to the total $S_3$
symmetry of $R_6(u,v,w)$, is equivalent to the line $(1,1,w)$
discussed in ref.~\cite{Dixon2013eka}.  As was the case at two and
three loops, we can express $R_6^{(4)}(u,1,1)$ solely in terms of HPLs
of a single argument.  In contrast to the line $(u,u,1)$,
here $\Delta(u,1,1) = (1-u)^2$ is non-vanishing.  The parity-odd
functions are non-vanishing on this line, and contribute
to the derivatives of the parity-even functions in the coproduct
representation.

Using the notation of~\sect{sec:uu1sec_R64}, the
two-loop result is,
\be
R_6^{(2)}(u,1,1) = \frac{1}{2}  h^{[4]}_{1}  + \frac{1}{4}  h^{[4]}_{5}
+ \frac{1}{2}  h^{[4]}_{9}  + \frac{1}{2}  h^{[4]}_{13}
- \frac{1}{2}\zeta_2\,h^{[2]}_{3} - \frac{5}{2}\zeta_4\, ,
\ee
the three-loop result is,
\be
\bsp
R_6^{(3)}(u,1,1) &= 
- \frac{3}{2}  h^{[6]}_{1}  + \frac{1}{2}  h^{[6]}_{3}  - \frac{1}{4}  h^{[6]}_{5}
- \frac{3}{4}  h^{[6]}_{9}  + \frac{1}{4}  h^{[6]}_{11}  - \frac{1}{4}  h^{[6]}_{13}
-   h^{[6]}_{17}  \\
&\quad+ \frac{1}{2}  h^{[6]}_{19}  - \frac{1}{2}  h^{[6]}_{21}
- \frac{1}{2}  h^{[6]}_{25}  + \frac{1}{2}  h^{[6]}_{27}
- \frac{3}{2}  h^{[6]}_{33}  + \frac{1}{2}  h^{[6]}_{35}
- \frac{1}{4}  h^{[6]}_{37}  \\
&\quad- \frac{3}{4}  h^{[6]}_{41}  + \frac{1}{2}  h^{[6]}_{43}
- \frac{5}{4}  h^{[6]}_{49}  + \frac{3}{4}  h^{[6]}_{51}
- \frac{1}{4}  h^{[6]}_{53}  - \frac{3}{4}  h^{[6]}_{57}
+ \frac{3}{4}  h^{[6]}_{59} \\
&\quad+\zeta_2\Bigl[ - \frac{1}{2}  h^{[4]}_{1}  + \frac{1}{2}  h^{[4]}_{3}
+ \frac{1}{2}  h^{[4]}_{5}  - \frac{1}{2}  h^{[4]}_{9}
- \frac{1}{2}  h^{[4]}_{13} \Bigr]\\
&\quad -\zeta_4\Bigl[h^{[2]}_{1}  - \frac{17}{4}  h^{[2]}_{3} \Bigr]
+ (\zeta_3)^2 + \frac{413}{24}\zeta_6\, ,
\esp
\ee
and the four-loop result is,
\be
\bsp
R_6^{(4)}(u,1,1) &= 
\frac{15}{2}  h^{[8]}_{1}  - \frac{13}{2}  h^{[8]}_{3}  - \frac{3}{4}  h^{[8]}_{5}
+ \frac{3}{4}  h^{[8]}_{7}  + \frac{9}{4}  h^{[8]}_{9}  - \frac{3}{4}  h^{[8]}_{11}
+ \frac{1}{2}  h^{[8]}_{13}  + \frac{15}{4}  h^{[8]}_{17}  \\
&\quad- \frac{5}{2}  h^{[8]}_{19}  + \frac{1}{2}  h^{[8]}_{21}
+ \frac{5}{8}  h^{[8]}_{23}  + \frac{5}{4}  h^{[8]}_{25}
- \frac{1}{2}  h^{[8]}_{27}  - \frac{1}{8}  h^{[8]}_{29}
+ \frac{9}{2}  h^{[8]}_{33}- \frac{17}{4}  h^{[8]}_{35}  \\
&\quad  - \frac{3}{8}  h^{[8]}_{37}  + \frac{3}{4}  h^{[8]}_{39}
+ \frac{11}{8}  h^{[8]}_{41}  - \frac{11}{8}  h^{[8]}_{43}
- \frac{5}{8}  h^{[8]}_{45}  + \frac{9}{4}  h^{[8]}_{49}
- \frac{9}{4}  h^{[8]}_{51}  - \frac{3}{4}  h^{[8]}_{53}   \\
&\quad + \frac{3}{4}  h^{[8]}_{55}  + \frac{3}{4}  h^{[8]}_{57}
+ \frac{21}{4}  h^{[8]}_{65}  - \frac{23}{4}  h^{[8]}_{67}
- \frac{7}{8}  h^{[8]}_{69}+ \frac{3}{4}  h^{[8]}_{71}
+ \frac{11}{8}  h^{[8]}_{73}  - \frac{13}{8}  h^{[8]}_{75}  \\
&\quad  - \frac{5}{8}  h^{[8]}_{77}  + \frac{23}{8}  h^{[8]}_{81}
- \frac{25}{8}  h^{[8]}_{83}  - \frac{5}{8}  h^{[8]}_{85}
+ \frac{7}{8}  h^{[8]}_{87}  + \frac{9}{8}  h^{[8]}_{89}
- \frac{3}{8}  h^{[8]}_{91}  + \frac{1}{8}  h^{[8]}_{93}\\
&\quad   + \frac{11}{4}  h^{[8]}_{97}   - 5  h^{[8]}_{99}
- \frac{11}{8}  h^{[8]}_{101}  + \frac{7}{8}  h^{[8]}_{103}
+ \frac{3}{4}  h^{[8]}_{105}  - \frac{5}{4}  h^{[8]}_{107}
- \frac{5}{8}  h^{[8]}_{109}  + \frac{7}{8}  h^{[8]}_{113} \\
&\quad - \frac{23}{8}  h^{[8]}_{115}  - \frac{9}{8}  h^{[8]}_{117}
+ \frac{7}{8}  h^{[8]}_{119} + \frac{15}{2}  h^{[8]}_{129}
- \frac{13}{2}  h^{[8]}_{131}  - \frac{3}{4}  h^{[8]}_{133}
+ \frac{3}{4}  h^{[8]}_{135} \\
&\quad  + \frac{9}{4}  h^{[8]}_{137}  -   h^{[8]}_{139}
+ \frac{1}{4}  h^{[8]}_{141}+ \frac{15}{4}  h^{[8]}_{145}
- 3  h^{[8]}_{147}  + \frac{1}{4}  h^{[8]}_{149}  +   h^{[8]}_{151}
+ \frac{5}{4}  h^{[8]}_{153}    \\
&\quad+ \frac{1}{4}  h^{[8]}_{157}  + \frac{9}{2}  h^{[8]}_{161}
- \frac{21}{4}  h^{[8]}_{163}- \frac{7}{8}  h^{[8]}_{165}
+ \frac{9}{8}  h^{[8]}_{167}  + \frac{9}{8}  h^{[8]}_{169}
- \frac{9}{8}  h^{[8]}_{171}  - \frac{1}{2}  h^{[8]}_{173}\\
&\quad  + 2  h^{[8]}_{177}  - \frac{11}{4}  h^{[8]}_{179}
- \frac{7}{8}  h^{[8]}_{181}  + \frac{9}{8}  h^{[8]}_{183}
+ \frac{3}{8}  h^{[8]}_{185}  + \frac{3}{8}  h^{[8]}_{187}  + 6  h^{[8]}_{193}
- 7  h^{[8]}_{195}  \\
&\quad - \frac{5}{4}  h^{[8]}_{197}  + \frac{9}{8}  h^{[8]}_{199}
+ \frac{3}{2}  h^{[8]}_{201}  - \frac{3}{2}  h^{[8]}_{203}
- \frac{3}{8}  h^{[8]}_{205}  + \frac{25}{8}  h^{[8]}_{209}
- \frac{31}{8}  h^{[8]}_{211}  - \frac{1}{4}  h^{[8]}_{213} 
\esp\eeq
\beq\bsp\nonumber
&\quad  + \frac{11}{8}  h^{[8]}_{215}+   h^{[8]}_{217}
+ \frac{1}{4}  h^{[8]}_{221}  + \frac{7}{2}  h^{[8]}_{225}  - 7  h^{[8]}_{227}
- \frac{17}{8}  h^{[8]}_{229}  + \frac{5}{4}  h^{[8]}_{231}
+ \frac{5}{8}  h^{[8]}_{233}   \\
&\quad - \frac{13}{8}  h^{[8]}_{235} - \frac{7}{8}  h^{[8]}_{237}
+ \frac{5}{4}  h^{[8]}_{241}  - \frac{19}{4}  h^{[8]}_{243}
- \frac{7}{4}  h^{[8]}_{245}  + \frac{5}{4}  h^{[8]}_{247}  \\
&\quad +  \zeta_2\Bigl[h^{[6]}_{1}  - 3  h^{[6]}_{3}  - \frac{7}{4}  h^{[6]}_{5}
+ \frac{1}{4}  h^{[6]}_{7}  - \frac{1}{4}  h^{[6]}_{9}  + \frac{1}{4}  h^{[6]}_{11}
+ \frac{1}{2}  h^{[6]}_{13}  + \frac{1}{4}  h^{[6]}_{17}- \frac{3}{4}  h^{[6]}_{19}\\
&\quad\qquad + \frac{1}{2}  h^{[6]}_{21}  - \frac{1}{4}  h^{[6]}_{23}
- \frac{3}{4}  h^{[6]}_{27}  - \frac{1}{2}  h^{[6]}_{29}  +   h^{[6]}_{33}
- \frac{5}{2}  h^{[6]}_{35}- \frac{3}{2}  h^{[6]}_{37}  - \frac{1}{2}  h^{[6]}_{39}\\
&\quad\qquad-   h^{[6]}_{43}  - \frac{1}{2}  h^{[6]}_{45}
+ \frac{3}{4}  h^{[6]}_{49}  - \frac{9}{4}  h^{[6]}_{51}  - \frac{5}{4}  h^{[6]}_{53}
- \frac{1}{2}  h^{[6]}_{55}  + \frac{3}{4}  h^{[6]}_{57}
- \frac{5}{4}  h^{[6]}_{59}\Bigr] \\
&\quad+\zeta_4\Bigl[\frac{15}{4}  h^{[4]}_{1}  - 5  h^{[4]}_{3}
- \frac{47}{8}  h^{[4]}_{5}  + \frac{3}{2}  h^{[4]}_{7}
+ \frac{15}{4}  h^{[4]}_{9}  + \frac{3}{2}  h^{[4]}_{11}
+ \frac{9}{2}  h^{[4]}_{13} \Bigr]\\
&\quad +\Bigl(\zeta_2\zeta_3-\frac{5}{2}\zeta_5\Bigr)
\Bigl[\frac{3}{2}h^{[3]}_{3}+h^{[3]}_{7}\Bigr]
+ \zeta_6\Bigl[\frac{73}{8}  h^{[2]}_{1}  - \frac{461}{16}  h^{[2]}_{3}\Bigr]
-\frac{1}{2} (\zeta_3)^2 \Bigl[h^{[2]}_{1}+ h^{[2]}_{3} \Bigr] \\
&\quad -\frac{3}{2}\zeta_2(\zeta_3)^2 - \frac{5}{2}\zeta_3 \zeta_5
- \frac{471}{4} \zeta_8 + \frac{3}{2}\zeta_{5,3}\,.
\esp
\ee
Using~\eqn{eq:check_ep}, it is easy to check that none of these
functions satisfies a property like \eqn{uu1_pure}, where the derivative
is expressed in terms of a single pure function multiplied by a rational
prefactor.  The reason is related to the nonvanishing contributions of
the parity-odd functions in the coproduct representation.

\begin{figure}
\begin{center}
\includegraphics[width=5.0in]{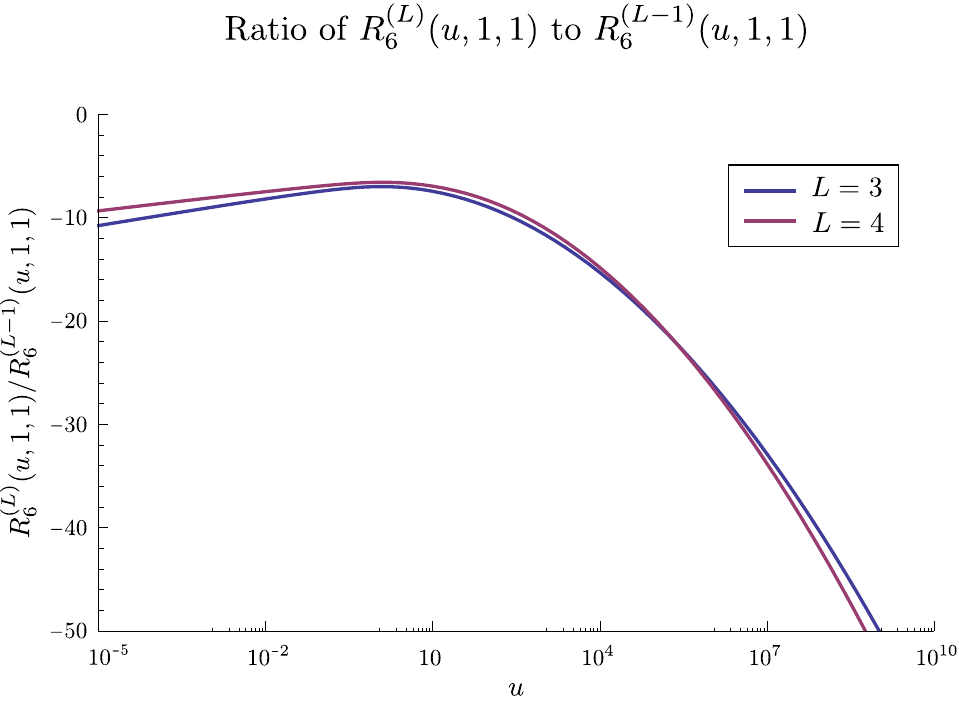}
\end{center}
\caption{The successive ratios $R_6^{(L)}/R_6^{(L-1)}$ on the line $(u,1,1)$.}
\label{fig:u11_R64}
\end{figure}

At both large and small $u$, these functions all diverge logarithmically.
At two and three loops, this was observed in ref.~\cite{Dixon2013eka}.
At four loops, we find at small $u$,
\be
\bsp
R_6^{(4)}(u,1,1) &=
\frac{1}{24}\Bigl(\frac{7}{2}\zeta_5-\zeta_2\zeta_3\Bigr)\ln^3u
- \frac{639}{256}\zeta_6\ln^2u
+ \Bigl(\frac{829}{64}\zeta_7 +\frac{69}{16} \zeta_3\zeta_4
+\frac{39}{8} \zeta_2\zeta_5\Big)\ln u\\
&\quad -\frac{3}{16}\zeta_2(\zeta_3)^2
- \frac{57}{16}\zeta_3 \zeta_5 - \frac{123523}{2880} \zeta_8 
+ \frac{19}{80}\zeta_{5,3} + \mathcal{O}(u)\,,
\esp
\ee
and at large $u$,
\be
\bsp
R_6^{(4)}(u,1,1) &= -\frac{37}{322560}\ln^8u - \frac{1}{80}\zeta_2\ln^6u 
+ \frac{7}{320}\zeta_3\ln^5u-\frac{533}{384}\zeta_4\ln^4u\\
&\quad +\Bigl(\frac{47}{48}\zeta_5 + \frac{53}{48}\zeta_2\zeta_3\Bigr)\ln^3u
- \Bigl(\frac{6019}{128}\zeta_6 + \frac{11}{16}(\zeta_3)^2\Bigr)\ln^2u\\
&\quad+ \Big(\frac{195}{8}\zeta_7 + \frac{923}{32}\zeta_3\zeta_4
+\frac{33}{2}\zeta_2\zeta_5\Bigr)\ln u \\
&\quad -3\zeta_2(\zeta_3)^2 - \frac{25}{2}\zeta_3 \zeta_5 
- \frac{1488641}{4608} \zeta_8 + \frac{1}{4}\zeta_{5,3} 
+ {\cal O}\left(\frac{1}{u}\right)\, .
\esp
\ee
The ratios $R_6^{(L)}(u,1,1)/R_6^{(L-1)}(u,1,1)$ also diverge in both limits,
\be
\bsp
\frac{R_6^{(3)}(u,1,1)}{R_6^{(2)}(u,1,1)}&\ \sim\ 
\Bigl(\frac{7 \pi^4}{1440\zeta_3}\Bigl)\,\ln u 
= \Bigl(0.393921796467\ldots\Bigr)\,\ln u,
\qquad \hbox{as $u\to0$\,,}\\
\frac{R_6^{(4)}(u,1,1)}{R_6^{(3)}(u,1,1)}&\ \sim\ 
\Bigl(\frac{60 \zeta_5}{\pi^4} -\frac{20 \zeta_3}{7\pi^2}\Bigr)\,\ln u
= \Bigl(0.290722549640\ldots\Bigr)\,\ln u,
\qquad \hbox{as $u\to0$\,,}
\label{ratiou11_smallu_R64}
\esp
\ee
and,
\be
\bsp
\frac{R_6^{(3)}(u,1,1)}{R_6^{(2)}(u,1,1)}&\ \sim\ -\frac{1}{10}\ln^2u,
\qquad \hbox{as $u\to\infty$\,,}\\
\frac{R_6^{(4)}(u,1,1)}{R_6^{(3)}(u,1,1)}&\ \sim\  -\frac{37}{336}\ln^2u,
\qquad \hbox{as $u\to\infty$\,.}
\label{ratiou11_largeu_R64}
\esp
\ee
In~\fig{fig:u11_R64}, we plot the ratios $R_6^{(L)}(u,1,1)/R_6^{(L-1)}(u,1,1)$
for a large range of $u$. The ratios are strikingly similar throughout
the entire region.

\subsection{The line $(u,u,u)$}
\label{sec:uuusec_R64}

At strong coupling, using the AdS/CFT correspondence,
gluon scattering amplitudes can be computed in the semi-classical
approximation by minimizing the area of a string world-sheet propagating
in AdS$_5 \times\,$S$^5$~\cite{AMStrong}.  The world-sheet boundary conditions
depend on the scattering kinematics. The amplitude has the generic
form,\footnote{%
It has recently been shown that another contribution has the same
dependence on $\lambda$ at strong coupling as the area term,
leading to a shift by an additive constant~\cite{Basso2014T2}.
We do not take this extra shift into account here.}
\beq
A_6 \propto
\exp\biggl( - \frac{\sqrt{\lambda}}{2\pi} \times {\rm Area} \biggr)
\propto 
\exp\biggl( \frac{\sqrt{\lambda}}{2\pi} \times R_6^{(\infty)} \biggr) \,,
\label{genstrong}
\eeq
where $\lambda = g_{\textrm{YM}}^2N_c = 8\pi^2\,a$.
As discussed in refs.~\cite{Alday2009dv,Dixon2013eka}, on the symmetrical
diagonal line $(u,u,u)$, the remainder function at strong coupling can
be written analytically,
\be
R_6^{(\infty)}(u,u,u) = - \frac{\pi}{6} + \frac{\phi^2}{3\pi}
+ \frac{3}{8} \, \Bigl[ \ln^2 u + 2 \, {\rm Li}_2(1-u) \Bigr]
- \frac{\pi^2}{12} \,,
\label{R6strong_R64}
\ee
where $\phi = 3 \, \cos^{-1}(1/\sqrt{4u})$.
The simplicity of this formula motivates us to evaluate the
four-loop remainder function on the line $(u,u,u)$,
as we did earlier at two and three loops~\cite{Dixon2013eka}.

In perturbation theory,
the function $R_6^{(L)}(u,u,u)$ cannot be written solely in terms of
HPLs with argument $(1-u)$. However, it is possible to use the
coproduct structure to derive differential equations which may be
solved by using series expansions around the three points $u=0$,
$u=1$, and $u=\infty$. This method was applied in
ref.~\cite{Dixon2013eka} at two and three loops, and here we extend it
to the four-loop case.

The expansion around $u=0$ takes the form,
\be
\bsp
R_6^{(4)}(u,u,u) &= 
\Bigl(\frac{1791}{32}\zeta_6-\frac{3}{4}(\zeta_3)^2\Bigr)\ln^2u
+\frac{32605}{512}\zeta_8-\frac{5}{2}\zeta_3\zeta_5
-\frac{9}{8}\zeta_2 (\zeta_3)^2
\\
&\hskip0.5cm\null
 + \, u \, \biggl[ \frac{5}{192}\ln^7u + \frac{5}{192}\ln^6u
        -\Bigl(\frac{19}{16}\zeta_2+\frac{5}{32}\Bigr)\, \ln^5u \\
&\hskip1.5cm\null
+\frac{5}{16}\Bigl(\zeta_3-3\zeta_2-\frac{3}{2}\Bigr)\, \ln^4u
+\Bigl(\frac{1129}{64}\zeta_4+\frac{5}{8}\zeta_3+3\zeta_2+\frac{15}{8}\Bigr)
\,\ln^3u\\
&\hskip1.5cm\null
-\Bigl(\frac{21}{8}\zeta_5 +\frac{3}{2}\zeta_2\zeta_3 
- \frac{669}{64}\zeta_4 + \frac{3}{2}\zeta_3 - 6\zeta_2 
- \frac{75}{8}\Bigr)\,\ln^2u\\
&\hskip1.5cm\null
+\Bigl(\frac{32073}{128}\zeta_6 -3 (\zeta_3)^2 - \frac{27}{4}\zeta_5
- \frac{3}{2}\zeta_2\zeta_3 - \frac{165}{32}\zeta_4 - \frac{15}{4}\zeta_3\\
&\hskip2.2cm\null
 - \frac{15}{2}\zeta_2 - \frac{75}{4}\Bigr)\,\ln u
+ \frac{3}{4}\zeta_2\zeta_5 -\frac{21}{16}  \zeta_3\zeta_4
+ \frac{7119}{128}\zeta_6 \\
&\hskip1.5cm\null
+ \frac{3}{4}(\zeta_3)^2 +\frac{27}{4}\zeta_5+\frac{3}{2}\zeta_2\zeta_3
+ \frac{45}{32}\zeta_4+\frac{21}{2}\zeta_3-\frac{15}{2}\zeta_2
-\frac{525}{4}\biggr]\\
&\hskip0.5cm\null
+ {\cal O}(u^2).
\esp
\label{R64_uuu_small_u}
\ee
The leading term at four loops diverges logarithmically, but, just
like at two and three loops, the divergence appears only as $\ln^2 u$.
This is another piece of evidence in support of the claim by
Alday, Gaiotto and Maldacena~\cite{Alday2009dv} that this property
should hold to all orders in perturbation theory. Because of this
fact, the ratios $R_6^{(3)}(u,u,u)/R_6^{(2)}(u,u,u)$ and
$R_6^{(4)}(u,u,u)/R_6^{(3)}(u,u,u)$ approach constants in the limit
$u\to0$,
\be
\bsp
\frac{R_6^{(3)}(u,u,u)}{R_6^{(2)}(u,u,u)}&\ \sim\ 
- \frac{7 \pi^2}{10} = -6.90872308076\ldots \, ,
\qquad \hbox{as $u\to0$\,,}\\
\frac{R_6^{(4)}(u,u,u)}{R_6^{(3)}(u,u,u)}&\ \sim\ 
-\frac{199 \pi^2}{294}+\frac{60 (\zeta_3)^2}{7 \pi^4} 
= -6.55330020271\ldots \, ,
\qquad \hbox{as $u\to0$\,.}
\label{ratiouuu_smallu_R64}
\esp
\ee

At large $u$, the expansion behaves as,
\be
\bsp
R_6^{(4)}(u,u,u) &= \frac{3}{2}\zeta_2 (\zeta_3)^2 -10 \zeta_3\zeta_5
+ \frac{1713}{64}\zeta_8 -\frac{3}{4}\zeta_{5,3} 
- \frac{4\pi^7}{5\, u^{1/2}}\\
& + \frac{1}{32\, u} \,\biggl[ \, \frac{1}{56} \ln^7u
+ \frac{5}{16}\ln^6u + \Bigl(\frac{51}{20}\zeta_2+\frac{33}{8}\Bigr)\ln^5u \\
&\hskip1.5cm\null - \Bigl(\frac{11}{2}\zeta_3-\frac{249}{8}\zeta_2
-\frac{345}{8}\Bigr)\ln^4u \\
&\hskip1.5cm\null+ \Bigl(\frac{1237}{4}\zeta_4-50 \zeta_3
+\frac{547}{2}\zeta_2+\frac{705}{2}\Bigr)\ln^3u\\
&\hskip1.5cm\null  - \Bigl(168 \zeta_5 +222 \zeta_2\zeta_3
- \frac{17607}{8}\zeta_4+330 \zeta_3-\frac{3441}{2}\zeta_2
-\frac{4275}{2}\Bigr)\ln^2u\\
&\hskip1.5cm\null + \Bigl(\frac{52347}{8}\zeta_6 + 144 (\zeta_3)^2
- 744 \zeta_5  - 1032 \zeta_2\zeta_3 + \frac{38397}{4}\zeta_4\\
&\hskip2.0cm\null   -1416 \zeta_3 + 7041 \zeta_2 + 8595\Bigr)\ln u
- 360 \zeta_7 -2499\zeta_3\zeta_4 \\
&\hskip1.5cm\null - 1200\zeta_2\zeta_5  + \frac{134553}{16}\zeta_6
+ 426 (\zeta_3)^2 -1596 \zeta_5 - 2292 \zeta_2\zeta_3\\
&\hskip1.5cm\null +\frac{80289}{4}\zeta_4 - 2976 \zeta_3
+ 14193 \zeta_2+17235\biggr]\\
&+ \frac{\pi^3}{32\, u^{3/2}}\,\biggl[\, 3 \ln^3u + \frac{45}{2}\ln^2u
+ \Bigl(306\zeta_2+99\Bigr)\ln u - 96 \zeta_4 + 36 \zeta_3\\
&\hskip1.9cm\null  +671 \zeta_2 + \frac{469}{2} \biggr]
+ {\cal O}\left(\frac{1}{u^2}\right) \, .
\esp
\ee
The ratios $R_6^{(3)}(u,u,u)/R_6^{(2)}(u,u,u)$ and
$R_6^{(4)}(u,u,u)/R_6^{(3)}(u,u,u)$ approach constants in the limit
$u\to\infty$,
\be
\bsp
\frac{R_6^{(3)}(u,u,u)}{R_6^{(2)}(u,u,u)}&\ \sim\ -1.22742782334\ldots\,,
\qquad \hbox{as $u\to\infty$\,,}\\
\frac{R_6^{(4)}(u,u,u)}{R_6^{(3)}(u,u,u)}&\ \sim\ 21.6155002540\ldots\,,
\qquad \hbox{as $u\to\infty$\,.}
\label{ratiouuu_largeu_R64}
\esp
\ee

In contrast to the expansions around $u=0$ and $u=\infty$,
the expansion around $u=1$ is regular,
\be
\bsp
R_6^{(4)}(u,u,u) &=  -\frac{3}{2}\zeta_2(\zeta_3)^2
- \frac{5}{2}\zeta_3 \zeta_5 - \frac{471}{4} \zeta_8 
+ \frac{3}{2}\zeta_{5,3} \\
&\quad + \Bigl(\frac{219}{8}\zeta_6 - \frac{3}{2}(\zeta_3)^2
+ \frac{45}{4}\zeta_4 + 3 \zeta_2 + \frac{45}{2}\Bigr)(1-u) 
+ \mathcal{O}\Bigl((1-u)^2\Bigr)\, .
\esp
\ee

\begin{figure}
\begin{center}
\includegraphics[width=5.0in]{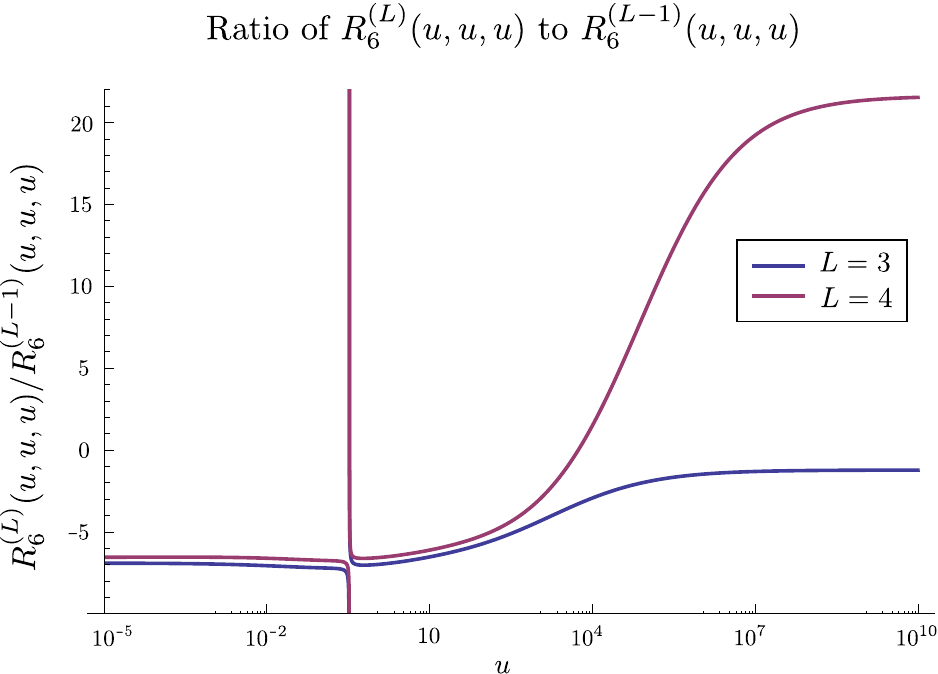}
\end{center}
\caption{The successive ratios $R_6^{(L)}/R_6^{(L-1)}$ on the line $(u,u,u)$.}
\label{fig:uuu_R64}
\end{figure}

We take 100 terms in each expansion, around 0, 1 and $\infty$
and piece them together to obtain a numerical representation for 
the function $R_6^{(4)}(u,u,u)$ that is valid along the entire line.
In the regions of overlap, we find agreement to at least 15 digits.
In~\fig{fig:uuu_R64}, we plot the ratios $R_6^{(L)}(u,u,u)/R_6^{(L-1)}(u,u,u)$
for a large range of $u$.  The spike in the plot is not a numerical
instability; it occurs because the denominators in the respective
ratios go through zero at a slightly different point from the numerators,
around $u=1/3$.

As noted in ref.~\cite{Dixon2013eka}, the two- and three-loop remainder
functions vanish along the line $(u,u,u)$, very close to the point
$u=1/3$.  More precisely, it was found that the
vanishing relation $R_6^{(L)}(u_0^{(L)},u_0^{(L)},u_0^{(L)}) = 0$ holds for
\be
u_0^{(2)} = 0.33245163\ldots, \qquad u_0^{(3)} = 0.3342763\ldots,
\label{uuu_zero_crossing_R62_R63}
\ee
for two and three loops, respectively.

The point $(u,v,w)=(1/3,1/3,1/3)$ is special because it
is where the line $(u,u,u)$ pierces the plane $u+v+w=1$.
This plane passes through all three of the lines marking
the collinear limits ($v=0$, $u+w=1$; and cyclic permutations thereof).
Because $R_6(u,v,w)$ vanishes on all three lines, one might expect
it to vanish close to the equilateral triangle that is bounded by them,
which lies in the plane $u+v+w=1$.  Indeed, that is what is seen at three
loops~\cite{Dixon2013eka}.  In this paper, we will not evaluate the four-loop
remainder function on this triangle, but we can verify that the
zero-crossing point remains close to $u=1/3$.   The precise zero-crossing
value at four loops is
\be
u_0^{(4)} = 0.33575561\ldots\,.
\label{uuu_zero_crossing_R64}
\ee
With respect to the three-loop value in \eqn{uuu_zero_crossing_R62_R63},
the zero-crossing point has shifted slightly further away from $u=1/3$.

As can be seen from~\fig{fig:uuu_R64}, $R_6^{(4)}(u,u,u)$ actually crosses
zero in a second place, at a very large value of $u$,
\be
\tilde{u}_{0}^{(4)} = 5529.65453\ldots\, .
\ee
This phenomenon does not happen at two or three loops:
$R_6^{(2)}(u,u,u)$ and $R_6^{(3)}(u,u,u)$ have unique zero crossings,
at the values given in \eqn{uuu_zero_crossing_R62_R63}.
Aside from the zero-crossing neighborhood, \fig{fig:uuu_R64}
shows excellent agreement between the two successive ratios
for relatively small $u$, say $u<1000$.
For large $u$, the ratios approach constant values that
differ by a factor of about $-17.6$ (see~\eqn{ratiouuu_largeu_R64}).

\begin{figure}
\begin{center}
\includegraphics[width=5.5in]{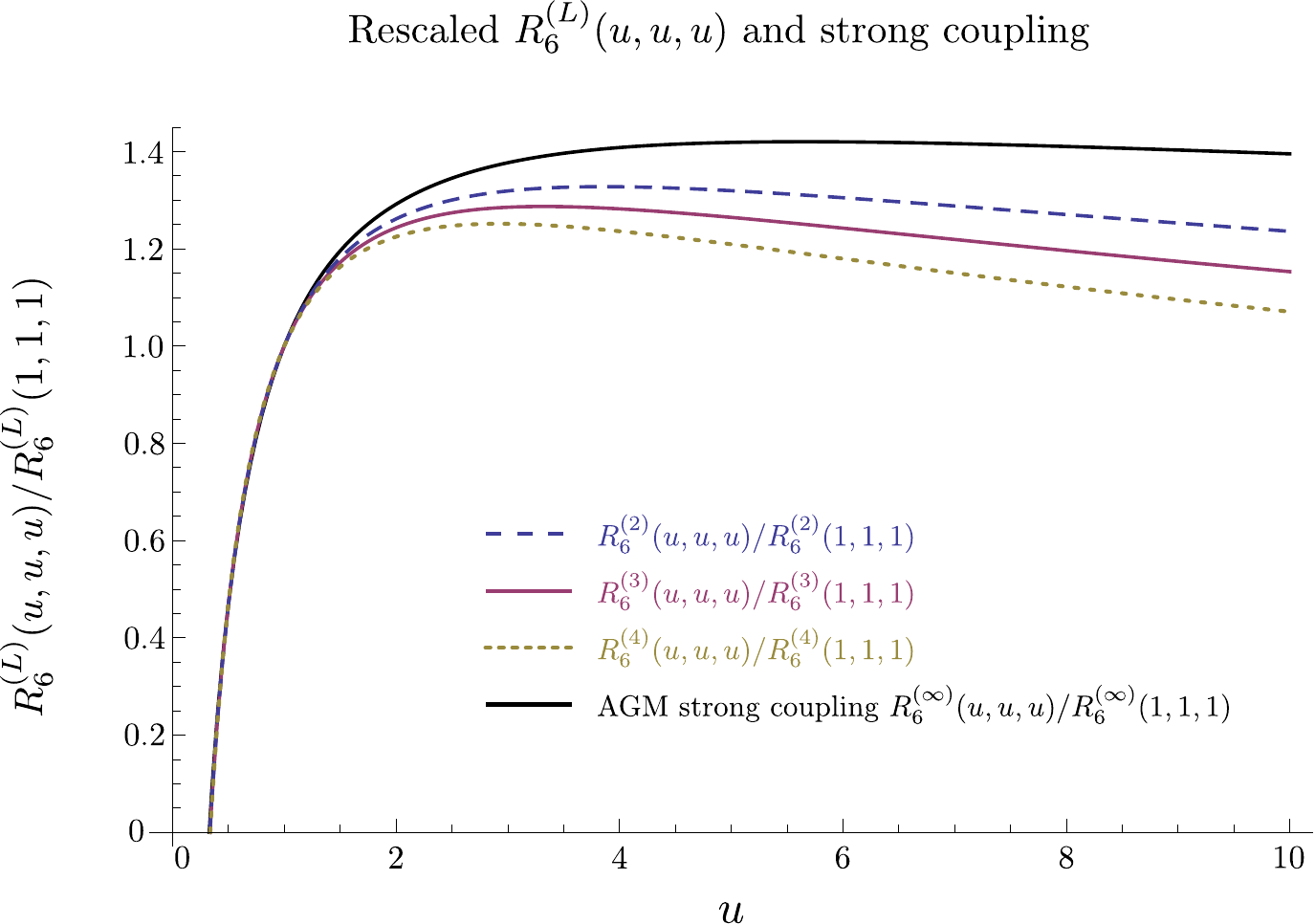}
\end{center}
\caption{The remainder function on the line $(u,u,u)$ plotted at two,
three, and four loops and at strong coupling. The functions have been
rescaled by their values at the point $(1,1,1)$. }
\label{fig:uuu_agm_R64}
\end{figure}

In \fig{fig:uuu_agm_R64}, we plot the two-, three-, and four-loop
and strong-coupling remainder functions on the line $(u,u,u)$.
In order to compare their relative shapes, we rescale each function
by its value at $(1,1,1)$. The remarkable similarity in shape that was
noticed at two loops~\cite{Hatsuda2012pb}\footnote{See
refs.~\cite{Brandhuber2009da,DelDuca2010zp,Hatsuda} for similar
observations for other kinematical configurations.}
and at three loops~\cite{Dixon2013eka} clearly
persists at four loops, particularly for the region $0<u<1$.

As discussed in ref.~\cite{Dixon2013eka}, a necessary condition for the
shapes to be so similar is that the limiting behavior of the ratios as
$u\to0$ is almost the same as the ratios' values at $u=1$.
Comparing~\eqn{ratiouuu_smallu_R64} to~\eqn{ratiouuu_u1_R64}, we find,
\be
\bsp
\frac{R_6^{(3)}(u,u,u)}{R_6^{(2)}(u,u,u)} \Bigg|_{u\to0}
\Big/\frac{R_6^{(3)}(1,1,1)}{R_6^{(2)}(1,1,1)}&\ =\
\biggl[ \frac{59}{63} + \frac{8}{147} \frac{(\zeta_3)^2}{\zeta_6} \biggr]^{-1}
\ \sim\ 
0.986\ldots\,,
\esp\eeq
\beq\bsp
\frac{R_6^{(4)}(u,u,u)}{R_6^{(3)}(u,u,u)}\Bigg|_{u\to0}
\Big/\frac{R_6^{(4)}(1,1,1)}{R_6^{(3)}(1,1,1)}&\ =\
\frac{\Bigl( 597 \zeta_6 - 8 (\zeta_3)^2 \Bigr) 
  \Bigl( \frac{413}{24} \zeta_6 + (\zeta_3)^2 \Bigr)}
{21 \, \zeta_4 \, \Bigl( - 6 \, \zeta_{5,3} + 10 \, \zeta_3 \, \zeta_5
                 + 6 \, \zeta_2 \, (\zeta_3)^2 + 471 \, \zeta_8 \Bigr)}
\ \sim\
0.995\ldots\,.
\esp
\ee
These ratios are indeed quite close to $1$, despite their complicated
representations in terms of $\zeta$ values. 
The agreement is slightly better for the double ratio between
four and three loops, than it is for the one between three and two loops.

We can also compute similar double ratios involving the 
perturbative and strong coupling coefficients,
\be 
\bsp
\frac{R_6^{(\infty)}(u,u,u)}{R_6^{(2)}(u,u,u)}\Bigg|_{u\to0}
\Big/\frac{R_6^{(\infty)}(1,1,1)}{R_6^{(2)}(1,1,1)}&\ \sim\ 1\phantom{.000} \,,\\
\frac{R_6^{(\infty)}(u,u,u)}{R_6^{(3)}(u,u,u)}\Bigg|_{u\to0}
\Big/\frac{R_6^{(\infty)}(1,1,1)}{R_6^{(3)}(1,1,1)}&\ \sim\ 1.014 \,,\\
\frac{R_6^{(\infty)}(u,u,u)}{R_6^{(4)}(u,u,u)}\Bigg|_{u\to0}
\Big/\frac{R_6^{(\infty)}(1,1,1)}{R_6^{(4)}(1,1,1)}&\ \sim\ 1.019 \,.
\esp
\ee
The ratio between the two-loop and strong-coupling points is exactly
$1$, while the corresponding ratios for three and four loops deviate
slightly from one. The deviations increase as $L$ increases,
suggesting that the shapes of the weak-coupling curves on the line
$(u,u,u)$ are getting slightly further from the shape of the strong
coupling curve, at least for small $L$. This observation is also
evident in~\fig{fig:uuu_agm_R64} at large $u$.

Let us conclude this section by making a comment on hexagon functions
on the line $(u,u,u)$. It is easy to check that on this line we have
\beq
u= \frac{y}{(1+y)^2}\,, \qquad y\equiv y_u\,,
\eeq
and the symbol of $R_6^{(4)}(u,u,u)$ has all its entries drawn from the
set $\{y,\Phi_2(y),\Phi_3(y)\}$, where 
\beq
\Phi_2(y) = 1+y {\rm~~and~~}\Phi_3(y) =1+y+y^2
\eeq
denote the second and third cyclotomic polynomials.  It follows then
that $R_6^{(4)}(u,u,u)$ can be entirely expressed through iterated
integrals over $d\ln$ forms with cyclotomic polynomials as
arguments. This class of iterated integrals is a generalization of
HPLs, called \emph{cyclotomic} HPLs, and was studied in detail in
ref.~\cite{Ablinger2011te}. Note that this observation only follows
from the entries in the symbol, and is by no means restricted to four
loops. As a consequence, we conclude that on the line $(u,u,u)$
hexagon functions, and thus the six-point remainder function, can
always be expressed in terms of cyclotomic HPLs.


\subsection{Approach to large orders}
\label{sec:large_Orders}

In the previous subsections, we have found that the portion of the
line $(u,u,u)$ with $0<u<1$ leads to quite constant ratios of
successive loop orders $L$.  We can also ask what this ratio should
become as $L\to\infty$.   Most quantum field theories have a zero
radius of convergence for their perturbative expansions; that is,
the series are asymptotic.  There are two generic reasons for this:
renormalons and instantons, each of which leads to factorial growth
of perturbative coefficients.
However, planar ${\cal N}=4$ super-Yang-Mills theory is free from
both of these phenomena.  Because it is conformally invariant, the
beta function vanishes and there are no renormalons.  Because the number
of colors $N_c$ is very large, at fixed 't Hooft coupling
$\lambda$ instantons are exponentially suppressed as $N_c\to\infty$
by a factor of $\exp(-8\pi^2/g_{\rm YM}^2) = \exp(-8\pi^2 \,N_c/\lambda)$.
Hence we should expect the perturbative expansion to have a
finite radius of convergence $r$.  The radius $r$ corresponds to a
growth rate of successive perturbative coefficients $c^{(L)}$, which
approaches a constant as $L$ becomes large,
\beq
\lim_{L\to\infty} \frac{c^{(L)}}{c^{(L-1)}} = -\frac{1}{r} \,.
\label{asymptotic_growth_rate}
\eeq
In \eqn{asymptotic_growth_rate} we have assumed an alternating series, which holds for $R_6^{(L)}$ for $L=2,3,4$ throughout Region I and on the lines $(u,u,1)$ and $(u,1,1)$, and for $L=2,3$ throughout almost all of the unit cube\footnote{As noted in ref.~\cite{Dixon2013eka}, there is a small region surrounding the plane $u+v+w=1$ in which $R_6^{(2)}$ and $R_6^{(3)}$ have the same sign.}.

\begin{table}[!ht]
\begin{center}
\begin{tabular}{|l|c|c|c|c|}
\hline\hline
\multicolumn{1}{|c|}{$L$} 
&\multicolumn{1}{c|}{$\gamma_K^{(L)} / \gamma_K^{(L-1)}$}
&\multicolumn{1}{c|}{$\bar{R}_6^{(L)}(1,1,1)$}
&\multicolumn{1}{c|}{$\overline{\ln{\cal W}}_{\rm hex}^{(L)}(\tfrac{3}{4},\tfrac{3}{4},\tfrac{3}{4})$}
&\multicolumn{1}{c|}{$\overline{\ln {\cal W}}_{\rm hex}^{(L)}(\tfrac{1}{4},\tfrac{1}{4},\tfrac{1}{4})$} \\
\hline\hline
2 & -1.6449340 & $\infty$ &-2.7697175&-2.8015275 \\
3  & -3.6188549 & -7.0040885  &-5.0036164 &-5.1380714 \\
4  & -4.9211827 & -6.5880519 & -5.8860842&-6.0359857 \\
5  & -5.6547494 & -- &-- &-- \\
6  & -6.0801089 & -- &-- &-- \\
7  & -6.3589220 & -- &-- &-- \\
8  & -6.5608621 & -- &-- &-- \\
9  & -6.7164600 & -- &-- &-- \\
10 & -6.8410049 & -- &-- &-- \\
11 & -6.9432839 & -- &-- &-- \\
12 & -7.0288902 & -- &-- &-- \\
13 & -7.1016320 & -- &-- &-- \\
14 & -7.1642208 & -- &-- &-- \\
15 & -7.2186492 & -- &-- &--\\
\hline\hline
\end{tabular}
\caption{\label{tab:large_orders} We list the ratio of loop order $L$
to the previous order through $L=15$ for the cusp anomalous dimension, and through $L=4$ for the remainder function
and the Wilson loop.
We introduced a bit of notation to save space: $\bar{R}_6^{(L)} \equiv R_6^{(L)} / R_6^{(L-1)}$ and $\overline{\ln {\cal W}}_{\rm hex}^{(L)} \equiv {\ln {\cal W}}_{\rm hex}^{(L)}/ {\ln {\cal W}}_{\rm hex}^{(L-1)}$.}
\end{center}
\end{table}

There is another quantity, closely related to the scattering amplitude,
which we can use as a simple benchmark for assessing large order behavior.
That quantity is the cusp anomalous dimension.
Its perturbative expansion can be computed to all orders using
the exact formula of Beisert, Eden and Staudacher (BES)~\cite{Beisert2006ez}.
Using this formula, we give the ratio of successive loop orders
in table~\ref{tab:large_orders}.  At very large loop orders,
the ratio approaches $-8$, corresponding to a radius of convergence
of $1/8$ when using the loop expansion parameter $a$.  (In terms of
the parameter used by BES, $g^2 = a/2$, the radius of convergence is $1/16$;
or $1/4$ in terms of $g$.)
However, the approach to this asymptotic value is quite slow.

Table~\ref{tab:large_orders} also shows the two nontrivial ratios
currently available for the remainder function at $(u,v,w)=(1,1,1)$,
as representative of the fairly constant region $(u,v,w)=(u,u,u)$ with
$0 < u \lesssim 1$.  We also give values for the three available ratios
for the Wilson loop ratio evaluated at two interior points,
$u=\tfrac{1}{4}$ and $u=\tfrac{3}{4}$. (The Wilson loop ratio diverges
at $(u,v,w)=(1,1,1)$.)  There is an extra ratio available for the Wilson
loop because its one-loop value is nonzero,
due to the function $X(u,v,w)$ appearing in \eqn{eq:WL_convert}.

Suppose that \eqn{asymptotic_growth_rate} holds for all observables in the
theory; {\it i.e.}, that the radius of convergence is the same
for all observables.
An optimist would say that the remainder-function ratios exhibit a
precocious approach to the expected asymptotic value of $-8$:
the cusp anomalous dimension ratio does not reach $-6.5$ until eight loops.
A pessimist would say that the trend is the wrong way: the ratio for $L=4$
is further from $-8$ than is the ratio for $L=3$. On the other hand, the
Wilson loop ratios are actually approaching $-8$ monotonically.
For both $u=\tfrac{1}{4}$ and $u=\tfrac{3}{4}$, they appear to be
converging more quickly to $-8$ than is the cusp anomalous dimension.

It is worth remarking that in Region I for $u=v$, the region shown
in \fig{fig:plot_uw}, the ratio $R_6^{(4)}/R_6^{(3)}$ lies between $-6.6$ and
$-7$ over the entire region shown.  More generally, sampling 1352
points in Region I, including ones with $u\neq v$, the ratio is
always between $-6.60$ and $-8.67$.
Clearly a computation of the remainder-function ratio at the next
loop order, $R_6^{(5)}/R_6^{(4)}$, would be
very illuminating in this regard.


\section{Conclusions}
\label{sec:conclusions}

In this article, we presented the four-loop remainder function, which
is a dual-conformally invariant function that describes six-point MHV
scattering amplitudes in planar $\mathcal{N}=4$ super Yang-Mills
theory. The result was bootstrapped from a limited set of assumptions
about the analytic properties of the relevant function
space. Following the strategy of ref.~\cite{Dixon2011pw}, we
constructed an ansatz for the symbol and constrained this ansatz using
various physical and mathematical consistency conditions. A unique
expression for the symbol was obtained by applying information from
the near-collinear expansion, as generated by the OPE for flux tube
excitations~\cite{Basso2013vsa}. The symbol, in turn, was lifted to a
full function, using the methods described in
ref.~\cite{Dixon2013eka}. In particular, a mathematically-consistent
ansatz for the function was obtained by applying the coproduct
bootstrap described in ref.~\cite{Dixon2013eka}. All of the
function-level parameters of this ansatz were fixed by again applying
information from the near-collinear expansion.

The final expression for the four-loop remainder function is quite
lengthy, but its functional form simplifies dramatically on various
one-dimensional lines in the three-dimensional space of cross
ratios. While the analytic form for the function on these lines is
rather different at two, three, and four loops, a numerical evaluation
shows that they are in fact quite similar for large portions of the
parameter space, at least up to an overall rescaling. On the line
where all three cross ratios are equal, an analytical result at strong
coupling is available. The perturbative coefficients are very similar in
shape to the strong-coupling one, particularly in the region where the
common cross ratio is less than one. This agreement suggests that an
interpolation from weak to strong coupling may depend rather weakly on
the kinematic variables, at least on this one-dimensional line.

Given the full functional form of the four-loop remainder function, it
is straightforward to extract its limit in multi-Regge
kinematics. This information allowed us to fix all of the previously
undetermined constants in the NNLLA BFKL eigenvalue and the N$^3$LLA
impact factor.  Although we used some multi-Regge factorization
information as input, the fact that we found a solution consistent
with all the OPE data suggests that factorization does hold beyond NLLA.
We also observed an intriguing correspondence between
the BFKL eigenvalue and the energy of a gluonic excitation of the GKP
string. It would be very interesting to better understand this
correspondence.

There are many avenues for future research.  For example, it would be
interesting
to try to understand the correspondence between the integrated results
found here (and at three loops) and the types of multi-loop integrals
that appear in recent formulations of the planar multi-loop
integrand~\cite{ArkaniHamed2010kv,ArkaniHamed2012nw,Amplituhedra}.

In implementing the kind of bootstrap used here beyond the six-point case,
it is important to have a good understanding of the relevant
space of functions from results at low loop order.  Progress is being
made on this front~\cite{CaronHuot2011ky,Golden2013lha},
most recently through the introduction of cluster
coordinates~\cite{Golden2013xva} and cluster
polylogarithms~\cite{Golden2014xqa}.

In principle, the methods
used in this work could be extended to five loops and beyond. The primary
limitation is computational power and the availability of boundary
data, such as the near-collinear limit, to fix the proliferation of
constants. It is remarkable that a fully nonperturbative formulation
of the near-collinear limit now exists. Ultimately, the hope is that
the full analytic structure of perturbative scattering amplitudes, as
exposed here through four loops for the six-point case, might in
some way pave the way for a nonperturbative formulation for generic
kinematics.


\vskip0.5cm
\noindent {\large\bf Acknowledgments}
\vskip0.3cm

We are extremely grateful to Benjamin Basso, Amit Sever and Pedro Vieira
for many useful discussions, for insightful comments on the manuscript,
and particularly for sharing their OPE results with us prior to publication.
We are also grateful to Nima Arkani-Hamed, John Joseph Carrasco,
Matt von Hippel, Yuji Satoh, Stephen Shenker, Matthias Staudacher
and Jaroslav Trnka for helpful discussions.
This research was supported by the US Department of Energy under
contract DE--AC02--76SF00515, by the Research Executive Agency (REA) of the
European Union under the Grant Agreement numbers PITN-GA-2010-264564
(LHCPhenoNet) and by the EU Initial Training Network in High-Energy Physics
and Mathematics: GATIS.


\appendix

\section{Sample coproducts}
\label{sec:coproduct}

As mentioned in \sect{sec:functionconstraints},
the construction of a complete set of hexagon functions at weight
five~\cite{Dixon2013eka} facilitated the construction
of $R_6^{(4)}$ at function level in the present paper.  We could identify
the symbols of all the coefficients $[R_6^{(4)}]^{s_i,s_j,s_k}$
of the $\{5,1,1,1\}$ coproduct,
$\Delta_{5,1,1,1}(R_6^{(4)})$ in \eqn{eq:R645111anz},
with linear combinations of the functions constituting the weight-five basis,
modulo the $\zeta$ value ambiguities listed in \tab{tab:R64_bts}.
Besides facilitating the construction, writing the $\{5,1,1,1\}$ coproduct
elements in terms of weight-five hexagon functions also provides a compact
way to {\it define} the final answer for $R_6^{(4)}$.
Essentially we are specifying the function via its derivatives.

In this appendix, we will list a few of the coproduct elements of
$R_6^{(4)}$ to give a flavor for this description, although they
are still too lengthy to list all of them here.  We will provide the
complete set as a computer-readable file~\cite{R64website}.

First, though, we briefly review the connection between the coproduct
and derivatives of hexagon functions~\cite{Dixon2013eka}.
A hexagon function $F$ of weight $n$ has a $\{n-1,1\}$
coproduct component of the form,
\be
\Delta_{n-1,1}(F) \equiv \sum_{i=1}^3 F^{u_i} \otimes \ln u_i 
+ F^{1-u_i} \otimes \ln (1-u_i) + F^{y_i} \otimes \ln y_i\, ,
\ee
where the nine functions $\{F^{u_i},F^{1-u_i}, F^{y_i}\}$ are of
weight $n-1$.  The first derivatives of $F$, in either the $u_i$
variables or the $y_i$ variables, are simple linear combinations
of these coproduct elements:
\be
\label{eq:der_F}
\bsp
\frac{\partial F}{\partial u}\bigg|_{v,w} &= 
\frac{F^u}{u} -\frac{F^{1-u}}{1-u} + \frac{1-u-v-w}{u\sqrt{\Delta}} F^{y_u}
+ \frac{1-u-v+w}{(1-u)\sqrt{\Delta}}F^{y_v}
+ \frac{1-u+v-w}{(1-u)\sqrt{\Delta}} F^{y_w}\,,\\
\sqrt{\Delta} \, y_u \frac{\partial F}{\partial y_u}\bigg|_{y_v,y_w}
&= \, (1-u)(1-v-w) F^u-u(1-v) F^v-u(1-w)F^w - u(1-v-w)F^{1-u} \\
&\quad + uv\,F^{1-v}+uw\,F^{1-w} + \sqrt{\Delta}\, F^{y_u}\, .
\esp
\ee
Derivatives with respect to $v$, $w$, $y_v$ and $y_w$ can be obtained from
the cyclic images of \eqn{eq:der_F}. 

As discussed extensively in ref.~\cite{Dixon2013eka},
the derivatives can be used to define various integral representations
for $F$, which can be evaluated numerically.  It is also possible to
integrate the differential equations analytically in various
kinematical limits.  For example, in the MRK limit,
the appropriate variables are $(\xi,w,\ws)$, where $\xi\equiv1-u_1$
is vanishing and $(w,\ws)$ are defined via \eqn{wdef}.
The differential equations in the MRK variables are~\cite{Dixon2013eka},
\be
\bsp
\frac{\partial F}{\partial \xi} 
&= -\frac{\partial F}{\partial u_1}
+ x\frac{\partial F}{\partial u_2} + y\frac{\partial F}{\partial u_3} \,,\\
\frac{\partial F}{\partial w} 
&= \frac{\xi}{w(1+w)}\left[
- w x\frac{\partial F}{\partial u_2} + y\frac{\partial F}{\partial u_3}\right]
\,,\\
\frac{\partial F}{\partial \ws} 
&= \frac{\xi}{\ws(1+\ws)}\left[
- \ws x\frac{\partial F}{\partial u_2} + y\frac{\partial F}{\partial u_3}\right]
\,.\\
\esp
\label{diff_xiwws}
\ee
Using \eqn{eq:der_F} and its cyclic images, we find that the $w$
derivative can be rewritten directly in terms of the coproduct
elements as,
\be
\frac{\partial F}{\partial w} =
\frac{1}{w} \Bigl( F^{u_3} - F^{y_3} \Bigr)
- \frac{1}{1+w} \Bigl( F^{u_2} + F^{u_3} + F^{y_2} - F^{y_3} \Bigr) \,.
\label{MRKcoprod}
\ee
This differential equation can be integrated up systematically
in terms of SVHPLs.

The MRK limiting behavior of all the weight-five hexagon functions was given in
ref.~\cite{Dixon2013eka}.  These results give directly the MRK limits
of all the independent elements $\Delta_{5,1,1,1}(R_6^{(4)})$.
Then we can integrate up \eqn{MRKcoprod} in order to get the MRK behavior
of all the $\Delta_{6,1,1}(R_6^{(4)})$ elements, integrate once more to get
the limiting behavior of the $\Delta_{7,1}(R_6^{(4)})$ elements,
and integrate a final time to get the desired MRK behavior of $R_6^{(4)}$ itself.

How many coproduct components have to be specified?  Thanks to the 
$S_3$ permutation symmetry of $R_6^{(4)}(u,v,w)$ and the differential
constraint corresponding to the final-entry condition,
the number is manageable.  First of all, there are only two
independent $\{7,1\}$ coproduct elements,
\be
 R^u \qquad \hbox{and} \qquad R^{y_u} \,,
\label{coprod71}
\ee
where we have suppressed the subscript 6 and superscript $(4)$ to
avoid clutter in subsequent equations.  The final-entry constraint becomes
\be
R^{1-u} = - R^u,\qquad R^{1-v} = - R^v,\qquad R^{1-w} = - R^w,
\label{Rextrapure}
\ee
for the coproduct.  The $S_3$ symmetry implies that the other elements can
be obtained by permuting the two elements given in \eqn{coprod71},
\be
\bsp
&R^v(u,v,w) = R^u(v,w,u), \quad R^w(u,v,w) = R^u(w,u,v), \\
&R^{y_v}(u,v,w) = R^{y_u}(v,w,u), \quad R^{y_w}(u,v,w) = R^{y_u}(w,u,v).
\label{perm71}
\esp
\ee

There are 11 independent $\{6,1,1\}$ coproduct elements:
\be
\bsp
&R^{u,u}, \quad R^{1-u,u}, \quad R^{y_u,u} = R^{u,y_u}, \quad R^{1-u,y_u}, 
\quad R^{y_u,y_u}, \\
&R^{v,u}, \quad R^{1-v,u}, \quad R^{y_v,u}, \quad R^{v,y_u}, \quad R^{1-v,y_u}, 
\quad R^{y_v,y_u}.
\label{coprod611}
\esp
\ee
The counting is as follows:  Using the cyclic symmetry, the last
entry can be rotated to be $u$, $1-u$ or $y_u$.  However,
the final-entry condition at function level
\eqn{Rextrapure} says that a last entry of $1-u$ can be exchanged for 
a last entry of $u$, at the price of a minus sign.
There is still a residual flip symmetry, 
exchanging $v \lr w$, which allows the next-to-last entry to
be forbidden from being $w$, $1-w$ or $y_w$.
That counting leaves 12 possibilities; however, we also find that
$R^{y_u,u} = R^{u,y_u}$, which presumably follows from integrability.

Here we will give the $\{5,1,1,1\}$ coproduct elements that
allow the construction of $R^{u,u}$.  In fact, the $\{5,1,1,1\}$ coproduct
entries allow us to construct the total derivative of $R^{u,u}$, so we
need to supplement them with a constant of integration, which we
specify at the point $(u,v,w)=(1,1,1)$:
\be
R^{u,u}(1,1,1) = \frac{73}{8} \zeta_6 - \frac{1}{2} (\zeta_3)^2.
\ee
Now, using the residual $v \lr w$
flip symmetry for $R^{u,u}$, the six independent elements required
to specify $R^{u,u}$ are:
$R^{u,u,u}$, $R^{1-u,u,u}$, $R^{v,u,u}$, $R^{1-v,u,u}$, $R^{y_u,u,u}$ and
$R^{y_v,u,u}$.
The parity-odd elements $R^{y_u,u,u}$ and $R^{y_v,u,u}$ are much simpler
to represent, because the basis of weight-five parity-odd functions
is much smaller than the parity-even basis.  They are given by,
\be
\bsp
R^{y_u,u,u} &= \frac{1}{128} \biggl[ 
- 3 \Bigl( H_1(u,v,w) + H_1(v,w,u) + H_1(w,u,v) \Bigr)
+ \frac{1}{4} \Bigl( 11 \, [ J_1(u,v,w) + J_1(v,w,u) ] \\
&\qquad + 7 \, J_1(w,u,v) \Bigr)
 + 2 \, H_1^u \, \Bigl( 2 \, F_1(u,v,w) - F_1(v,w,u) - F_1(w,u,v) \Bigr)\\
&\qquad + \Bigl( 2 \, H_2^u - 14 \, ( H_2^v + H_2^w ) - 7 \, (H_1^u)^2
   - 3 \, [ (H_1^v)^2 + (H_1^w)^2 ] - 8 \, H_1^v \, H_1^w \\
&\qquad\qquad
      - 2 \, H_1^u \, ( H_1^v + H_1^w ) + 74 \, \zeta_2 \Bigl) 
 \, \tilde{\Phi}_6(u,v,w) \biggr] \,,
\label{R_yu_u_u}
\esp
\ee
\be
\bsp
R^{y_v,u,u} &= \frac{1}{256} \biggl[ 
- 5 \, [ H_1(u,v,w) + H_1(v,w,u) ] - 13 \, H_1(w,u,v)
+ \frac{1}{4} \Bigl( 5 \, J_1(u,v,w) + 25 \, J_1(v,w,u)\\
&\qquad 
+ 9 \, J_1(w,u,v) \Bigr)
+ 4 \, H_1^u \, \Bigl( 3 \, F_1(u,v,w) - F_1(v,w,u) - 2 \, F_1(w,u,v) \Bigr)\\
&\qquad 
+ \Bigl( 6 \, H_2^u - 26 \, ( H_2^v + H_2^w )
  - 9 \, (H_1^u)^2 - 5 \, [ (H_1^v)^2 + (H_1^w)^2 ]
 - 4 \, H_1^u \, H_1^v - 8 \, H_1^u \, H_1^w \\
&\qquad\qquad - 16 \, H_1^v \, H_1^w
  + 110 \,\zeta_2 \Bigr) \, \tilde{\Phi}_6(u,v,w) \biggr] \,.
\label{R_yv_u_u}
\esp
\ee
The four parity-even elements are given by,
\newcommand{\LogSu}[0]{\ln u}
\newcommand{\LogSv}[0]{\ln v}
\newcommand{\LogSw}[0]{\ln w}
\newcommand{\Logu}[1]{\ln^{#1}u}
\newcommand{\Logv}[1]{\ln^{#1}v}
\newcommand{\Logw}[1]{\ln^{#1}w}
\newcommand{\hplsA}[2]{H_{#2}^{#1}}
\newcommand{\hplsB}[3]{H_{#2,#3}^{#1}}
\newcommand{\hplsC}[4]{H_{#2,#3,#4}^{#1}}
\newcommand{\hplsD}[5]{H_{#2,#3,#4,#5}^{#1}}
\newcommand{\hplsE}[6]{H_{#2,#3,#4,#5,#6}^{#1}}
\newcommand{\hplA}[3]{(H_{#3}^{#2})^{#1}}
\newcommand{\hplB}[4]{(H_{#3,#4}^{#2})^{#1}}
\newcommand{\hplC}[5]{(H_{#3,#4,#5}^{#2})^{#1}}
\newcommand{\hplD}[6]{(H_{#3,#4,#5,#6}^{#2})^{#1}}
\newcommand{\hplE}[7]{(H_{#3,#4,#5,#6,#7}^{#2})^{#1}}
\beq\bsp\nonumber
R^{u,u,u} &\,=\frac{11}{384}\biggl[M_1(u,v,w)+M_1(u,w,v)-M_1(v,u,w)-M_1(w,u,v)\biggr]\\
&\,-\frac{1}{12}\biggl[\Qep(u,v,w)+\Qep(u,w,v)\biggr]-\frac{17}{18}\biggl[\Qep(v,u,w)+\Qep(w,u,v)\biggr]\\
&\,+\frac{19}{36}\biggl[\Qep(v,w,u)+\Qep(w,v,u)\biggr]+\frac{1}{96}\,N(u,v,w) + \frac{1}{96}\,O(u,v,w)
\esp\eeq
\beq\bsp\nonumber
&\,+\ln u\, \biggl[\frac{1}{6}\Omega^{(2)}(u,v,w)+\frac{5}{192}\Omega^{(2)}(v,w,u)+\frac{1}{6}\Omega^{(2)}(w,u,v)\biggr]\\
&\,+\frac{1}{384}\,\ln v\, \biggl[15\,\Omega^{(2)}(v,w,u)+\Omega^{(2)}(w,u,v)-\Omega^{(2)}(u,v,w)\biggr]\\
&\,+\frac{1}{384}\,\ln w\, \biggl[15\,\Omega^{(2)}(v,w,u)+\Omega^{(2)}(u,v,w)-\Omega^{(2)}(w,u,v)\biggr]\\
&\,-\frac{47}{1152}\,\hplsB{u}{2}{1}\,\hplsA{v}{2}+\frac{121}{2304}\,\hplsA{u}{2}\,\hplsB{v}{2}{1}-\frac{47}{1152}\,\hplsB{u}{2}{1}\,\hplsA{w}{2}+\frac{121}{2304}\,\hplsA{u}{2}\,\hplsB{w}{2}{1}-\frac{191}{192}\,\hplsA{u}{2}\,\hplsB{u}{2}{1}\\
&\,+\frac{47}{256}\,\hplsA{v}{2}\,\hplsB{w}{2}{1}+\frac{47}{256}\,\hplsB{v}{2}{1}\,\hplsA{w}{2}+\frac{1}{192}\,\hplsA{v}{2}\,\hplsB{v}{2}{1}+\frac{1}{192}\,\hplsA{w}{2}\,\hplsB{w}{2}{1}-\frac{47}{2304}\,\hplsA{u}{2}\,\hplsA{v}{3}\\
&\,-\frac{53}{1152}\,\hplsA{u}{3}\,\hplsA{v}{2}-\frac{47}{2304}\,\hplsA{u}{2}\,\hplsA{w}{3}-\frac{53}{1152}\,\hplsA{u}{3}\,\hplsA{w}{2}-\frac{89}{24}\,\hplsA{u}{5}+\frac{61}{96}\,\hplsA{u}{2}\,\hplsA{u}{3}+\frac{11}{256}\,\hplsA{v}{2}\,\hplsA{w}{3}\\
&\,+\frac{11}{256}\,\hplsA{v}{3}\,\hplsA{w}{2}-\frac{35}{96}\,\hplsA{v}{5}-\frac{1}{48}\,\hplsA{v}{2}\,\hplsA{v}{3}-\frac{35}{96}\,\hplsA{w}{5}-\frac{1}{48}\,\hplsA{w}{2}\,\hplsA{w}{3}+\frac{13}{64}\,\hplsB{u}{3}{2}+\frac{79}{24}\,\hplsB{u}{4}{1}\\
&\,+\frac{5}{32}\,\hplsB{v}{3}{2}+\frac{23}{48}\,\hplsB{v}{4}{1}+\frac{5}{32}\,\hplsB{w}{3}{2}+\frac{23}{48}\,\hplsB{w}{4}{1}-\frac{53}{32}\,\hplsC{u}{3}{1}{1}+\frac{71}{192}\,\hplsC{u}{2}{2}{1}+\frac{15}{64}\,\hplsC{v}{3}{1}{1}\\
&\,+\frac{17}{192}\,\hplsC{v}{2}{2}{1}+\frac{15}{64}\,\hplsC{w}{3}{1}{1}+\frac{17}{192}\,\hplsC{w}{2}{2}{1}+\frac{15}{4}\,\hplsD{u}{2}{1}{1}{1}-\frac{9}{16}\,\hplsD{v}{2}{1}{1}{1}-\frac{9}{16}\,\hplsD{w}{2}{1}{1}{1}\\
&\,+\frac{1}{6}\,\LogSv\,\hplsA{w}{4}+\frac{1}{6}\,\LogSw\,\hplsA{v}{4}-\frac{1}{16}\,\LogSv\,\hplsB{v}{3}{1}-\frac{1}{16}\,\LogSw\,\hplsB{w}{3}{1}+\frac{1}{128}\,\Logv{3}\,\hplsA{v}{2}\\
&\,+\frac{1}{128}\,\Logw{3}\,\hplsA{w}{2}-\frac{1}{768}\,\LogSv\,\hplA{2}{v}{2}-\frac{1}{768}\,\LogSw\,\hplA{2}{w}{2}+\frac{3}{16}\,\LogSv\,\hplsC{w}{2}{1}{1}\\
&\,+\frac{3}{16}\,\LogSw\,\hplsC{v}{2}{1}{1}-\frac{3}{64}\,\LogSu\,\hplsC{v}{2}{1}{1}-\frac{3}{64}\,\LogSu\,\hplsC{w}{2}{1}{1}-\frac{3}{64}\,\Logv{2}\,\hplsB{v}{2}{1}\\
\phantom{R^{u,u,u}}
&\,-\frac{3}{64}\,\Logw{2}\,\hplsB{w}{2}{1}-\frac{5}{16}\,\LogSu\,\hplA{2}{v}{2}-\frac{5}{16}\,\LogSu\,\hplA{2}{w}{2}+\frac{7}{192}\,\LogSv\,\hplsB{u}{3}{1}+\frac{7}{192}\,\LogSw\,\hplsB{u}{3}{1}\\
&\,+\frac{7}{384}\,\Logu{3}\,\hplsA{u}{2}-\frac{7}{768}\,\LogSv\,\hplA{2}{u}{2}-\frac{7}{768}\,\LogSw\,\hplA{2}{u}{2}-\frac{9}{8}\,\LogSu\,\hplsB{u}{3}{1}-\frac{11}{64}\,\LogSu\,\hplsB{v}{3}{1}\\
&\,-\frac{11}{64}\,\LogSu\,\hplsB{w}{3}{1}+\frac{11}{1536}\,\Logv{3}\,\hplsA{u}{2}+\frac{11}{1536}\,\Logw{3}\,\hplsA{u}{2}-\frac{11}{2304}\,\Logv{2}\,\hplsB{u}{2}{1}\\
&\,-\frac{11}{2304}\,\Logw{2}\,\hplsB{u}{2}{1}-\frac{13}{192}\,\LogSv\,\hplsB{w}{3}{1}-\frac{13}{192}\,\LogSw\,\hplsB{v}{3}{1}-\frac{21}{64}\,\LogSv\,\hplsC{v}{2}{1}{1}-\frac{21}{64}\,\LogSw\,\hplsC{w}{2}{1}{1}\\
\phantom{R^{u,u,u}}&\,-\frac{23}{1536}\,\Logv{3}\,\hplsA{w}{2}-\frac{23}{1536}\,\Logw{3}\,\hplsA{v}{2}+\frac{25}{12}\,\LogSu\,\hplsA{u}{4}-\frac{29}{384}\,\Logv{2}\,\hplsA{v}{3}-\frac{29}{384}\,\Logw{2}\,\hplsA{w}{3}\\
&\,-\frac{31}{48}\,\Logu{2}\,\hplsA{u}{3}+\frac{49}{192}\,\LogSv\,\hplsA{v}{4}+\frac{49}{192}\,\LogSw\,\hplsA{w}{4}+\frac{53}{64}\,\LogSu\,\hplsC{u}{2}{1}{1}-\frac{67}{768}\,\Logu{3}\,\hplsA{v}{2}\\
&\,-\frac{67}{768}\,\Logu{3}\,\hplsA{w}{2}+\frac{67}{2304}\,\Logv{2}\,\hplsA{u}{3}+\frac{67}{2304}\,\Logw{2}\,\hplsA{u}{3}-\frac{83}{768}\,\LogSv\,\hplA{2}{w}{2}\\
&\,-\frac{83}{768}\,\LogSw\,\hplA{2}{v}{2}-\frac{83}{1536}\,\Logv{2}\,\hplsB{w}{2}{1}-\frac{83}{1536}\,\Logw{2}\,\hplsB{v}{2}{1}+\frac{89}{1536}\,\Logv{2}\,\hplsA{w}{3}\\
&\,+\frac{89}{1536}\,\Logw{2}\,\hplsA{v}{3}+\frac{103}{192}\,\LogSu\,\hplsA{v}{4}+\frac{103}{192}\,\LogSu\,\hplsA{w}{4}-\frac{109}{192}\,\LogSu\,\hplA{2}{u}{2}+\frac{361}{4608}\,\Logu{2}\,\hplsB{v}{2}{1}
\esp\eeq
\beq\bsp
\phantom{R^{u,u,u}}
&\,+\frac{361}{4608}\,\Logu{2}\,\hplsB{w}{2}{1}+\frac{769}{4608}\,\Logu{2}\,\hplsA{v}{3}+\frac{769}{4608}\,\Logu{2}\,\hplsA{w}{3}+\frac{1}{12}\,\LogSu\,\LogSv\,\hplsB{w}{2}{1}\\
&\,+\frac{1}{12}\,\LogSu\,\LogSw\,\hplsB{v}{2}{1}-\frac{1}{24}\,\LogSv\,\LogSw\,\hplsB{u}{2}{1}+\frac{3}{64}\,\LogSu\,\Logv{2}\,\hplsA{v}{2}+\frac{3}{64}\,\LogSu\,\Logw{2}\,\hplsA{w}{2}\\
&\,+\frac{3}{64}\,\LogSv\,\Logw{2}\,\hplsA{w}{2}+\frac{3}{64}\,\Logv{2}\,\LogSw\,\hplsA{v}{2}-\frac{5}{48}\,\LogSv\,\LogSw\,\hplsA{u}{3}-\frac{5}{192}\,\LogSv\,\hplsA{u}{2}\,\hplsA{w}{2}\\
&\,-\frac{5}{192}\,\LogSw\,\hplsA{u}{2}\,\hplsA{v}{2}-\frac{5}{384}\,\LogSu\,\Logv{2}\,\hplsA{w}{2}-\frac{5}{384}\,\LogSu\,\Logw{2}\,\hplsA{v}{2}+\frac{5}{2304}\,\LogSu\,\Logv{2}\,\hplsA{u}{2}\\
&\,+\frac{5}{2304}\,\LogSu\,\Logw{2}\,\hplsA{u}{2}+\frac{11}{768}\,\LogSv\,\hplsA{v}{2}\,\hplsA{w}{2}+\frac{11}{768}\,\LogSw\,\hplsA{v}{2}\,\hplsA{w}{2}+\frac{17}{192}\,\LogSu\,\LogSv\,\hplsB{v}{2}{1}\\
&\,+\frac{17}{192}\,\LogSu\,\LogSw\,\hplsB{w}{2}{1}-\frac{23}{96}\,\LogSv\,\LogSw\,\hplsA{v}{3}-\frac{23}{96}\,\LogSv\,\LogSw\,\hplsA{w}{3}+\frac{23}{2304}\,\LogSv\,\hplsA{u}{2}\,\hplsA{v}{2}\\
&\,+\frac{23}{2304}\,\LogSw\,\hplsA{u}{2}\,\hplsA{w}{2}-\frac{25}{768}\,\LogSv\,\Logw{2}\,\hplsA{u}{2}-\frac{25}{768}\,\Logv{2}\,\LogSw\,\hplsA{u}{2}-\frac{31}{96}\,\LogSu\,\LogSv\,\hplsA{v}{3}\\
&\,-\frac{31}{96}\,\LogSu\,\LogSw\,\hplsA{w}{3}+\frac{37}{192}\,\LogSv\,\LogSw\,\hplsB{v}{2}{1}+\frac{37}{192}\,\LogSv\,\LogSw\,\hplsB{w}{2}{1}-\frac{41}{192}\,\LogSu\,\LogSv\,\hplsA{w}{3}\\
&\,-\frac{41}{192}\,\LogSu\,\LogSw\,\hplsA{v}{3}+\frac{53}{1152}\,\LogSu\,\hplsA{u}{2}\,\hplsA{v}{2}+\frac{53}{1152}\,\LogSu\,\hplsA{u}{2}\,\hplsA{w}{2}-\frac{61}{96}\,\LogSu\,\hplsA{v}{2}\,\hplsA{w}{2}\\
&\,-\frac{97}{768}\,\Logu{2}\,\LogSv\,\hplsA{w}{2}-\frac{97}{768}\,\Logu{2}\,\LogSw\,\hplsA{v}{2}-\frac{103}{1536}\,\LogSv\,\Logw{2}\,\hplsA{v}{2}-\frac{103}{1536}\,\Logv{2}\,\LogSw\,\hplsA{w}{2}\\
&\,-\frac{187}{4608}\,\Logu{2}\,\LogSv\,\hplsA{v}{2}-\frac{187}{4608}\,\Logu{2}\,\LogSw\,\hplsA{w}{2}+\frac{1}{24}\,\LogSu\,\LogSv\,\LogSw\,\hplsA{u}{2}\\
&\,-\frac{19}{48}\,\LogSu\,\LogSv\,\LogSw\,\hplsA{v}{2}-\frac{19}{48}\,\LogSu\,\LogSv\,\LogSw\,\hplsA{w}{2}+\frac{17}{24}\,\zeta_3\,\hplsA{u}{2}+\frac{87}{128}\,\zeta_2\,\hplsA{u}{3}-\frac{7}{64}\,\zeta_2\,\hplsA{v}{3}\\
&\,-\frac{53}{192}\,\zeta_3\,\hplsA{v}{2}-\frac{7}{64}\,\zeta_2\,\hplsA{w}{3}-\frac{53}{192}\,\zeta_3\,\hplsA{w}{2}+\frac{13}{96}\,\zeta_2\,\hplsB{u}{2}{1}-\frac{29}{96}\,\zeta_2\,\hplsB{v}{2}{1}-\frac{29}{96}\,\zeta_2\,\hplsB{w}{2}{1}\\
&\,-\frac{73}{384}\,\zeta_2\,\LogSu\,\hplsA{u}{2}+\frac{53}{48}\,\zeta_2\,\LogSu\,\hplsA{v}{2}+\frac{53}{48}\,\zeta_2\,\LogSu\,\hplsA{w}{2}+\frac{7}{192}\,\zeta_2\,\LogSv\,\hplsA{u}{2}-\frac{13}{96}\,\zeta_2\,\LogSv\,\hplsA{v}{2}\\
&\,+\frac{7}{64}\,\zeta_2\,\LogSv\,\hplsA{w}{2}+\frac{7}{192}\,\zeta_2\,\LogSw\,\hplsA{u}{2}+\frac{7}{64}\,\zeta_2\,\LogSw\,\hplsA{v}{2}-\frac{13}{96}\,\zeta_2\,\LogSw\,\hplsA{w}{2}\\
&\,-\frac{1}{192}\,\LogSu\,\LogSv\,\Logw{3}-\frac{1}{192}\,\LogSu\,\Logv{3}\,\LogSw-\frac{167}{768}\,\LogSu\,\Logv{2}\,\Logw{2}\\
&\,-\frac{5}{192}\,\Logu{3}\,\LogSv\,\LogSw-\frac{97}{1536}\,\Logu{2}\,\LogSv\,\Logw{2}-\frac{97}{1536}\,\Logu{2}\,\Logv{2}\,\LogSw\\
&\,-\frac{17}{3072}\,\Logu{2}\,\Logv{3}-\frac{47}{1536}\,\Logu{3}\,\Logv{2}-\frac{17}{3072}\,\Logu{2}\,\Logw{3}-\frac{47}{1536}\,\Logu{3}\,\Logw{2}\\
&\,-\frac{17}{1024}\,\Logv{2}\,\Logw{3}-\frac{17}{1024}\,\Logv{3}\,\Logw{2}-\frac{57}{16}\,\zeta_4\,\LogSu+\frac{31}{24}\,\zeta_2\,\LogSu\,\LogSv\,\LogSw\\
&\,-\frac{3}{32}\,\zeta_2\,\LogSu\,\Logv{2}-\frac{3}{32}\,\zeta_2\,\LogSu\,\Logw{2}+\frac{15}{64}\,\zeta_4\,\LogSv+\frac{3}{32}\,\zeta_2\,\LogSv\,\Logw{2}+\frac{15}{64}\,\zeta_4\,\LogSw\\
&\,+\frac{3}{32}\,\zeta_2\,\Logv{2}\,\LogSw+\frac{1}{128}\,\zeta_2\,\Logu{3}+\frac{23}{96}\,\zeta_3\,\Logu{2}-\frac{5}{192}\,\zeta_2\,\Logv{3}-\frac{31}{384}\,\zeta_3\,\Logv{2}\\
&\,-\frac{5}{192}\,\zeta_2\,\Logw{3}-\frac{31}{384}\,\zeta_3\,\Logw{2}+\frac{5}{8}\,\zeta_5-\frac{1}{4}\,\zeta_2\,\zeta_3\,,
\esp\eeq

\beq\bsp\nonumber
R^{1-u,u,u} &\,=\frac{3}{64}\,\ln u\,\biggl[\Omega^{(2)}(u,v,w)+2\Omega^{(2)}(v,w,u)+\Omega^{(2)}(w,u,v)\biggr]\\
&\,+\frac{5}{12}\biggl[\Qep(v,u,w)-\Qep(v,w,u)+\Qep(w,u,v)-\Qep(w,v,u)\biggr]\\
&\,-\frac{35}{768}\,\hplsA{u}{2}\,\hplsB{v}{2}{1}+\frac{395}{768}\,\hplsB{u}{2}{1}\,\hplsA{v}{2}-\frac{35}{768}\,\hplsA{u}{2}\,\hplsB{w}{2}{1}+\frac{395}{768}\,\hplsB{u}{2}{1}\,\hplsA{w}{2}+\frac{167}{128}\,\hplsA{u}{2}\,\hplsB{u}{2}{1}\\
&\,+\frac{45}{256}\,\hplsA{v}{2}\,\hplsB{v}{2}{1}+\frac{45}{256}\,\hplsA{w}{2}\,\hplsB{w}{2}{1}+\frac{25}{768}\,\hplsA{u}{2}\,\hplsA{v}{3}-\frac{361}{768}\,\hplsA{u}{3}\,\hplsA{v}{2}+\frac{25}{768}\,\hplsA{u}{2}\,\hplsA{w}{3}\\
&\,-\frac{361}{768}\,\hplsA{u}{3}\,\hplsA{w}{2}+6\,\hplsA{u}{5}-\frac{85}{64}\,\hplsA{u}{2}\,\hplsA{u}{3}+\frac{5}{128}\,\hplsA{v}{2}\,\hplsA{v}{3}+\frac{5}{128}\,\hplsA{w}{2}\,\hplsA{w}{3}+\frac{31}{128}\,\hplsB{u}{3}{2}\\
&\,-\frac{123}{32}\,\hplsB{u}{4}{1}-\frac{5}{64}\,\hplsB{v}{4}{1}-\frac{15}{256}\,\hplsB{v}{3}{2}-\frac{5}{64}\,\hplsB{w}{4}{1}-\frac{15}{256}\,\hplsB{w}{3}{2}-\frac{17}{128}\,\hplsC{u}{2}{2}{1}+\frac{111}{32}\,\hplsC{u}{3}{1}{1}\\
&\,-\frac{75}{64}\,\hplsC{v}{3}{1}{1}-\frac{95}{256}\,\hplsC{v}{2}{2}{1}-\frac{75}{64}\,\hplsC{w}{3}{1}{1}-\frac{95}{256}\,\hplsC{w}{2}{2}{1}-\frac{81}{16}\,\hplsD{u}{2}{1}{1}{1}+\frac{5}{32}\,\hplsD{v}{2}{1}{1}{1}\\
&\,+\frac{5}{32}\,\hplsD{w}{2}{1}{1}{1}-\frac{1}{4}\,\LogSu\,\hplsC{v}{2}{1}{1}-\frac{1}{4}\,\LogSu\,\hplsC{w}{2}{1}{1}+\frac{3}{16}\,\LogSu\,\hplsA{v}{4}+\frac{3}{16}\,\LogSu\,\hplsA{w}{4}\\
&\,+\frac{3}{64}\,\LogSu\,\hplA{2}{v}{2}+\frac{3}{64}\,\LogSu\,\hplA{2}{w}{2}-\frac{5}{16}\,\Logu{2}\,\hplsB{u}{2}{1}+\frac{5}{64}\,\LogSv\,\hplsA{v}{4}-\frac{5}{64}\,\LogSv\,\hplsA{w}{4}\\
&\,-\frac{5}{64}\,\LogSv\,\hplsC{w}{2}{1}{1}-\frac{5}{64}\,\LogSw\,\hplsA{v}{4}+\frac{5}{64}\,\LogSw\,\hplsA{w}{4}-\frac{5}{64}\,\LogSw\,\hplsC{v}{2}{1}{1}+\frac{5}{64}\,\Logv{2}\,\hplsB{v}{2}{1}\\
&\,-\frac{5}{64}\,\Logv{2}\,\hplsB{w}{2}{1}-\frac{5}{64}\,\Logw{2}\,\hplsB{v}{2}{1}+\frac{5}{64}\,\Logw{2}\,\hplsB{w}{2}{1}-\frac{5}{128}\,\LogSv\,\hplA{2}{w}{2}\\
&\,-\frac{5}{128}\,\LogSw\,\hplA{2}{v}{2}+\frac{5}{384}\,\Logv{3}\,\hplsA{w}{2}+\frac{5}{384}\,\Logw{3}\,\hplsA{v}{2}-\frac{5}{1536}\,\Logv{3}\,\hplsA{u}{2}\\
&\,-\frac{5}{1536}\,\Logw{3}\,\hplsA{u}{2}+\frac{7}{256}\,\Logu{3}\,\hplsA{u}{2}+\frac{7}{512}\,\Logu{3}\,\hplsA{v}{2}+\frac{7}{512}\,\Logu{3}\,\hplsA{w}{2}-\frac{9}{32}\,\LogSu\,\hplsB{v}{3}{1}\\
&\,-\frac{9}{32}\,\LogSu\,\hplsB{w}{3}{1}+\frac{15}{64}\,\LogSv\,\hplsB{w}{3}{1}+\frac{15}{64}\,\LogSw\,\hplsB{v}{3}{1}-\frac{15}{256}\,\Logv{2}\,\hplsA{v}{3}-\frac{15}{256}\,\Logw{2}\,\hplsA{w}{3}\\
&\,+\frac{15}{512}\,\LogSv\,\hplA{2}{v}{2}+\frac{15}{512}\,\LogSw\,\hplA{2}{w}{2}-\frac{35}{128}\,\LogSv\,\hplsB{v}{3}{1}-\frac{35}{128}\,\LogSw\,\hplsB{w}{3}{1}\\
&\,+\frac{35}{256}\,\LogSv\,\hplsC{v}{2}{1}{1}+\frac{35}{256}\,\LogSw\,\hplsC{w}{2}{1}{1}+\frac{35}{1536}\,\Logv{3}\,\hplsA{v}{2}+\frac{35}{1536}\,\Logw{3}\,\hplsA{w}{2}\\
&\,-\frac{49}{16}\,\LogSu\,\hplsA{u}{4}+\frac{59}{128}\,\Logu{2}\,\hplsA{u}{3}+\frac{83}{1536}\,\Logv{2}\,\hplsB{u}{2}{1}+\frac{83}{1536}\,\Logw{2}\,\hplsB{u}{2}{1}\\
&\,-\frac{95}{1536}\,\Logu{2}\,\hplsA{v}{3}-\frac{95}{1536}\,\Logu{2}\,\hplsA{w}{3}+\frac{145}{256}\,\LogSu\,\hplA{2}{u}{2}+\frac{149}{64}\,\LogSu\,\hplsB{u}{3}{1}\\
&\,-\frac{217}{1536}\,\Logv{2}\,\hplsA{u}{3}-\frac{217}{1536}\,\Logw{2}\,\hplsA{u}{3}-\frac{275}{1536}\,\Logu{2}\,\hplsB{v}{2}{1}-\frac{275}{1536}\,\Logu{2}\,\hplsB{w}{2}{1}\\
&\,-\frac{303}{128}\,\LogSu\,\hplsC{u}{2}{1}{1}+\frac{1}{8}\,\LogSu\,\hplsA{v}{2}\,\hplsA{w}{2}-\frac{3}{16}\,\LogSv\,\LogSw\,\hplsA{u}{3}+\frac{3}{64}\,\LogSu\,\LogSv\,\hplsB{v}{2}{1}
\esp\eeq
\beq\bsp
\phantom{R^{1-u,u,u}}
&\,+\frac{3}{64}\,\LogSu\,\LogSw\,\hplsB{w}{2}{1}+\frac{5}{64}\,\LogSu\,\LogSv\,\hplsA{w}{3}+\frac{5}{64}\,\LogSu\,\LogSw\,\hplsA{v}{3}+\frac{5}{64}\,\LogSv\,\LogSw\,\hplsA{v}{3}\\
&\,+\frac{5}{64}\,\LogSv\,\LogSw\,\hplsA{w}{3}-\frac{5}{64}\,\LogSv\,\LogSw\,\hplsB{v}{2}{1}-\frac{5}{64}\,\LogSv\,\LogSw\,\hplsB{w}{2}{1}+\frac{5}{128}\,\LogSu\,\Logv{2}\,\hplsA{v}{2}\\
&\,+\frac{5}{128}\,\LogSu\,\Logw{2}\,\hplsA{w}{2}-\frac{5}{128}\,\LogSv\,\Logw{2}\,\hplsA{v}{2}-\frac{5}{128}\,\LogSv\,\Logw{2}\,\hplsA{w}{2}\\
&\,-\frac{5}{128}\,\Logv{2}\,\LogSw\,\hplsA{v}{2}-\frac{5}{128}\,\Logv{2}\,\LogSw\,\hplsA{w}{2}+\frac{13}{32}\,\LogSv\,\LogSw\,\hplsB{u}{2}{1}+\frac{15}{64}\,\LogSu\,\LogSv\,\hplsB{w}{2}{1}\\
&\,+\frac{15}{64}\,\LogSu\,\LogSw\,\hplsB{v}{2}{1}-\frac{15}{128}\,\LogSu\,\Logv{2}\,\hplsA{w}{2}-\frac{15}{128}\,\LogSu\,\Logw{2}\,\hplsA{v}{2}+\frac{15}{128}\,\Logu{2}\,\LogSv\,\hplsA{w}{2}\\
&\,+\frac{15}{128}\,\Logu{2}\,\LogSw\,\hplsA{v}{2}-\frac{17}{64}\,\LogSu\,\LogSv\,\hplsA{v}{3}-\frac{17}{64}\,\LogSu\,\LogSw\,\hplsA{w}{3}-\frac{25}{768}\,\LogSv\,\hplsA{u}{2}\,\hplsA{v}{2}\\
&\,-\frac{25}{768}\,\LogSw\,\hplsA{u}{2}\,\hplsA{w}{2}-\frac{35}{1536}\,\LogSu\,\Logv{2}\,\hplsA{u}{2}-\frac{35}{1536}\,\LogSu\,\Logw{2}\,\hplsA{u}{2}\\
&\,-\frac{85}{1536}\,\Logu{2}\,\LogSv\,\hplsA{v}{2}-\frac{85}{1536}\,\Logu{2}\,\LogSw\,\hplsA{w}{2}-\frac{107}{768}\,\LogSu\,\hplsA{u}{2}\,\hplsA{v}{2}-\frac{107}{768}\,\LogSu\,\hplsA{u}{2}\,\hplsA{w}{2}\\
&\,-\frac{3}{32}\,\LogSu\,\LogSv\,\LogSw\,\hplsA{u}{2}+\frac{9}{32}\,\LogSu\,\LogSv\,\LogSw\,\hplsA{v}{2}+\frac{9}{32}\,\LogSu\,\LogSv\,\LogSw\,\hplsA{w}{2}-\frac{15}{64}\,\zeta_3\,\hplsA{u}{2}\\
&\,+\frac{91}{256}\,\zeta_2\,\hplsA{u}{3}+\frac{5}{512}\,\zeta_2\,\hplsA{v}{3}+\frac{15}{128}\,\zeta_3\,\hplsA{v}{2}+\frac{5}{512}\,\zeta_2\,\hplsA{w}{3}+\frac{15}{128}\,\zeta_3\,\hplsA{w}{2}-\frac{57}{64}\,\zeta_2\,\hplsB{u}{2}{1}\\
&\,-\frac{35}{128}\,\zeta_2\,\hplsB{v}{2}{1}-\frac{35}{128}\,\zeta_2\,\hplsB{w}{2}{1}+\frac{53}{256}\,\zeta_2\,\LogSu\,\hplsA{u}{2}-\frac{23}{32}\,\zeta_2\,\LogSu\,\hplsA{v}{2}-\frac{23}{32}\,\zeta_2\,\LogSu\,\hplsA{w}{2}\\
&\,-\frac{85}{512}\,\zeta_2\,\LogSv\,\hplsA{v}{2}+\frac{5}{32}\,\zeta_2\,\LogSv\,\hplsA{w}{2}+\frac{5}{32}\,\zeta_2\,\LogSw\,\hplsA{v}{2}-\frac{85}{512}\,\zeta_2\,\LogSw\,\hplsA{w}{2}\\
&\,-\frac{5}{192}\,\LogSu\,\LogSv\,\Logw{3}-\frac{5}{192}\,\LogSu\,\Logv{3}\,\LogSw+\frac{1}{16}\,\LogSu\,\Logv{2}\,\Logw{2}\\
&\,+\frac{3}{64}\,\Logu{3}\,\LogSv\,\LogSw+\frac{15}{256}\,\Logu{2}\,\LogSv\,\Logw{2}+\frac{15}{256}\,\Logu{2}\,\Logv{2}\,\LogSw\\
&\,-\frac{17}{1024}\,\Logu{3}\,\Logv{2}-\frac{25}{3072}\,\Logu{2}\,\Logv{3}-\frac{17}{1024}\,\Logu{3}\,\Logw{2}-\frac{25}{3072}\,\Logu{2}\,\Logw{3}\\
&\,+\frac{161}{32}\,\zeta_4\,\LogSu-\frac{19}{16}\,\zeta_2\,\LogSu\,\LogSv\,\LogSw+\frac{15}{64}\,\zeta_2\,\LogSu\,\Logv{2}+\frac{15}{64}\,\zeta_2\,\LogSu\,\Logw{2}\\
&\,+\frac{5}{64}\,\zeta_2\,\LogSv\,\Logw{2}+\frac{5}{64}\,\zeta_2\,\Logv{2}\,\LogSw-\frac{15}{128}\,\zeta_3\,\Logu{2}-\frac{19}{256}\,\zeta_2\,\Logu{3}+\frac{15}{256}\,\zeta_3\,\Logv{2}\\
&\,-\frac{55}{1536}\,\zeta_2\,\Logv{3}+\frac{15}{256}\,\zeta_3\,\Logw{2}-\frac{55}{1536}\,\zeta_2\,\Logw{3}\,,
\esp\eeq

\beq\bsp\nonumber
R^{v,u,u}&\,= \frac{1}{12}\biggl[2\Qep(u,v,w)-2\Qep(u,w,v)+3\Qep(v,w,u)+5\Qep(w,u,v)-8\Qep(w,v,u)\biggr]\\
&\,+\frac{1}{128}\biggl[-M_1(u,v,w)+M_1(u,w,v)+M_1(v,u,w)-M_1(w,u,v)\biggr]\\
&\,+\frac{1}{128}\ln v\,\biggl[9\Omega^{(2)}(u,v,w)+3\Omega^{(2)}(v,w,u)+7\Omega^{(2)}(w,u,v)\biggr]
\esp\eeq
\beq\bsp\nonumber
&\,+\frac{1}{128}\ln w\,\biggl[7\Omega^{(2)}(u,v,w)+9\Omega^{(2)}(v,w,u)+5\Omega^{(2)}(w,u,v)\biggr]\\
&\,+\frac{1}{32}\ln u\,\biggl[\Omega^{(2)}(v,w,u)+\Omega^{(2)}(w,u,v)\biggr]%
-\frac{5}{48}\,\hplsA{u}{2}\,\hplsB{v}{2}{1}+\frac{71}{768}\,\hplsB{u}{2}{1}\,\hplsA{v}{2}\\
&\,-\frac{1}{64}\,\hplsB{u}{2}{1}\,\hplsA{w}{2}-\frac{9}{256}\,\hplsA{u}{2}\,\hplsB{w}{2}{1}-\frac{1}{256}\,\hplsA{u}{2}\,\hplsB{u}{2}{1}-\frac{43}{384}\,\hplsA{v}{2}\,\hplsB{w}{2}{1}+\frac{103}{384}\,\hplsB{v}{2}{1}\,\hplsA{w}{2}\\
&\,+\frac{27}{128}\,\hplsA{v}{2}\,\hplsB{v}{2}{1}-\frac{129}{256}\,\hplsA{w}{2}\,\hplsB{w}{2}{1}+\frac{5}{96}\,\hplsA{u}{2}\,\hplsA{v}{3}-\frac{37}{768}\,\hplsA{u}{3}\,\hplsA{v}{2}+\frac{1}{64}\,\hplsA{u}{3}\,\hplsA{w}{2}\\
&\,-\frac{13}{256}\,\hplsA{u}{2}\,\hplsA{w}{3}-\frac{1}{128}\,\hplsA{u}{2}\,\hplsA{u}{3}-\frac{13}{384}\,\hplsA{v}{2}\,\hplsA{w}{3}-\frac{95}{384}\,\hplsA{v}{3}\,\hplsA{w}{2}+\frac{25}{32}\,\hplsA{v}{5}-\frac{15}{64}\,\hplsA{v}{2}\,\hplsA{v}{3}\\
&\,+\frac{11}{32}\,\hplsA{w}{5}-\frac{1}{128}\,\hplsA{w}{2}\,\hplsA{w}{3}-\frac{1}{256}\,\hplsB{u}{3}{2}-\frac{3}{64}\,\hplsB{u}{4}{1}+\frac{7}{128}\,\hplsB{v}{3}{2}-\frac{13}{32}\,\hplsB{v}{4}{1}-\frac{19}{64}\,\hplsB{w}{4}{1}\\
&\,-\frac{29}{256}\,\hplsB{w}{3}{2}+\frac{15}{64}\,\hplsC{u}{3}{1}{1}+\frac{15}{256}\,\hplsC{u}{2}{2}{1}-\frac{3}{128}\,\hplsC{v}{2}{2}{1}+\frac{29}{64}\,\hplsC{v}{3}{1}{1}+\frac{23}{8}\,\hplsC{w}{3}{1}{1}\\
&\,+\frac{231}{256}\,\hplsC{w}{2}{2}{1}-\frac{5}{32}\,\hplsD{u}{2}{1}{1}{1}-\frac{3}{4}\,\hplsD{v}{2}{1}{1}{1}-\frac{3}{32}\,\hplsD{w}{2}{1}{1}{1}+\frac{1}{4}\,\LogSw\,\hplsB{u}{3}{1}+\frac{1}{8}\,\LogSw\,\hplsA{v}{4}\\
&\,+\frac{1}{16}\,\LogSv\,\hplA{2}{v}{2}-\frac{1}{16}\,\LogSv\,\hplsC{w}{2}{1}{1}-\frac{1}{16}\,\Logu{2}\,\hplsB{u}{2}{1}+\frac{1}{32}\,\LogSw\,\hplsA{w}{4}-\frac{1}{64}\,\LogSu\,\hplsA{u}{4}\\
&\,-\frac{1}{64}\,\LogSw\,\hplsA{u}{4}-\frac{1}{64}\,\Logv{2}\,\hplsA{v}{3}+\frac{1}{64}\,\Logw{2}\,\hplsB{u}{2}{1}+\frac{1}{64}\,\Logw{2}\,\hplsB{w}{2}{1}+\frac{1}{128}\,\Logw{2}\,\hplsA{u}{3}\\
&\,-\frac{1}{512}\,\Logw{3}\,\hplsA{u}{2}+\frac{1}{768}\,\Logw{2}\,\hplsB{v}{2}{1}+\frac{1}{1536}\,\Logu{3}\,\hplsA{u}{2}+\frac{1}{1536}\,\Logu{3}\,\hplsA{v}{2}+\frac{3}{8}\,\LogSv\,\hplsB{v}{3}{1}\\
&\,-\frac{3}{16}\,\LogSu\,\hplsC{v}{2}{1}{1}-\frac{3}{32}\,\LogSw\,\hplsC{v}{2}{1}{1}-\frac{3}{32}\,\Logv{2}\,\hplsB{v}{2}{1}-\frac{3}{512}\,\LogSu\,\hplA{2}{u}{2}+\frac{5}{32}\,\LogSv\,\hplsA{w}{4}\\
&\,+\frac{5}{64}\,\LogSu\,\hplsA{w}{4}+\frac{5}{128}\,\LogSu\,\hplA{2}{v}{2}-\frac{5}{192}\,\Logu{3}\,\hplsA{w}{2}-\frac{5}{256}\,\Logu{2}\,\hplsA{u}{3}-\frac{5}{384}\,\Logv{3}\,\hplsA{w}{2}\\
&\,-\frac{5}{384}\,\Logw{3}\,\hplsA{v}{2}+\frac{7}{32}\,\LogSu\,\hplsA{v}{4}+\frac{7}{32}\,\LogSv\,\hplsB{u}{3}{1}-\frac{7}{64}\,\LogSu\,\hplA{2}{w}{2}+\frac{7}{64}\,\LogSu\,\hplsC{w}{2}{1}{1}\\
&\,-\frac{9}{32}\,\LogSu\,\hplsB{v}{3}{1}+\frac{9}{64}\,\LogSu\,\hplsB{w}{3}{1}-\frac{9}{256}\,\LogSv\,\hplA{2}{w}{2}-\frac{11}{32}\,\LogSv\,\hplsA{v}{4}+\frac{11}{128}\,\LogSu\,\hplsB{u}{3}{1}\\
&\,+\frac{11}{768}\,\Logv{3}\,\hplsA{v}{2}-\frac{13}{64}\,\LogSv\,\hplsA{u}{4}-\frac{13}{768}\,\Logv{2}\,\hplsA{w}{3}-\frac{13}{768}\,\Logv{3}\,\hplsA{u}{2}-\frac{15}{64}\,\LogSw\,\hplsC{u}{2}{1}{1}\\
\phantom{R^{v,u,u}}&\,-\frac{19}{192}\,\Logu{2}\,\hplsA{v}{3}-\frac{21}{64}\,\LogSv\,\hplsB{w}{3}{1}+\frac{21}{64}\,\LogSw\,\hplsB{v}{3}{1}-\frac{21}{512}\,\Logu{2}\,\hplsA{w}{3}-\frac{23}{768}\,\Logw{2}\,\hplsA{v}{3}\\
&\,+\frac{29}{256}\,\LogSv\,\hplA{2}{u}{2}-\frac{33}{256}\,\LogSw\,\hplA{2}{u}{2}-\frac{35}{256}\,\LogSw\,\hplsC{w}{2}{1}{1}-\frac{37}{1536}\,\Logv{2}\,\hplsA{u}{3}\\
&\,-\frac{39}{64}\,\LogSv\,\hplsC{u}{2}{1}{1}+\frac{41}{1536}\,\Logw{3}\,\hplsA{w}{2}-\frac{43}{256}\,\Logw{2}\,\hplsA{w}{3}-\frac{47}{256}\,\LogSw\,\hplA{2}{v}{2}\\
&\,+\frac{55}{384}\,\Logu{2}\,\hplsB{v}{2}{1}-\frac{59}{256}\,\LogSu\,\hplsC{u}{2}{1}{1}+\frac{59}{1536}\,\Logv{2}\,\hplsB{u}{2}{1}+\frac{63}{128}\,\LogSw\,\hplsB{w}{3}{1}\\
&\,-\frac{71}{128}\,\LogSv\,\hplsC{v}{2}{1}{1}+\frac{83}{512}\,\Logu{2}\,\hplsB{w}{2}{1}-\frac{109}{512}\,\LogSw\,\hplA{2}{w}{2}+\frac{143}{768}\,\Logv{2}\,\hplsB{w}{2}{1}\\
&\,-\frac{1}{16}\,\LogSu\,\LogSv\,\hplsB{u}{2}{1}+\frac{1}{16}\,\LogSu\,\LogSw\,\hplsB{u}{2}{1}+\frac{1}{16}\,\LogSu\,\LogSw\,\hplsB{v}{2}{1}-\frac{1}{32}\,\LogSu\,\hplsA{v}{2}\,\hplsA{w}{2}
\esp\eeq
\beq\bsp
\phantom{R^{v,u,u}}
&\,-\frac{1}{32}\,\LogSu\,\Logw{2}\,\hplsA{u}{2}+\frac{1}{64}\,\LogSu\,\LogSv\,\hplsA{w}{3}+\frac{1}{64}\,\LogSu\,\LogSv\,\hplsB{v}{2}{1}-\frac{1}{64}\,\LogSu\,\Logv{2}\,\hplsA{w}{2}\\
&\,-\frac{1}{64}\,\LogSv\,\LogSw\,\hplsB{u}{2}{1}+\frac{1}{64}\,\LogSv\,\LogSw\,\hplsB{w}{2}{1}+\frac{1}{64}\,\LogSv\,\hplsA{u}{2}\,\hplsA{w}{2}-\frac{1}{128}\,\LogSv\,\Logw{2}\,\hplsA{w}{2}\\
&\,-\frac{3}{32}\,\LogSu\,\LogSw\,\hplsA{v}{3}-\frac{3}{64}\,\LogSu\,\hplsA{u}{2}\,\hplsA{w}{2}+\frac{3}{128}\,\LogSu\,\Logw{2}\,\hplsA{w}{2}-\frac{3}{256}\,\Logv{2}\,\LogSw\,\hplsA{u}{2}\\
&\,+\frac{5}{64}\,\LogSu\,\LogSv\,\hplsA{u}{3}-\frac{5}{64}\,\LogSu\,\Logw{2}\,\hplsA{v}{2}+\frac{5}{128}\,\LogSu\,\Logv{2}\,\hplsA{v}{2}+\frac{5}{128}\,\Logu{2}\,\LogSw\,\hplsA{u}{2}\\
&\,+\frac{5}{384}\,\Logv{2}\,\LogSw\,\hplsA{w}{2}-\frac{7}{64}\,\LogSu\,\LogSw\,\hplsA{w}{3}+\frac{7}{64}\,\LogSv\,\LogSw\,\hplsB{v}{2}{1}-\frac{7}{64}\,\LogSw\,\hplsA{u}{2}\,\hplsA{v}{2}\\
&\,+\frac{7}{128}\,\Logu{2}\,\LogSv\,\hplsA{u}{2}-\frac{9}{64}\,\LogSu\,\LogSv\,\hplsA{v}{3}+\frac{9}{64}\,\LogSu\,\LogSw\,\hplsB{w}{2}{1}-\frac{9}{64}\,\LogSv\,\LogSw\,\hplsA{w}{3}\\
&\,+\frac{9}{128}\,\Logv{2}\,\LogSw\,\hplsA{v}{2}-\frac{11}{384}\,\LogSv\,\Logw{2}\,\hplsA{v}{2}+\frac{13}{1536}\,\LogSu\,\Logv{2}\,\hplsA{u}{2}-\frac{17}{256}\,\LogSv\,\Logw{2}\,\hplsA{u}{2}\\
&\,-\frac{17}{256}\,\Logu{2}\,\LogSv\,\hplsA{w}{2}+\frac{17}{384}\,\LogSv\,\hplsA{v}{2}\,\hplsA{w}{2}-\frac{19}{64}\,\LogSv\,\LogSw\,\hplsA{v}{3}-\frac{19}{192}\,\LogSv\,\hplsA{u}{2}\,\hplsA{v}{2}\\
&\,-\frac{23}{256}\,\LogSw\,\hplsA{u}{2}\,\hplsA{w}{2}-\frac{25}{64}\,\LogSu\,\LogSw\,\hplsA{u}{3}-\frac{27}{64}\,\LogSu\,\LogSv\,\hplsB{w}{2}{1}+\frac{37}{768}\,\Logu{2}\,\LogSv\,\hplsA{v}{2}\\
&\,-\frac{43}{256}\,\Logu{2}\,\LogSw\,\hplsA{v}{2}-\frac{49}{512}\,\Logu{2}\,\LogSw\,\hplsA{w}{2}+\frac{49}{768}\,\LogSu\,\hplsA{u}{2}\,\hplsA{v}{2}-\frac{65}{384}\,\LogSw\,\hplsA{v}{2}\,\hplsA{w}{2}\\
&\,-\frac{3}{8}\,\LogSu\,\LogSv\,\LogSw\,\hplsA{w}{2}-\frac{7}{64}\,\LogSu\,\LogSv\,\LogSw\,\hplsA{v}{2}-\frac{5}{512}\,\zeta_2\,\hplsA{u}{3}-\frac{15}{128}\,\zeta_3\,\hplsA{u}{2}+\frac{5}{32}\,\zeta_3\,\hplsA{v}{2}\\
&\,+\frac{111}{256}\,\zeta_2\,\hplsA{v}{3}-\frac{5}{128}\,\zeta_3\,\hplsA{w}{2}+\frac{71}{512}\,\zeta_2\,\hplsA{w}{3}-\frac{1}{128}\,\zeta_2\,\hplsB{u}{2}{1}-\frac{25}{64}\,\zeta_2\,\hplsB{v}{2}{1}+\frac{135}{128}\,\zeta_2\,\hplsB{w}{2}{1}\\
&\,-\frac{27}{512}\,\zeta_2\,\LogSu\,\hplsA{u}{2}-\frac{9}{32}\,\zeta_2\,\LogSu\,\hplsA{v}{2}+\frac{7}{32}\,\zeta_2\,\LogSu\,\hplsA{w}{2}-\frac{17}{64}\,\zeta_2\,\LogSv\,\hplsA{u}{2}-\frac{43}{256}\,\zeta_2\,\LogSv\,\hplsA{v}{2}\\
&\,-\frac{7}{64}\,\zeta_2\,\LogSv\,\hplsA{w}{2}+\frac{33}{64}\,\zeta_2\,\LogSw\,\hplsA{u}{2}+\frac{47}{64}\,\zeta_2\,\LogSw\,\hplsA{v}{2}+\frac{433}{512}\,\zeta_2\,\LogSw\,\hplsA{w}{2}\\
&\,-\frac{5}{192}\,\LogSu\,\LogSv\,\Logw{3}+\frac{1}{384}\,\LogSu\,\Logv{3}\,\LogSw-\frac{23}{256}\,\LogSu\,\Logv{2}\,\Logw{2}\\
&\,-\frac{7}{192}\,\Logu{3}\,\LogSv\,\LogSw-\frac{95}{512}\,\Logu{2}\,\LogSv\,\Logw{2}-\frac{29}{512}\,\Logu{2}\,\Logv{2}\,\LogSw\\
&\,-\frac{5}{512}\,\Logu{2}\,\Logv{3}-\frac{5}{1024}\,\Logu{3}\,\Logv{2}-\frac{7}{384}\,\Logu{3}\,\Logw{2}-\frac{47}{3072}\,\Logu{2}\,\Logw{3}\\
&\,-\frac{1}{48}\,\Logv{2}\,\Logw{3}-\frac{1}{128}\,\Logv{3}\,\Logw{2}+\frac{45}{64}\,\zeta_4\,\LogSu+\frac{25}{32}\,\zeta_2\,\LogSu\,\LogSv\,\LogSw\\
&\,-\frac{5}{64}\,\zeta_2\,\LogSu\,\Logv{2}+\frac{9}{64}\,\zeta_2\,\LogSu\,\Logw{2}+\frac{79}{32}\,\zeta_4\,\LogSv-\frac{7}{64}\,\zeta_2\,\Logu{2}\,\LogSv\\
&\,+\frac{17}{64}\,\zeta_2\,\LogSv\,\Logw{2}-\frac{201}{32}\,\zeta_4\,\LogSw+\frac{19}{64}\,\zeta_2\,\Logu{2}\,\LogSw-\frac{9}{64}\,\zeta_2\,\Logv{2}\,\LogSw\\
&\,-\frac{15}{256}\,\zeta_3\,\Logu{2}-\frac{17}{1536}\,\zeta_2\,\Logu{3}+\frac{1}{16}\,\zeta_3\,\Logv{2}-\frac{13}{768}\,\zeta_2\,\Logv{3}-\frac{1}{256}\,\zeta_3\,\Logw{2}\\
&\,-\frac{85}{1536}\,\zeta_2\,\Logw{3}\,,
\esp\eeq


\beq\bsp\nonumber
R^{1-v,u,u}&\,= \frac{1}{64}\biggl[M_1(u,v,w)-M_1(u,w,v)-M_1(v,u,w)+M_1(w,u,v)\biggr]\\
&\,+\frac{1}{12}\biggl[-3\Qep(u,v,w)+3\Qep(u,w,v)+3\Qep(v,u,w)\\
&\,\qquad-3\Qep(v,w,u)-8\Qep(w,u,v)+8\Qep(w,v,u)\biggr]\\
&\,+\frac{1}{32}\,\ln u\,\biggl[\Omega^{(2)}(u,v,w)-\Omega^{(2)}(v,w,u)-2\Omega^{(2)}(w,u,v)\biggr]\\
&\,+\frac{1}{64}\,\ln v\,\biggl[-8\Omega^{(2)}(u,v,w)-5\Omega^{(2)}(v,w,u)-6\Omega^{(2)}(w,u,v)\biggr]\\
&\,+\frac{1}{64}\,\ln w\,\biggl[-\Omega^{(2)}(u,v,w)-\Omega^{(2)}(v,w,u)+\Omega^{(2)}(w,u,v)\biggr]\\
&\,-\frac{17}{192}\,\hplsB{u}{2}{1}\,\hplsA{v}{2}+\frac{41}{192}\,\hplsA{u}{2}\,\hplsB{v}{2}{1}+\frac{3}{256}\,\hplsB{u}{2}{1}\,\hplsA{w}{2}-\frac{19}{256}\,\hplsA{u}{2}\,\hplsB{w}{2}{1}+\frac{1}{256}\,\hplsA{u}{2}\,\hplsB{u}{2}{1}\\
&\,+\frac{27}{256}\,\hplsA{v}{2}\,\hplsB{w}{2}{1}-\frac{67}{256}\,\hplsB{v}{2}{1}\,\hplsA{w}{2}+\frac{31}{256}\,\hplsA{v}{2}\,\hplsB{v}{2}{1}+\frac{11}{64}\,\hplsA{w}{2}\,\hplsB{w}{2}{1}-\frac{1}{48}\,\hplsA{u}{2}\,\hplsA{v}{3}\\
&\,+\frac{5}{96}\,\hplsA{u}{3}\,\hplsA{v}{2}+\frac{5}{256}\,\hplsA{u}{2}\,\hplsA{w}{3}-\frac{5}{256}\,\hplsA{u}{3}\,\hplsA{w}{2}+\frac{1}{128}\,\hplsA{u}{2}\,\hplsA{u}{3}-\frac{5}{256}\,\hplsA{v}{2}\,\hplsA{w}{3}\\
&\,+\frac{77}{256}\,\hplsA{v}{3}\,\hplsA{w}{2}-\frac{9}{8}\,\hplsA{v}{5}+\frac{31}{128}\,\hplsA{v}{2}\,\hplsA{v}{3}+\frac{1}{256}\,\hplsB{u}{3}{2}+\frac{3}{64}\,\hplsB{u}{4}{1}+\frac{15}{256}\,\hplsB{v}{3}{2}+\frac{49}{64}\,\hplsB{v}{4}{1}\\
&\,-\frac{1}{16}\,\hplsB{w}{4}{1}-\frac{15}{64}\,\hplsC{u}{3}{1}{1}-\frac{15}{256}\,\hplsC{u}{2}{2}{1}-\frac{145}{256}\,\hplsC{v}{2}{2}{1}-\frac{153}{64}\,\hplsC{v}{3}{1}{1}-\frac{5}{16}\,\hplsC{w}{2}{2}{1}\\
&\,-\frac{15}{16}\,\hplsC{w}{3}{1}{1}+\frac{5}{32}\,\hplsD{u}{2}{1}{1}{1}+\frac{23}{32}\,\hplsD{v}{2}{1}{1}{1}+\frac{1}{8}\,\hplsD{w}{2}{1}{1}{1}-\frac{1}{8}\,\LogSw\,\hplsB{w}{3}{1}+\frac{1}{16}\,\LogSu\,\hplA{2}{v}{2}\\
&\,-\frac{1}{16}\,\LogSu\,\hplsA{w}{4}+\frac{1}{16}\,\LogSw\,\hplsC{w}{2}{1}{1}+\frac{1}{16}\,\Logu{2}\,\hplsB{u}{2}{1}+\frac{1}{16}\,\Logv{2}\,\hplsB{v}{2}{1}-\frac{1}{32}\,\LogSw\,\hplsA{w}{4}\\
&\,-\frac{1}{48}\,\Logv{2}\,\hplsB{u}{2}{1}+\frac{1}{64}\,\LogSu\,\hplsA{u}{4}+\frac{1}{64}\,\Logw{2}\,\hplsA{w}{3}+\frac{1}{64}\,\Logw{2}\,\hplsB{w}{2}{1}+\frac{1}{96}\,\Logu{3}\,\hplsA{v}{2}\\
&\,+\frac{1}{128}\,\LogSu\,\hplA{2}{w}{2}+\frac{1}{128}\,\LogSv\,\hplA{2}{u}{2}+\frac{1}{128}\,\LogSw\,\hplA{2}{u}{2}+\frac{1}{384}\,\Logw{3}\,\hplsA{w}{2}\\
&\,-\frac{1}{512}\,\Logv{2}\,\hplsB{w}{2}{1}-\frac{1}{1536}\,\Logu{3}\,\hplsA{u}{2}+\frac{1}{1536}\,\Logw{3}\,\hplsA{u}{2}-\frac{1}{1536}\,\Logw{3}\,\hplsA{v}{2}\\
&\,+\frac{3}{64}\,\LogSw\,\hplA{2}{w}{2}-\frac{3}{64}\,\LogSw\,\hplsA{u}{4}+\frac{3}{512}\,\LogSu\,\hplA{2}{u}{2}+\frac{3}{512}\,\Logv{2}\,\hplsA{w}{3}-\frac{5}{64}\,\LogSv\,\hplsC{w}{2}{1}{1}\\
&\,-\frac{5}{64}\,\LogSw\,\hplsC{u}{2}{1}{1}-\frac{5}{128}\,\LogSw\,\hplA{2}{v}{2}+\frac{5}{192}\,\Logv{2}\,\hplsA{u}{3}+\frac{5}{256}\,\Logu{2}\,\hplsA{u}{3}\\
&\,-\frac{5}{512}\,\Logw{2}\,\hplsA{u}{3}+\frac{7}{32}\,\LogSu\,\hplsC{w}{2}{1}{1}+\frac{7}{384}\,\Logv{3}\,\hplsA{u}{2}+\frac{9}{64}\,\LogSu\,\hplsB{v}{3}{1}-\frac{9}{64}\,\LogSu\,\hplsC{v}{2}{1}{1}\\
&\,+\frac{9}{64}\,\LogSv\,\hplsB{w}{3}{1}-\frac{9}{64}\,\LogSw\,\hplsB{v}{3}{1}+\frac{11}{32}\,\LogSv\,\hplsA{v}{4}+\frac{11}{64}\,\LogSw\,\hplsB{u}{3}{1}-\frac{11}{128}\,\LogSu\,\hplsB{u}{3}{1}
\esp\eeq
\beq\bsp\nonumber
&\,-\frac{11}{512}\,\Logu{2}\,\hplsA{w}{3}+\frac{13}{64}\,\LogSw\,\hplsA{v}{4}-\frac{15}{64}\,\LogSu\,\hplsA{v}{4}+\frac{15}{64}\,\LogSw\,\hplsC{v}{2}{1}{1}+\frac{17}{64}\,\LogSv\,\hplsA{u}{4}\\
&\,-\frac{17}{512}\,\Logw{2}\,\hplsB{u}{2}{1}+\frac{21}{512}\,\Logw{2}\,\hplsA{v}{3}-\frac{23}{512}\,\Logu{2}\,\hplsB{w}{2}{1}+\frac{23}{1536}\,\Logu{3}\,\hplsA{w}{2}\\
&\,-\frac{25}{96}\,\Logu{2}\,\hplsB{v}{2}{1}-\frac{31}{64}\,\LogSv\,\hplsA{w}{4}+\frac{31}{192}\,\Logu{2}\,\hplsA{v}{3}+\frac{33}{128}\,\LogSv\,\hplA{2}{w}{2}-\frac{41}{64}\,\LogSv\,\hplsB{u}{3}{1}\\
&\,+\frac{41}{1536}\,\Logv{3}\,\hplsA{w}{2}+\frac{43}{256}\,\Logv{2}\,\hplsA{v}{3}+\frac{53}{512}\,\LogSv\,\hplA{2}{v}{2}+\frac{59}{64}\,\LogSv\,\hplsC{u}{2}{1}{1}\\
&\,+\frac{59}{256}\,\LogSu\,\hplsC{u}{2}{1}{1}-\frac{67}{1536}\,\Logv{3}\,\hplsA{v}{2}-\frac{95}{128}\,\LogSv\,\hplsB{v}{3}{1}-\frac{95}{512}\,\Logw{2}\,\hplsB{v}{2}{1}\\
&\,+\frac{161}{256}\,\LogSv\,\hplsC{v}{2}{1}{1}+\frac{1}{8}\,\LogSv\,\hplsA{u}{2}\,\hplsA{w}{2}+\frac{1}{16}\,\LogSu\,\LogSv\,\hplsB{w}{2}{1}+\frac{1}{16}\,\LogSu\,\LogSw\,\hplsA{v}{3}\\
&\,+\frac{1}{16}\,\LogSu\,\LogSw\,\hplsA{w}{3}+\frac{1}{32}\,\LogSu\,\hplsA{v}{2}\,\hplsA{w}{2}-\frac{1}{32}\,\LogSw\,\hplsA{u}{2}\,\hplsA{v}{2}+\frac{1}{32}\,\Logv{2}\,\LogSw\,\hplsA{v}{2}\\
&\,-\frac{1}{48}\,\LogSu\,\hplsA{u}{2}\,\hplsA{v}{2}+\frac{1}{48}\,\LogSu\,\Logv{2}\,\hplsA{u}{2}+\frac{1}{64}\,\LogSu\,\LogSv\,\hplsA{w}{3}-\frac{1}{64}\,\LogSu\,\Logw{2}\,\hplsA{w}{2}\\
&\,+\frac{1}{64}\,\LogSv\,\LogSw\,\hplsB{u}{2}{1}-\frac{1}{128}\,\LogSu\,\Logw{2}\,\hplsA{v}{2}-\frac{1}{128}\,\LogSv\,\Logw{2}\,\hplsA{u}{2}+\frac{1}{256}\,\LogSu\,\hplsA{u}{2}\,\hplsA{w}{2}\\
&\,+\frac{1}{512}\,\LogSu\,\Logw{2}\,\hplsA{u}{2}+\frac{3}{16}\,\LogSu\,\LogSv\,\hplsA{v}{3}+\frac{3}{32}\,\LogSu\,\LogSv\,\hplsB{u}{2}{1}-\frac{3}{32}\,\LogSu\,\LogSw\,\hplsB{u}{2}{1}\\
&\,-\frac{3}{32}\,\LogSv\,\Logw{2}\,\hplsA{w}{2}+\frac{3}{64}\,\LogSu\,\LogSw\,\hplsB{w}{2}{1}-\frac{3}{64}\,\LogSu\,\Logv{2}\,\hplsA{v}{2}-\frac{3}{128}\,\Logu{2}\,\LogSw\,\hplsA{u}{2}\\
&\,+\frac{5}{64}\,\LogSu\,\LogSw\,\hplsA{u}{3}+\frac{5}{64}\,\LogSv\,\LogSw\,\hplsB{v}{2}{1}+\frac{5}{256}\,\LogSw\,\hplsA{v}{2}\,\hplsA{w}{2}+\frac{7}{16}\,\LogSv\,\LogSw\,\hplsA{w}{3}\\
&\,-\frac{9}{128}\,\Logu{2}\,\LogSv\,\hplsA{u}{2}-\frac{9}{256}\,\LogSw\,\hplsA{u}{2}\,\hplsA{w}{2}-\frac{9}{512}\,\Logu{2}\,\LogSw\,\hplsA{w}{2}+\frac{11}{128}\,\Logv{2}\,\LogSw\,\hplsA{u}{2}\\
&\,-\frac{13}{64}\,\LogSu\,\LogSv\,\hplsB{v}{2}{1}-\frac{13}{64}\,\LogSv\,\LogSw\,\hplsB{w}{2}{1}+\frac{13}{128}\,\LogSu\,\Logv{2}\,\hplsA{w}{2}+\frac{15}{64}\,\LogSu\,\LogSv\,\hplsA{u}{3}\\
&\,+\frac{15}{64}\,\Logu{2}\,\LogSv\,\hplsA{w}{2}+\frac{19}{64}\,\LogSu\,\LogSw\,\hplsB{v}{2}{1}+\frac{25}{384}\,\Logu{2}\,\LogSv\,\hplsA{v}{2}+\frac{27}{256}\,\LogSv\,\hplsA{v}{2}\,\hplsA{w}{2}\\
&\,-\frac{37}{512}\,\LogSv\,\Logw{2}\,\hplsA{v}{2}+\frac{43}{192}\,\LogSv\,\hplsA{u}{2}\,\hplsA{v}{2}+\frac{45}{512}\,\Logv{2}\,\LogSw\,\hplsA{w}{2}+\frac{3}{16}\,\LogSu\,\LogSv\,\LogSw\,\hplsA{w}{2}\\
&\,+\frac{19}{64}\,\LogSu\,\LogSv\,\LogSw\,\hplsA{v}{2}+\frac{5}{512}\,\zeta_2\,\hplsA{u}{3}+\frac{15}{128}\,\zeta_3\,\hplsA{u}{2}-\frac{23}{128}\,\zeta_3\,\hplsA{v}{2}-\frac{293}{512}\,\zeta_2\,\hplsA{v}{3}\\
&\,+\frac{1}{16}\,\zeta_3\,\hplsA{w}{2}+\frac{1}{128}\,\zeta_2\,\hplsB{u}{2}{1}-\frac{49}{128}\,\zeta_2\,\hplsB{v}{2}{1}-\frac{9}{32}\,\zeta_2\,\hplsB{w}{2}{1}+\frac{27}{512}\,\zeta_2\,\LogSu\,\hplsA{u}{2}\\
&\,-\frac{1}{32}\,\zeta_2\,\LogSu\,\hplsA{v}{2}+\frac{3}{32}\,\zeta_2\,\LogSu\,\hplsA{w}{2}-\frac{11}{32}\,\zeta_2\,\LogSv\,\hplsA{u}{2}-\frac{283}{512}\,\zeta_2\,\LogSv\,\hplsA{v}{2}-\frac{11}{16}\,\zeta_2\,\LogSv\,\hplsA{w}{2}\\
&\,+\frac{3}{32}\,\zeta_2\,\LogSw\,\hplsA{u}{2}+\frac{1}{16}\,\zeta_2\,\LogSw\,\hplsA{v}{2}-\frac{1}{8}\,\zeta_2\,\LogSw\,\hplsA{w}{2}-\frac{1}{192}\,\LogSu\,\LogSv\,\Logw{3}\\
&\,+\frac{11}{384}\,\LogSu\,\Logv{3}\,\LogSw+\frac{23}{256}\,\LogSu\,\Logv{2}\,\Logw{2}+\frac{7}{192}\,\Logu{3}\,\LogSv\,\LogSw\\
&\,+\frac{15}{256}\,\Logu{2}\,\LogSv\,\Logw{2}+\frac{47}{256}\,\Logu{2}\,\Logv{2}\,\LogSw+\frac{1}{96}\,\Logu{3}\,\Logv{2}+\frac{5}{192}\,\Logu{2}\,\Logv{3}
\esp\eeq
\beq\bsp
\phantom{R^{1-v,u,u}}
&\,-\frac{1}{1024}\,\Logu{2}\,\Logw{3}+\frac{13}{1024}\,\Logu{3}\,\Logw{2}-\frac{5}{3072}\,\Logv{2}\,\Logw{3}+\frac{31}{1024}\,\Logv{3}\,\Logw{2}\\
&\,-\frac{45}{64}\,\zeta_4\,\LogSu-\frac{25}{32}\,\zeta_2\,\LogSu\,\LogSv\,\LogSw-\frac{3}{32}\,\zeta_2\,\LogSu\,\Logv{2}+\frac{1}{32}\,\zeta_2\,\LogSu\,\Logw{2}\\
&\,+\frac{229}{64}\,\zeta_4\,\LogSv-\frac{15}{64}\,\zeta_2\,\Logu{2}\,\LogSv+\frac{1}{4}\,\zeta_2\,\LogSv\,\Logw{2}+\frac{15}{64}\,\zeta_4\,\LogSw+\frac{3}{64}\,\zeta_2\,\Logu{2}\,\LogSw\\
&\,-\frac{3}{8}\,\zeta_2\,\Logv{2}\,\LogSw+\frac{15}{256}\,\zeta_3\,\Logu{2}+\frac{17}{1536}\,\zeta_2\,\Logu{3}-\frac{15}{256}\,\zeta_3\,\Logv{2}+\frac{119}{1536}\,\zeta_2\,\Logv{3}\\
&\,-\frac{1}{192}\,\zeta_2\,\Logw{3}\,.
\esp\eeq


\section{Logarithmic divergences on the surface $u=v$}
\label{app:LLUV}
In this appendix we show the coefficients of the
leading logarithmic divergence up to four loops on the surface $u=v$ defined
in eq.~\eqref{eq:UWLLdef}. The results are given in terms of HPLs
$H^z_{\vec m}\equiv H_{\vec m}(1-z)$.
\begin{align}
\mathcal{W}^{(2)}_2(w) &\, = \frac{1}{4}\,H_2^{w}\,,\\
\mathcal{W}^{(3)}_3(w) &\, = \frac{1}{8}\,H_3^{w} - \frac{1}{8}\,H_{2,1}^{w} \,,\\
\mathcal{W}^{(4)}_4(w) &\, = \frac{5}{96}\, H_4^{w}-\frac{7}{96}\, H_{2,2}^{w}-\frac{19}{96} \,H_{3,1}^{w}+\frac{5}{96} \,H_{2,1,1}^{w}\,,\\
\mathcal{U}^{(2)}_1(y) &\, =\frac{1}{2}\, \ln^2y\,H_1^{1-y} +\ln y\, \Big[H_{-2}^{1-y}-H_2^{1-y}+\frac{1}{2}\,\zeta_2\Big]
-H_{-2,-1}^{1-y}-H_{1,-2}^{1-y}-2 H_{-3}^{1-y}\\
&\,+H_3^{1-y}+\frac{1}{2}\, \zeta _2\, H_1^{1-y} \nonumber\,,\\
\mathcal{U}^{(3)}_2(y) &\, = \frac{1}{16}\,\ln^3y\,H_1^{1-y} + \ln^2y\Big[-\frac{3 }{8}\,H_{1,1}^{1-y}+\frac{3 }{8}\,H_{-2}^{1-y}-\frac{7 }{16}\,H_2^{1-y}\Big]
+\ln y\Big[-H_{-2,-1}^{1-y}\\
&\,-H_{1,-2}^{1-y}+\frac{3 }{4}\,H_{1,2}^{1-y}+\frac{3 }{4}\,H_{2,1}^{1-y}-\frac{1}{4} \,\zeta_2\, H_1^{1-y}-\frac{3 }{2}\,H_{-3}^{1-y}+\frac{11 }{8}\,H_3^{1-y}\Big]+2\, H_{-3,-1}^{1-y}+\frac{1}{4} \,H_{-2,-2}^{1-y}\nonumber\\
&\,+\frac{13}{8}\, H_{1,-3}^{1-y}-\frac{3 }{4}\,H_{1,3}^{1-y}+\frac{3}{2} \,H_{2,-2}^{1-y}-\frac{3 }{4}\,H_{2,2}^{1-y}-\frac{3 }{4}\,H_{3,1}^{1-y}+H_{-2,-1,-1}^{1-y}+H_{1,-2,-1}^{1-y}\nonumber\\
&\,+\frac{3}{4} \,H_{1,1,-2}^{1-y}+\frac{9 }{4}\,H_{-4}^{1-y}-\frac{15 }{8}\,H_4^{1-y}-\frac{3}{8}\, \zeta _2 \,H_{1,1}^{1-y}-\frac{1}{8}\, \zeta _3\, H_1^{1-y}-\frac{5 }{16}\,\zeta _4\,,\nonumber\\
\mathcal{U}^{(4)}_3(y) &\, = \frac{1}{288}\,\ln^4y\,H_1^{1-y}
+\ln^3y\,\Big[\frac{5 }{72}\,H_{-2}^{1-y}-\frac{1}{12}\,H_{1,1}^{1-y}-\frac{11 }{144}\,H_2^{1-y}\Big]
+\ln^2y\,\Big[\frac{5 }{12}\,H_3^{1-y}\\
&\,-\frac{5 }{12}\,H_{-3}^{1-y}-\frac{3}{8} \,H_{-2,-1}^{1-y}-\frac{3}{8} \,H_{1,-2}^{1-y}+\frac{1}{3}\,H_{1,2}^{1-y}+\frac{3 }{8}\,H_{2,1}^{1-y}+\frac{5}{24}\, H_{1,1,1}^{1-y}-\frac{1}{16}\, \zeta _2\, H_1^{1-y}\Big]\nonumber\\
&\,+\ln y\,\Big[\frac{5 }{4}\,H_{-4}^{1-y}-\frac{29 }{24}\,H_4^{1-y}+\frac{3}{2}\, H_{-3,-1}^{1-y}+\frac{9}{8} \,H_{1,-3}^{1-y}-\frac{5}{6}\, H_{1,3}^{1-y}+\frac{5}{4} \,H_{2,-2}^{1-y}-\frac{11 }{12}\,H_{2,2}^{1-y}-H_{3,1}^{1-y}\nonumber\\
&\,+H_{-2,-1,-1}^{1-y}+H_{1,-2,-1}^{1-y}+\frac{3}{4}\, H_{1,1,-2}^{1-y}-\frac{5}{12}\, H_{1,1,2}^{1-y}-\frac{5}{12}\, H_{1,2,1}^{1-y}-\frac{5}{12}\, H_{2,1,1}^{1-y}+\frac{1}{24}\, \zeta _2\, H_{1,1}^{1-y}\nonumber\\
&\,-\frac{1}{12}\, \zeta _2\, H_{-2}^{1-y}+\frac{1}{8}\, \zeta _2\, H_2^{1-y}-\frac{1}{12} \,\zeta _3 \,H_1^{1-y}\Big] %
-\frac{5 }{3}\,H_{-5}^{1-y}+\frac{19 }{12}\,H_5^{1-y}-\frac{9}{4}\, H_{-4,-1}^{1-y}+\frac{1}{3} \,H_{-2,-3}^{1-y}\nonumber\\
&\,-\frac{29}{24}\, H_{1,-4}^{1-y}+H_{1,4}^{1-y}-\frac{7}{4} \,H_{2,-3}^{1-y}+\frac{13 }{12}\,H_{2,3}^{1-y}-\frac{7}{4}\, H_{3,-2}^{1-y}+\frac{7}{6}\,H_{3,2}^{1-y}+\frac{5}{4}\,H_{4,1}^{1-y}-2\, H_{-3,-1,-1}^{1-y}\nonumber\\
&\,-\frac{1}{4} \,H_{-2,-2,-1}^{1-y}-\frac{1}{4} \,H_{-2,-1,-2}^{1-y}-\frac{13}{8} \,H_{1,-3,-1}^{1-y}-\frac{1}{4}\, H_{1,-2,-2}^{1-y}-H_{1,1,-3}^{1-y}+\frac{5}{12} \,H_{1,1,3}^{1-y} -\frac{11}{12}\, H_{1,2,-2}^{1-y}\nonumber\\
&\,+\frac{5}{12} \,H_{1,2,2}^{1-y}+\frac{5}{12}\, H_{1,3,1}^{1-y}-\frac{3}{2} \,H_{2,-2,-1}^{1-y}-H_{2,1,-2}^{1-y}+\frac{5}{12}\, H_{2,1,2}^{1-y}+\frac{5}{12} \,H_{2,2,1}^{1-y}+\frac{5}{12} \,H_{3,1,1}^{1-y}\nonumber\\
&\,-H_{-2,-1,-1,-1}^{1-y}-H_{1,-2,-1,-1}^{1-y}-\frac{3}{4} \,H_{1,1,-2,-1}^{1-y}-\frac{5}{12} \,H_{1,1,1,-2}^{1-y}+ \frac{1}{24}\, \zeta _2\, H_{1,2}^{1-y}+\frac{1}{12}\, \zeta _2 \,H_{2,1}^{1-y}\nonumber\\
&\,+\frac{5}{24}\, \zeta _2\, H_{1,1,1}^{1-y}+\frac{1}{6}\, \zeta _2 \,H_{-3}^{1-y}-\frac{1}{8} \,\zeta _2 \,H_3^{1-y}+\frac{1}{6}\, \zeta _3 \,H_{1,1}^{1-y}-\frac{1}{4}\, \zeta _3 \,H_{-2}^{1-y}+\frac{5}{24} \,\zeta _3\, H_2^{1-y}+\frac{7}{96} \,\zeta _4 \,H_1^{1-y}\,.\nonumber
\end{align}


\end{document}